\documentclass[11pt,letterpaper]{article}

\usepackage[margin=1in]{geometry}
\usepackage[T1]{fontenc}
\usepackage{setspace}
\usepackage{amsmath,amssymb,amsthm,mathtools}

\usepackage{graphicx}
\usepackage{booktabs}
\usepackage{caption}
\usepackage{float}
\usepackage{array,tabularx}
\usepackage{xltabular}

\newcolumntype{L}[1]{>{\raggedright\arraybackslash}p{#1}}
\newcolumntype{Y}{>{\raggedright\arraybackslash}X}

\usepackage{enumitem}
\usepackage{xcolor}
\usepackage{authblk}
\usepackage[super,sort&compress]{natbib}
\usepackage{xurl}
\usepackage[colorlinks,
            citecolor=blue!60!black,
            linkcolor=blue!60!black,
            urlcolor=blue!60!black]{hyperref}

\providecommand{\doi}[1]{%
  \href{https://doi.org/#1}{doi:\nolinkurl{#1}}%
}

\usepackage{cleveref}

\usepackage{tikz}
\usetikzlibrary{arrows.meta,positioning,shapes.geometric,
                decorations.pathreplacing,calc,fit,backgrounds}

\newtheorem{proposition}{Proposition}
\theoremstyle{definition}

\newcommand{\F}{\mathcal{F}}
\newcommand{\Apre}{A_{\mathrm{pre}}}
\newcommand{\tauL}{\tau_L}
\newcommand{\UCB}{\mathrm{UCB}_{0.95}}
\newcommand{\LCB}{\mathrm{LCB}_{0.95}}

\providecommand{\LevelIIARunHashRaw}{0cd4cac11153c546}
\providecommand{\LevelIIARunHash}{\texttt{\LevelIIARunHashRaw}}

\providecommand{\LevelIIAM}{1200}

\providecommand{\LevelIIAAnchorSupportCountRate}{0/1200=0.000}

\providecommand{\LevelIIAAnchorAdequateCountRate}{1073/1200=0.894}

\providecommand{\LevelIIAAnchorSelectionLimitedCountRate}{9/1200=0.008}

\providecommand{\LevelIIAAnchorInconclusiveCountRate}{118/1200=0.098}

\providecommand{\LevelIIAInjectedSupportCountRate}{1189/1200=0.991}

\providecommand{\LevelIIAInjectedAdequateCountRate}{0/1200=0.000}

\providecommand{\LevelIIAInjectedSelectionLimitedCountRate}{11/1200=0.009}

\providecommand{\LevelIIALeakageDiagnosticFailureCountRate}{1200/1200=1.000}

\providecommand{\LevelIIASelectionSelectionLimitedCountRate}{1196/1200=0.997}

\providecommand{\LevelIIASelectionDiagnosticFailureCountRate}{4/1200=0.003}

\providecommand{\LevelIIAColliderSelectionLimitedCountRate}{1199/1200=0.999}

\providecommand{\LevelIIAColliderDiagnosticFailureCountRate}{1/1200=0.001}

\providecommand{\LevelIIAAdversarialSupportCountRate}{0/1200=0.000}

\providecommand{\LevelIIAAdversarialAdequateCountRate}{1185/1200=0.988}

\providecommand{\LevelIIAAdversarialSelectionLimitedCountRate}{12/1200=0.010}

\providecommand{\LevelIIAAdversarialDiagnosticFailureCountRate}{1/1200=0.001}

\providecommand{\LevelIIAAdversarialInconclusiveCountRate}{2/1200=0.002}

\providecommand{\LevelIIAOppositeOppositeDirectionCountRate}{1191/1200=0.993}

\title{\textbf{Testing the limits of past-adapted explanations\\[4pt] by post-endpoint randomisation:\\[4pt]
anticipatory EEG as a worked case}}

\author[1]{George Sopasakis}
\author[2]{Alexandros Sopasakis\thanks{Corresponding author:
           \texttt{alexandros.sopasakis@math.lth.se}}}
\affil[1]{Research and Development Department, Ximantis AB, Onsala, Sweden}
\affil[2]{Department of Mathematics, Lund University, Lund, Sweden}

\date{}

\begin{document}
\maketitle

\begin{abstract}
\noindent
A predictive model can fit its data even when its information set is insufficient; fit alone cannot establish sufficiency. This Perspective introduces Level~II-A, a new design-based inference framework to test this distinction, illustrated in anticipatory EEG using contingent negative variation. A pre-event endpoint is committed before the delay to the imperative event is randomised. That later-assigned delay thereby becomes a negative-control probe of whether past-adapted information was sufficient for an already committed result. Under the past-adapted factorisation, accounts using only pre-commitment information cannot systematically order the endpoint by that delay. Leakage-safe preprocessing, a frozen label-blind comparator and retained-sample qualifications carry the exclusion to the confirmatory residual. A qualified material negative ordering supports conditional insufficiency without identifying a mechanism; an adequately sensitive null supports a bounded affirmative conclusion calibrated by the pipeline's false-adequacy rate. A non-compensatory rule separates these from diagnostic failure, selection-limited, opposite-direction and inconclusive outcomes. No human EEG data are analysed. In the synthetic benchmark, grid-based false-adequacy boundaries are \(15\,\mu\mathrm{V\,s^{-1}}\) for assignment isolation and \(30\,\mu\mathrm{V\,s^{-1}}\) for the sequential e-value route, in both directions. The design transfers wherever endpoint commitment precedes an exogenous label, probing sufficiency only where a declared alternative predicts ordering by it. It turns “the past explains it” from a working explanatory assumption into a magnitude-qualified, testable claim.

\end{abstract}

\section{Introduction}
\label{sec:introduction}

The brain prepares for events before they occur. Slow cortical potentials build
during a foreperiod, motor systems pre-activate, and pre-stimulus excitability
rises and falls with temporal expectation
\citep{Walter1964,Macar2004,Trillenberg2000,Brunia2012,Nobre2010,Coull2009}.
Researchers usually explain this preparation from information already available:
elapsed time, cue probability, conditional hazard, arousal and recent foreperiod
history
\citep{Niemi1981,Los2001,Miniussi1999,Vanrullen2013,Steinborn2008}.
We call an account past-adapted when every current-trial quantity it
specifies is generated from information available when the pre-event window closes at
\(t_1\). The account may also include ordinary disturbances, but the collection of
disturbances affecting the committed quantity must not systematically differ according
to the scheduler draw made afterwards. Formally, their joint probability distribution
must be independent of the current post-commitment draw conditional on the declared
randomisation stratum. Temporal measurability is therefore necessary but not sufficient:
the past-adapted class must also exclude systematic statistical dependence between the committed
quantity and the scheduler draw made afterwards. This is the usual explanatory form in
cognitive neuroscience, and it predicts many anticipatory signals well. Showing
that hazard predicts the contingent negative variation (CNV) establishes fit.
Sufficiency is a stronger claim: hazard, arousal, motor preparation and trial
history must together exhaust the systematic variation in the chosen EEG
endpoint within the declared design envelope.

The design developed in this Perspective separates these questions by generating
a randomised probe only after the EEG quantity has been fixed. In its motivating
EEG instantiation, the protocol first locks a terminal CNV-like amplitude over a
fronto-central cluster. This fixed
summary is the committed endpoint. Only then does the scheduler draw the delay
to the imperative event from its declared law, using only the registered
scheduler inputs and under concealment. That operational mechanism establishes
the post-commitment probe and is qualified by scheduler, timestamp, logging and
concealment audits.
Under a past-adapted account, the later assigned delay should therefore
carry no systematic information about the already committed endpoint before selection.
Formally, this is the statistical factorisation developed below. The assigned delay
therefore functions as a randomised negative-control exposure: a later-generated factor
that should not systematically relate to the already committed endpoint under the
past-adapted account. Later analyses may remove recorded anticipatory
structure using a comparator fixed without access to the confirmatory labels. The
trialwise remainder left by that fixed comparator is the frozen residual.
Delivery, retention, preprocessing and inclusion can still create an association,
so the retained sample and frozen analysis objects require separate conditions
below. Together, endpoint commitment at \(t_1\) and post-commitment randomisation
create an information boundary: the assigned delay is generated only after the
quantity being tested has been fixed. Testing the exclusion implied by that
boundary draws on negative controls and randomisation inference
\citep{Lipsitch2010,Shi2020,Candes2018,Freedman2008,Lin2013}. This boundary test
forms the core of the design-based inference framework we call Level~II-A. Its
target is class-level adequacy.

The design can support a positive result even when it finds no residual ordering.
If the registered analysis is sensitive enough to detect a material linear
association and finds none, it can bound what remains compatible with the data.
We call this bounded result an affirmative null. It supports the statement that
the complete registered analysis found no material linear assigned-delay ordering
at or above the simulation-qualified magnitude. Nonlinear, time-varying and
subgroup-specific features remain outside its scope.

Simulation determines the effect magnitude at which such an
adequacy claim is warranted. The benchmark adds a simulated linear assigned-delay
effect of magnitude \(\delta\) directly to the endpoint, making the endpoint
change systematically with assigned delay. It then asks how often the full
analysis nevertheless returns an adequacy
classification at that effect magnitude, despite the simulated ordering that was
deliberately inserted. This proportion is the false-adequacy rate.
The analysis may affirm adequacy only where that rate and
its Monte Carlo uncertainty both meet a tolerance fixed in advance. The benchmark
therefore yields a separate qualified magnitude for each inferential route and
direction. Below the relevant magnitude,
the article makes no adequacy claim in that direction.
This calibration gives the affirmative claim its magnitude-qualified meaning:
it turns the working phrase ``the past explains it'' into an auditable statement
with an explicit minimum effect magnitude, expressed on the native measurement
scale, and a reported false-adequacy rate.
The construction joins a wider family of equivalence, severity and evidence-of-absence
arguments \citep{AltmanBland1995,Lakens2017,MayoSpanos2006,Keysers2020}. Its wider value
is an informative, calibrated null that complements searches for exceptional
effects.

The directional alternative, a qualified negative departure, is deliberately
harder to reach. The past-adapted class is defined structurally across all
admissible fitted models. It contains linear, nonlinear, latent-state,
mechanistic and data-adaptive accounts, provided that their pre-assignment outputs
obey the registered information boundary and the required relation to the later
scheduler draw. This breadth is essential: every
account in the class must imply that the later scheduler draw cannot
systematically order the already committed endpoint or its qualified frozen
residual. The experiment can test that implication
only when the operational scheduler, temporal integrity, frozen-residual
construction and retained-sample conditions all pass their registered checks.
The operational audits qualify whether the probe was generated and concealed
as declared; they do not establish the past-adapted factorisation interrogated
by the confirmatory contrast.

When those conditions hold, no past-adapted account can generate a systematic
ordering of the committed endpoint or frozen residual by the assigned delay.
A material negative ordering can therefore support conditional insufficiency
only when the full decision rule is satisfied and within the registered endpoint,
contrast, delay range and experimental regime. Failures of randomisation,
concealment, temporal processing, delivery, residual construction or selection are
premise failures that prevent the test from adjudicating the past-adapted class and
are not counted as evidence against it. The audits qualify the conditional
interpretation within the declared envelope. A qualified negative departure supports
class-level insufficiency and remains neutral across mechanism, cause and
foreknowledge interpretations alike; those questions belong to a downstream
programme.

Level~II-A embeds that boundary test within a non-compensatory
decision architecture implemented through estimators, benchmarks, audits and
reporting qualifications. Its estimators, benchmarks and interpretations are
independent of any downstream reciprocal, dynamical or physical-substrate model.
A negative slope, a favourable result from one inferential route, or a visually
ordered residual is insufficient on its own. Failure of any required operational
audit, selection gate, collider diagnostic or participant-level estimability check
assigns the dataset to a diagnostic, selection-limited or inconclusive category.
The reporting categories in \Cref{tab:decision} keep these outcomes distinct.

The present EEG instantiation uses a terminal CNV-like endpoint, but the Level~II-A
logic is not EEG-specific. It applies to any statistic irrevocably fixed before
generation of a scientifically meaningful randomised or otherwise defensibly
exogenous label. We use the term label to emphasise its analytical role: it indexes
trials only after the endpoint being tested has already been fixed. In the present
design that label is generated by post-endpoint randomisation. More generally, a
non-randomised label could serve the same role only if its generation were defensibly
external to the processes that produced the committed endpoint. Endpoints with higher
throughput or lower noise may support more informative affirmative nulls than
slow-potential EEG. \Cref{sec:broader} sets out what transfer beyond neuroscience
would require.

The article first explains the forward-only problem and the anticipatory
processes that motivate the comparator (\Cref{sec:anticipatory-eeg}). It then
develops the post-endpoint randomisation boundary together with the
retained-sample conditions required after delivery, retention, preprocessing and
inclusion (\Cref{sec:post-endpoint-randomisation}), before turning to the
committed endpoint, forward-only comparator and participant-level estimand
(\Cref{sec:endpoint-estimand-comparator}). The inference, decision rule, audits
and omitted-pathway sensitivity analysis are presented in
\Cref{sec:decision-audits-sensitivity}, followed by synthetic validation and
practical implementation (\Cref{sec:synthetic-validation}). The discussion
considers scope and places mechanistic interpretation downstream of the
evidential conclusion (\Cref{sec:discussion}).

\section{Anticipatory EEG and the forward-only problem}
\label{sec:anticipatory-eeg}

The strongest conventional candidates for explaining an apparent assigned-delay
ordering are the anticipatory processes already known to shape pre-event EEG. The contingent
negative variation is a slow negative potential that develops between a warning
cue and an imperative event and tracks temporal expectation and preparation
\citep{Walter1964,Macar2004,Trillenberg2000,Brunia2012,NobreVanEde2018}.
Foreperiod and conditional hazard shape reaction time and preparatory activity
because elapsed time changes the conditional probability that the event is
about to occur
\citep{Niemi1981,Los2001,Nobre2010,Coull2009}. Temporal attention, arousal and
ongoing oscillatory state modulate pre-stimulus excitability
\citep{Miniussi1999,Vanrullen2013}.
Sequential foreperiod effects make current-trial preparation depend on the
foreperiod history of preceding trials \citep{Steinborn2008}.
Predictive-coding and active-inference accounts place
these effects within a broader view: the nervous system uses an internal model
and currently available information to anticipate what comes next
\citep{Friston2010,Clark2013,Hohwy2013,Millidge2021,FristonKiebel2009}.

Despite mechanistic diversity, these accounts share one defining property: every
current-trial output produced before delay assignment is past-adapted. A model
may predict a future event, yet it must form that
prediction from a neural state and information already available. Predictive
coding is no exception. A generative model may anticipate a distribution of
events and may use delays from earlier trials, but it cannot condition the
committed endpoint on a random label that does not yet exist. Under protected
post-endpoint assignment, the entire past-adapted class therefore implies the
same exclusion of the assigned delay, formalised in
\Cref{sec:post-endpoint-randomisation}.

The forward-only comparator has a narrower and more practical role. In plain
terms, it predicts the committed EEG endpoint from ordinary information available
before the later delay is assigned. It uses prospectively declared measures of
foreperiod, hazard, slow-potential development, arousal, preparation, sequential
history and session structure, and is locked before confirmatory access to the
delay label (\Cref{sec:endpoint-estimand-comparator}). Its prediction is removed
from the observed endpoint to form the frozen residual used in the confirmatory
analysis. The comparator therefore removes recorded anticipatory structure; it
does not attempt exhaustive recovery of latent neural processes and does not
establish the assigned-delay exclusion. The pre-selection exclusion instead
follows from the past-adapted factorisation under the protected post-endpoint
scheduler. For that exclusion to extend to the confirmatory residual, temporal
integrity, retained-sample delay-neutrality, retained-support positivity and
frozen-comparator independence must also hold. Unrecorded past-adapted processes
may remain in that residual, increasing variance and reducing precision, but
under those identifying conditions they cannot systematically track the later
label. By reducing variation attributable to familiar pre-assignment processes,
the comparator makes any surviving assigned-delay ordering harder to explain as
ordinary anticipatory variation.

The synthetic benchmark also tests whether the comparator is too flexible: a
known departure of the form targeted by the confirmatory test is first inserted
into the synthetic endpoint, and the comparator is then fitted without access
to the confirmatory delay labels. The full pipeline therefore tests whether that
inserted signal survives rather than being absorbed by the comparator, providing
a direct check that variance reduction does not routinely erase the departure
the analysis is designed to detect.

This architecture also reflects methodological lessons from adjacent contested
literatures on pre-event physiology and later nominally unpredictable outcomes
\citep{Mossbridge2012,Schwarzkopf2014,Kekecs2023}. Those lessons reinforce the
need for fixed endpoints, verified assignment chronology, leakage-safe
processing, prospective selection rules and independent replication. This
Perspective makes no evidential claim about the reported phenomena. It
generalises these requirements into a graded post-endpoint randomisation test
combining a locked cross-fitted comparator, an equal-participant estimand,
route-specific assignment calibration, retained-sample qualification and a
magnitude-indexed false-adequacy benchmark. Together, these components yield a
non-compensatory classifier capable of supporting either a qualified
directional departure or a bounded affirmative null.

The same prospective-locking logic also addresses circular analysis in
neuroscience and machine learning.
If researchers choose endpoint windows, channel clusters, covariates,
preprocessing rules, folds, nuisance models, tuning settings or exclusions after
viewing the contrast to be tested, the confirmatory statistic is no longer
protected from the assigned labels. This changes its information base and can
invalidate assignment-based calibration. Endpoint construction, comparator
fitting, fold formation, tuning, exclusions and fallback rules are therefore
fixed before confirmatory delay labels or delay-labelled residual diagnostics
can influence the analysis \citep{Kriegeskorte2009,VarmaSimon2006}. A pilot is
restricted to label-blind feasibility estimation; endpoint-by-delay optimisation
is excluded from confirmatory development.

The proposal therefore belongs to the design-based tradition of protected
randomisation, covariate adjustment, leakage-safe signal processing and negative
controls. The assigned delay acts as a randomised negative-control exposure
\citep{Lipsitch2010,Shi2020}. Its null does not follow from chronology alone.
The scheduler must generate the current label after endpoint commitment, from a
declared conditional law, and the label must remain outside endpoint and
comparator construction. Randomisation exogeneity and temporal integrity then
exclude a systematic pre-selection association. Delivery, preprocessing and
selection occur later, so carrying that exclusion into the analysed sample
requires the additional retained-sample and frozen-object conditions developed
below.

The assigned delay can then test whether a committed statistic respects
the information restriction that defines the past-adapted class. A qualified
material association would leave two
possibilities: either an identifying or assignment-to-analysis condition failed,
or the past-adapted class failed a necessary implication for the declared test.
The audits, selection analyses and non-compensatory rule are designed to keep
those conclusions apart.

\section{Post-endpoint randomisation as a boundary test}
\label{sec:post-endpoint-randomisation}

The design turns on one temporal order: the pre-event endpoint is committed
before the delay to the imperative event is assigned. This creates a
pre-selection boundary. Grouping trials later by their assigned labels cannot, by
itself, make an earlier fixed statistic systematically follow those labels. The
analysed sample is different. It is formed after assignment through delivery,
trigger timing, artefact rejection, missingness, preprocessing and inclusion.
The analysis must therefore distinguish the exclusion created before selection
from the conditions that preserve it after selection.

\subsection{Endpoint commitment, assignment and past-adapted information}
\label{sec:boundary-setup}

Consider trial \(j\) in a warned-foreperiod paradigm, in which a warning
cue signals that an imperative event will follow after a preparatory interval
(the foreperiod). A prospectively fixed
endpoint window \(I_{\mathrm{pre}}=[t_0,t_1]\) closes at \(t_1\). The
single-trial endpoint \(\Apre^{(j)}\) uses a locked channel or source set,
reference, baseline, temporal operation, aggregation rule and sign convention.
It uses only samples available by \(t_1\). The acquisition system records a
commit acknowledgement before it requests the trial's assigned delay. Endpoint
closure, commitment and assignment are therefore implemented as auditable
sequential operations. After that acknowledgement, the registered randomisation service selects the
assigned delay \(\tauL^{(j)}\) from the prospectively declared scheduler law
\[
P_{\mathrm{sch}}
\!\left(\mathrm{d}\tauL\mid\mathcal{R}_j\right)
\]
within the declared randomisation stratum \(\mathcal{R}_j\).
Here \(P_{\mathrm{sch}}(\,\cdot\mid\mathcal{R}_j)\) is the declared
stratum-conditional probability law of the assigned delay; probabilities and
expectations carrying the subscript \(P_{\mathrm{sch}}\) are taken under that
law.

The empirical delay support begins at a strictly positive, logged and auditable interval after endpoint commitment. A zero-valued synthetic bin denotes the analysis origin; empirical assignment and delivery begin after a strictly positive logged interval. Here \(\tauL^{(j)}\) is the post-commitment added delay from \(t_1\) to the commanded imperative event, not the entire warning-to-imperative foreperiod. Any elapsed foreperiod or protocol-derived hazard available by \(t_1\) belongs to the past-adapted information set and may enter the frozen comparator; the current post-commitment draw does not. The scheduler may use only declared pre-assignment information. These permitted inputs together form the randomisation stratum \(\mathcal{R}_j\). They include every factor used by the scheduler and every factor capable of changing its conditional law: participant, session, block, device route, scheduler state, fixed-multiset position and any planned timing condition. In the worked design, the stratum contains no current-trial endpoint, neural or behavioural quantity. A more general covariate-adaptive design could use a pre-assignment quantity provided that its use were declared prospectively and the analysis conditioned on the resulting scheduler law. The worked implementation uses a fixed stratum-conditional scheduler law determined entirely by the declared inputs listed above. A confirmatory reassignment must preserve the strata and assignment constraints that generated the observed labels.

The current assigned delay is not generated, revealed or accessible to the
endpoint, comparator or analysis pipeline before endpoint commitment. Once the
service generates it, no component may use it to revise the endpoint,
comparator, folds, residuals, eligibility rules or other confirmatory objects.
The apparatus then attempts delivery under the assigned command and records
\(\widetilde{\tau}_L^{(j)}
=t_{\mathrm{event}}^{(j)}-t_1\), where
\(t_{\mathrm{event}}^{(j)}\) is the logged onset time of the delivered
imperative event. This measured latency enters the delivery and
timing-compliance audits. The intention-to-treat primary analysis retains the
randomisation-protected assigned label \(\tauL^{(j)}\).

Let \(\{\F_t^{(j)}:t\leq t_1\}\) denote the trial-\(j\) filtration,
the increasing family of information sets generated by everything available by
time \(t\). This information includes the observed neural and physiological
history, protocol state, elapsed foreperiod, protocol-derived hazard, arousal,
motor preparation, block and session state, time on task, previous-trial
variables and any scheduler state realised before the current draw. It does not
include \(\tauL^{(j)}\), because that label has not yet been generated.
Measurability with respect to this filtration is the temporal part of
past-adaptedness, but it is not sufficient by itself: a pre-assignment quantity
could be measurable at \(t_1\) and still be statistically associated with a later
draw under a non-forward alternative. An account belongs to the declared
past-adapted class only when its pre-assignment outputs are
\(\F_t^{(j)}\)-measurable at the relevant time and its joint law with the
current scheduler draw obeys the class-level factorisation below. Nonlinear
state-space, recurrent, hierarchical-latent, predictive-coding and
machine-learning predictors all qualify when they satisfy both requirements.

The operational scheduler and the past-adapted statistical restriction must
be distinguished. Condition~\textup{(R1)} concerns the mechanism: after endpoint
commitment, the registered service selects the current delay from
\(P_{\mathrm{sch}}(\cdot\mid\mathcal{R}_j)\), using only the declared
scheduler inputs and under prospective concealment. Under a past-adapted account,
that mechanism additionally entails, for every Borel set \(D\) in the declared
delay support,
\begin{equation}
\label{eq:assignment-law}
\mathbb{P}
\!\left\{
\tauL^{(j)}\in D
\mid
\F_{t_1}^{(j)}
\right\}
=
P_{\mathrm{sch}}
\!\left(D\mid\mathcal{R}_j\right)
\quad\text{almost surely.}
\end{equation}
Equivalently,
\begin{equation}
\label{eq:tau-filtration-independence}
\tauL^{(j)}
\ \perp\!\!\!\perp\
\F_{t_1}^{(j)}
\ \big|\
\mathcal{R}_j .
\end{equation}
Equations~\eqref{eq:assignment-law}--\eqref{eq:tau-filtration-independence}
state the class-level forward-only factorisation implied by every past-adapted
account, not an operational audit result. The operational scheduler that
generates the probe, the audits that qualify it, and the confirmatory role of
this factorisation are set out with the design conditions in
\Cref{sec:conditions}.

\begin{figure}[t]
\centering
\begin{tikzpicture}[
  >=Stealth,
  font=\small,
  box/.style={
    draw,
    rounded corners=2pt,
    inner sep=5pt,
    align=center,
    minimum height=0.95cm,
    line width=0.85pt
  },
  lab/.style={
    font=\scriptsize\itshape,
    align=center
  },
  note/.style={
    font=\scriptsize,
    align=center
  }
]

\draw[very thick,->] (-0.4,0) -- (15.8,0);

\foreach \x/\t in {
  0/$t_0$,
  4.8/$t_1$,
  8.8/assignment,
  13.2/$t_{\mathrm{event}}$
}{
  \draw (\x,0.12) -- (\x,-0.12);
  \node[below=2pt] at (\x,-0.12) {\t};
}

\node[
  box,
  draw=blue!65!black,
  fill=blue!22,
  minimum width=4.65cm
] (prebox) at (2.40,0.76)
  {\footnotesize committed endpoint\\[-1pt]
   \footnotesize \(\Apre\) fixed at \(t_1\)};

\node[lab] at (2.35,-0.56)
  {past-adapted information \(\F_{t_1}\)};

\draw[dashed,line width=0.9pt]
  (4.8,-0.18) -- (4.8,1.66);

\node[
  lab,
  anchor=south east,
  text width=1.8cm
] at (4.60,1.46)
  {logged commit\\barrier};

\node[
  box,
  draw=orange!65!black,
  fill=orange!18,
  minimum width=3.15cm
] (rand) at (8.8,2.12)
  {randomise \(\tauL\)\\[-1pt]
   \footnotesize after commit};

\node[lab] at (8.8,2.77)
  {within declared stratum \(\mathcal{R}_j\)};

\draw[->,thick]
  (rand.south) -- (8.8,0.18);

\draw[
  ->,
  dashed,
  red!70!black,
  line width=0.95pt
]
  (rand.west)
  .. controls (7.20,1.60) and (5.65,1.60) ..
  (prebox.north east);

\node[
  lab,
  text=red!70!black,
  inner sep=1.0pt,
  text width=2.15cm
] at (6.00,2.28)
  {excluded\\pre-selection ordering};

\node[
  text=red!70!black,
  font=\bfseries\small,
  inner sep=0.3pt
] at (6.85,1.48)
  {\(\times\)};

\node[
  box,
  draw=black,
  fill=green!35,
  minimum width=2.25cm
] (event) at (13.2,1.05)
  {imperative\\[-1pt]event};

\draw[<->,line width=0.8pt]
  (4.95,-1.00)
  --
  node[
    lab,
    fill=white,
    inner sep=1.2pt,
    above=2pt
  ]
  {assigned delay \(\tauL\)}
  (13.05,-1.00);

\node[
  note,
  anchor=north
] at (9.0,-1.22)
  {measured latency \(\widetilde{\tau}_L\) audited separately};

\end{tikzpicture}

\caption{Post-endpoint randomisation boundary. The single-trial endpoint
\(\Apre\) is committed over \(I_{\mathrm{pre}}=[t_0,t_1]\) using only
pre-assignment information \(\F_{t_1}\). The current trial's assigned delay
\(\tauL\) is generated only after a logged commitment barrier by the
operational scheduler in \textup{(R1)}. Under the past-adapted factorisation and
\textup{(R2)}, conditional on the declared randomisation stratum, the committed
endpoint cannot be systematically ordered by the subsequently assigned label before selection.
Delivery, preprocessing, retention and inclusion occur after assignment and
require the separate retained-sample condition \(\mathrm{(R3^\star)}\).
Measured latency \(\widetilde{\tau}_L\) is used for compliance auditing; the
confirmatory estimand remains indexed by the assigned label.}
\label{fig:boundary}
\end{figure}
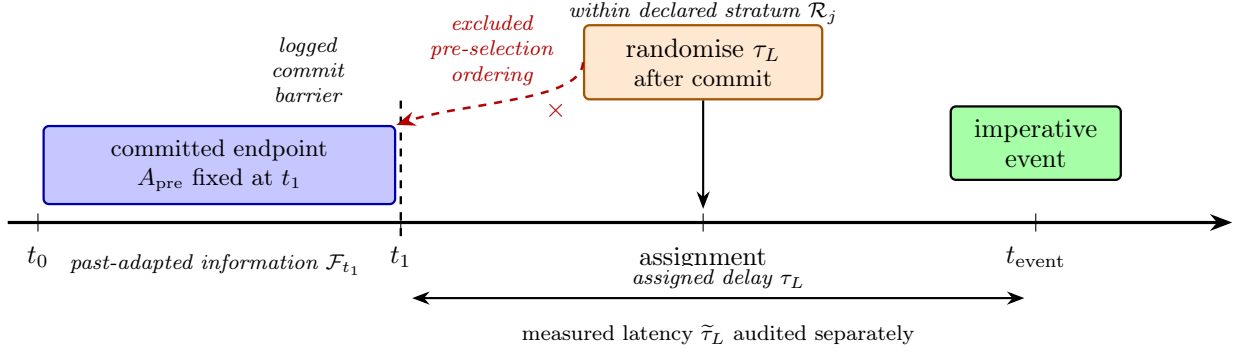

\subsection{Design conditions \textup{(R1)}--\textup{(R3)}}
\label{sec:conditions}

The boundary rests on three declared conditions. The operational parts of
\textup{(R1)}--\textup{(R2)} are audited directly; the statistical and
retained-sample restrictions are qualified by the route-specific construction,
diagnostics and sensitivity analyses below.

\paragraph{\textup{(R1)} Post-endpoint scheduler and concealment.}
The system requests the current assignment only after acknowledging endpoint
commitment. The registered service selects the delay from the declared scheduler
law within \(\mathcal{R}_j\), using no input beyond the prospectively declared
scheduler state. The label is concealed from every operation that could revise
the committed endpoint, comparator, folds, residuals, eligibility rules or other
confirmatory objects. The study retains scheduler states, seeds or other
verifiable randomisation records, request and response timestamps, and command
logs. Where declared in advance, an independent implementation or command-path
swap tests whether an association depends on one assignment implementation.
Condition~\textup{(R1)} is an operational mechanism condition. It does not
assert that the current draw is independent of the entire pre-assignment
filtration under every hypothesis; that class-level statistical restriction is
\Cref{eq:assignment-law,eq:tau-filtration-independence} under past-adapted
sufficiency.

\paragraph{\textup{(R2)} Temporal integrity of the endpoint and comparator.}
The confirmatory endpoint, its comparator covariates and every rule that
constructs them may use only information available by \(t_1\). More strictly,
an output assigned to time \(t\leq t_1\) may depend only on samples acquired at
\(u\leq t\). This excludes zero-phase, bidirectional, forward--reverse,
symmetric and time-shift-corrected filtering from confirmatory endpoint and
covariate construction. Such operations can smear later activity across the
commitment boundary and distort apparent timing
\citep{Rousselet2012,Widmann2015,deCheveigne2019}.

Any data-adaptive preprocessing object used confirmatorily must be frozen from an independent calibration sample or updated causally using information available by the relevant time. A whole-session transform satisfies strict causality only when its fitted parameters exclude later samples, even when the transform is label blind. Rules governing endpoint eligibility before \(t_1\) are fixed as part of the endpoint specification. Delivery failure, artefact rejection, missingness and analysis inclusion occur after assignment and therefore enter \textup{(R3)} as selection processes rather than as temporal-integrity violations. A prospective audit challenges the pipeline with controlled post-\(t_1\) impulses, steps, ringing patterns and event-locked transients. Audit passage requires every resulting committed pre-event output to remain within the declared tolerance.

\paragraph{\textup{(R3)} Assignment-to-analysis integrity.}
\textup{(R3)} requires that the endpoint-level no-ordering implication in
\Cref{eq:fwd-null} survive the passage from the pre-selection population to the
retained confirmatory sample. The operational scheduler may remain fully intact
while delivery, retention, preprocessing or inclusion breaks this retained-sample
factorisation. Unlike the scheduler and temporal operations in \textup{(R1)} and
\textup{(R2)}, retained-sample delay-neutrality is not directly enforceable by an
operational procedure, and the analysis therefore proceeds through a sufficient
condition for it. Under the past-adapted factorisation, the pre-selection endpoint
is conditionally unordered by the assigned label; retaining only some trials can
break that implication. Let
\(S^{(j)}=1\) indicate that trial \(j\) enters the confirmatory sample, and
define
\(\mathcal{Z}_j:=
(\Apre^{(j)},\mathbf{X}_j)\),
where \(\mathbf{X}_j\) is the locked vector of past-adapted comparator
covariates. Comparator construction also uses analysis objects shared across trials:
locked folds, nuisance-fitting and tuning rules, fitted nuisance
functions and residual-construction objects.
We collect them in the frozen comparator object
\(\mathcal{G}_{\mathrm{frz}}\), defined formally in SI~\S1. The SI gives the
following sufficient retained-sample condition:
\begin{equation}
\label{eq:r3star-main}
\mathrm{(R3^\star)}
\qquad
S^{(j)}
\ \perp\!\!\!\perp\
\tauL^{(j)}
\ \big|\
\bigl(
\mathcal{Z}_j,
\mathcal{R}_j,
\mathcal{G}_{\mathrm{frz}}
\bigr).
\end{equation}
It is used together with retained-support positivity and the
frozen-comparator independence condition stated in SI~\S1. In plain terms,
once the endpoint, covariates, stratum and frozen comparator are held fixed,
retention must remain possible and must not depend on the assigned delay.

Condition~\eqref{eq:r3star-main} provides a sufficient criterion for
retained-sample exclusion; alternative sufficient conditions may exist, and
randomisation alone leaves \(\mathrm{(R3^\star)}\) unestablished. Under \textup{(R1)}--\textup{(R2)}, the pre-selection past-adapted
factorisation, \(\mathrm{(R3^\star)}\), retained-support positivity and
frozen-comparator independence, Lemma~S1 in SI~\S1 extends the exclusion result
to the retained endpoint and frozen residual. Selection through
delivery, trigger timing, artefact rejection, missingness, preprocessing or
inclusion can induce endpoint--delay dependence by conditioning on a common
effect of the endpoint and assigned delay. Balanced marginal retention rates
remain compatible with this form of selection.

Intention-to-treat indexing, delivery and retention audits, auxiliary negative
controls, implementation checks and the endpoint-by-delay collider diagnostic
qualify reliance on \(\mathrm{(R3^\star)}\) within the declared sensitivity
envelope. For an already resolved material departure, the scalar
selection-sensitivity gate additionally tests whether the declared marginal
selection pathway could account for that departure; failure of the applicable
gate yields a selection-limited outcome. These checks do not establish
\(\mathrm{(R3^\star)}\) globally. Nor does the trial-level condition determine
which participants retain sufficient assigned-delay information for slope
estimation; that separate selection problem is addressed by the
participant-level estimability analysis.

\subsection{What the boundary excludes}
\label{sec:exclusion}

\begin{proposition}[Endpoint exclusion under the past-adapted factorisation]
\label{prop:exclusion}
Let \(Y^{(j)}\) be an integrable real-valued statistic and let
\(\mathcal{R}_j\) be the randomisation stratum.
\begin{enumerate}[leftmargin=2.2em,itemsep=2pt]
\item If \(Y^{(j)}\) is \(\F_{t_1}^{(j)}\)-measurable and the past-adapted
factorisation in \Cref{eq:assignment-law} holds, then
\[
Y^{(j)}
\ \perp\!\!\!\perp\
\tauL^{(j)}
\ \big|\
\mathcal{R}_j,
\]
and therefore
\begin{equation}
\label{eq:fwd-null}
\mathbb{E}
\!\left[
Y^{(j)}
\mid
\sigma\!\left(\tauL^{(j)}\right),
\mathcal{R}_j
\right]
=
\mathbb{E}
\!\left[
Y^{(j)}
\mid
\mathcal{R}_j
\right]
\quad\text{almost surely.}
\end{equation}
Here \(\sigma(\tauL^{(j)})\) denotes the information generated by the current
assigned-delay label.

\item Under the operational scheduler in \textup{(R1)}, temporal integrity in
\textup{(R2)} and past-adapted sufficiency, the factorisation applies to the
committed endpoint \(Y^{(j)}=\Apre^{(j)}\) and to any prospectively fixed
trialwise transform of pre-\(t_1\) information whose definition and parameters
remain independent of the assignment vector.

\item Equation~\eqref{eq:fwd-null} is a pre-selection statement. Retained-sample
exclusion for the committed endpoint and frozen residual follows only under the
conditions of SI~\S1, Lemma~S1, including
\(\mathrm{(R3^\star)}\), positivity and frozen-comparator independence.
\end{enumerate}
\end{proposition}

\noindent \Cref{prop:exclusion} states the endpoint-level implication of the
past-adapted factorisation under the declared scheduler. Following the
operational-versus-statistical separation of \Cref{sec:conditions}, a qualified
negative ordering rejects that forward-only implication while the audited
scheduler mechanism remains intact, subject to temporal integrity and the
frozen-object and retained-sample conditions. This implication applies
independently of whether a fitted comparator captures every neural cause of
anticipation. Extending the
result to a residual requires an additional construction argument because
nuisance fitting can transmit endpoint information across trials. In the
assignment-isolation route, the comparator, folds, tuning rule and residual array
are frozen using information that excludes assigned labels, delay-labelled
diagnostics and assignment descendants capable of invalidating the admissible
reassignment law.
In the sequential route, prequential nuisance fitting can make the current
residual measurable directly from the current pre-assignment history.
Participant-disjoint fitting instead conditions on a frozen nuisance object
trained outside the participant's fold; validity then requires that this
external-fold object not alter the current participant's declared scheduler law
(SI~\S1, SI~\S3).

Sufficiency is class-relative and contrast-specific here. At the population
level, every past-adapted account must satisfy the factorisation in
\Cref{eq:assignment-law,eq:fwd-null} and, for retained confirmatory objects, its
qualified retained-sample counterpart.
At the dataset level, adequacy requires the complete decision architecture to
qualify. The two-directional classifier, applicable audits, retained-sample
qualifications and magnitude-indexed false-adequacy benchmarks must all meet
their declared criteria. A qualified material negative ordering has the
opposite meaning: it supports conditional insufficiency of the past-adapted class for the
declared endpoint, linear contrast, delay support and regime.

The substantive alternative is a restricted subset of non-factorising
joint-history processes. It preserves the operational scheduler mechanism,
concealment, temporal integrity, frozen-object invariance and the declared
retained-sample procedures, while violating the past-adapted factorisation in
\Cref{eq:assignment-law,eq:fwd-null}. In addition, it predicts the prespecified
negative assigned-delay ordering in the confirmatory statistic.

This directional requirement is important. Non-factorisation alone is broader:
it need not produce a negative linear ordering, and it may instead appear as a
positive, nonlinear, time-varying or otherwise different dependence. The tested
alternative is therefore the part of the non-factorising class that predicts the
declared negative contrast.

The inferential distinction is between an intact operational assignment
mechanism and failure of the statistical factorisation implied by the
past-adapted class. Conditional on the remaining identifying qualifications, a
qualified negative ordering violates a necessary implication of that class.

A qualified result in the prespecified negative direction therefore excludes
the declared past-adapted class for the tested endpoint, contrast, delay support
and regime. The inference remains mechanism-neutral across physical, dynamical
and foreknowledge interpretations. Comparison among restricted non-past-adapted joint-history
models is reserved for a downstream Level~II-B programme, which would begin only
after independent replication of a Level~II-A class-level departure.

One convention governs the remainder. Passage of an audit, gate or diagnostic
licenses interpretation of the contrast within its declared operational and
sensitivity envelope; the relevant independence condition remains unestablished
beyond that envelope. This convention applies to every statement of support,
every affirmative null and every passed audit. A dataset that fails to qualify
is classified as a diagnostic failure, a selection limitation or an
inconclusive result. Each of these classifications precludes both support and a
clean affirmative null.

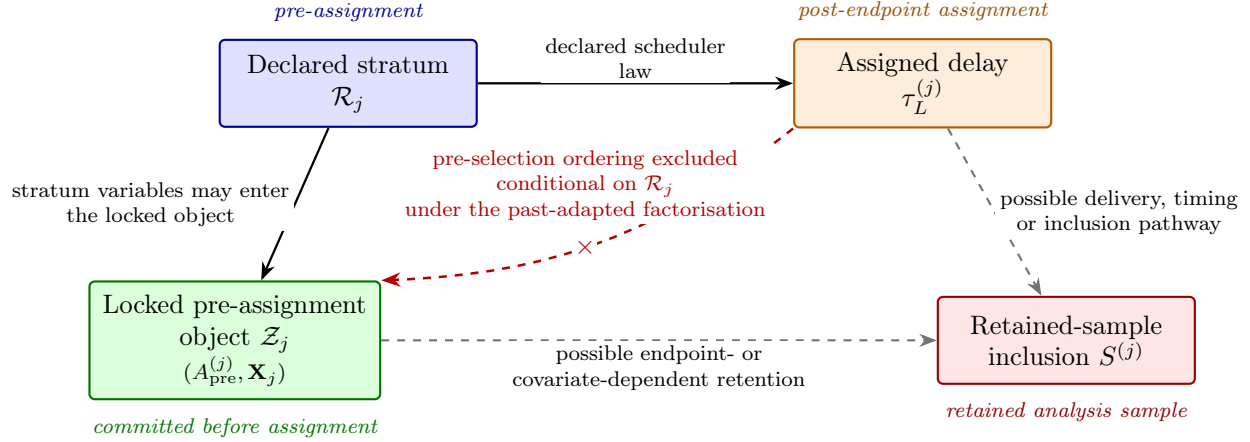
\begin{figure}[t]
\centering
\begin{tikzpicture}[
  >=Stealth,
  font=\small,
  box/.style={
    draw,
    rounded corners=2pt,
    align=center,
    inner sep=5pt,
    minimum width=3.4cm,
    minimum height=1.15cm,
    line width=0.8pt
  },
  edgelab/.style={
    font=\scriptsize,
    fill=white,
    inner sep=1.6pt,
    align=center
  },
  ann/.style={
    font=\scriptsize\itshape,
    text=black!70
  },
  design/.style={
    draw,
    ->,
    line width=0.85pt
  },
  threat/.style={
    draw=black!55,
    ->,
    dashed,
    line width=0.8pt
  },
  forbidden/.style={
    draw=red!70!black,
    ->,
    dashed,
    line width=0.95pt
  }
]

\node[
  box,
  draw=blue!60!black,
  fill=blue!12
] (R) at (0,3.4)
  {Declared stratum\\ \(\mathcal{R}_j\)};

\node[
  box,
  draw=orange!70!black,
  fill=orange!14
] (tau) at (7.6,3.4)
  {Assigned delay\\ \(\tauL^{(j)}\)};

\node[
  box,
  draw=green!45!black,
  fill=green!14
] (Z) at (-1.5,0)
  {Locked pre-assignment\\ object \(\mathcal{Z}_j\)\\[1pt]
   {\scriptsize \((\Apre^{(j)},\mathbf{X}_j)\)}};

\node[
  box,
  draw=red!60!black,
  fill=red!10
] (S) at (9.5,0)
  {Retained-sample\\ inclusion \(S^{(j)}\)};

\node[ann,text=blue!55!black,above=1mm of R]
  {pre-assignment};

\node[ann,text=orange!60!black,above=1mm of tau]
  {post-endpoint assignment};

\node[ann,text=green!45!black,below=1mm of Z]
  {committed before assignment};

\node[ann,text=red!55!black,below=1mm of S]
  {retained analysis sample};

\draw[design]
  (R) --
  node[edgelab,above]
  {declared scheduler\\law}
  (tau);

\draw[design]
  (R) --
  node[edgelab,left]
  {stratum variables may enter\\the locked object}
  (Z);

\draw[threat]
  (Z) --
  node[edgelab,below]
  {possible endpoint- or\\covariate-dependent retention}
  (S);

\draw[threat]
  (tau) --
  node[edgelab,right]
  {possible delivery, timing\\or inclusion pathway}
  (S);

\draw[forbidden]
  (tau.south west)
  to[bend left=18]
  coordinate[pos=0.54] (forbidmid)
  (Z.north east);

\node[
  edgelab,
  text=red!70!black,
  fill=white,
  above=3mm of forbidmid
]
  {pre-selection ordering excluded\\conditional on \(\mathcal{R}_j\)\\under the past-adapted factorisation};

\node[
  text=red!70!black,
  font=\bfseries\small,
  fill=white,
  inner sep=0.3pt
] at (forbidmid)
  {\(\times\)};

\end{tikzpicture}

\caption{Schematic design and selection-threat graph for the
committed-endpoint test. The declared stratum \((\mathcal{R}_j)\) contains the
observed pre-assignment scheduler state used by the declared assignment law;
the exogenous randomisation
source is omitted from the schematic. The locked object
\(\mathcal{Z}_j=(\Apre^{(j)},\mathbf{X}_j)\) is built from information
available by endpoint commitment. Solid arrows show design-permitted
dependence.
The red crossed relation denotes the pre-selection no-ordering restriction:
under the past-adapted factorisation and temporal integrity in \textup{(R2)},
\(\mathcal{Z}_j\) is not systematically ordered by the subsequently assigned
\(\tauL^{(j)}\), conditional on \(\mathcal{R}_j\). Grey dashed arrows show possible
post-assignment selection pathways into \(S^{(j)}\). If both pathways operate,
conditioning on \(S^{(j)}=1\) can open the collider path
\(\mathcal{Z}_j\to S^{(j)}\leftarrow\tauL^{(j)}\).
Condition~\(\mathrm{(R3^\star)}\) requires inclusion to be delay-neutral after
conditioning on \(\mathcal{Z}_j\), \(\mathcal{R}_j\) and the frozen comparator
object. The graph is schematic and omits
\(\mathcal{G}_{\mathrm{frz}}\), possible direct effects of
\(\mathcal{R}_j\) on retention, participant-level estimability and other
recorded causes for readability.}
\label{fig:dag}
\end{figure}

The retained sample therefore requires a distinct justification. The operational
scheduler establishes the post-commitment probe before selection; the
past-adapted factorisation supplies its pre-selection no-ordering implication.
Identification in the analysed sample additionally depends on the
assignment-to-analysis chain through delivery, trigger timing, artefact
rejection, missingness, preprocessing and inclusion. The locked audits and
sensitivity analyses qualify whether that chain has opened a selection path.
Failure of an operational integrity check, including delivery,
temporal-integrity or auxiliary negative-control qualification, yields a
diagnostic-failure outcome. Failure of the retention qualification or firing of
the endpoint-by-delay collider diagnostic yields a selection-limited outcome.
For an already resolved material departure, failure of the applicable scalar
selection-sensitivity gate likewise yields a selection-limited outcome. These
classifications block directional support; the operational, retention and
collider qualifications also determine eligibility for a clean affirmative
null, whereas the scalar selection-sensitivity gate is not an
affirmative-null gate.

The forward-only comparator has a separate role. It removes recorded
forward-accessible structure, including foreperiod, protocol-derived hazard,
arousal, CNV-like activity, sequential history and time on task, and can thereby
improve precision. Identification derives from the operational scheduler, the
past-adapted factorisation, temporal integrity, frozen-object construction and
assignment-to-analysis conditions.

Comparator flexibility creates asymmetric risks, treated in full in
\Cref{sec:comparator}: under the qualified forward-only factorisation a locked
label-blind comparator cannot systematically induce an assigned-delay ordering,
whereas a non-factorising alternative that also orders comparator inputs may be
partly attenuated. Leakage, comparator instability or post-result retuning
constitutes a failure of the protected construction and yields a diagnostic-failure outcome.

\section{Endpoint, estimand and comparator}
\label{sec:endpoint-estimand-comparator}

This section fixes the three objects that must be declared before confirmatory
assigned-delay access: the committed pre-event endpoint, the forward-only
comparator, and the participant-level estimand. The lock has temporal and
procedural components. The numerical endpoint for trial \(j\) is committed at
\(t_1\). The endpoint rule, comparator specification, fold construction,
estimand and eligibility rules are fixed prospectively, with delay-labelled
results excluded from any subsequent revision.

\subsection{Committed pre-event endpoint}
\label{sec:endpoint}

The endpoint \(\Apre^{(j)}\) is a prospectively locked, signed,
baseline-corrected single-trial scalar feature computed from a committed
pre-event window and electrode or source set \citep{Luck2014,Blankertz2011}.
The specification fixes the recording modality, sampling rate, channel or
source set, reference, baseline interval, endpoint window, temporal operation,
channel aggregation, sign convention, units, missing-data rule and
pre-\(t_1\) trial-eligibility criteria. Criteria requiring post-\(t_1\)
information belong to post-assignment retention operations and receive the
selection qualifications specified in \Cref{sec:sensitivity}.

Confirmatory preprocessing is strictly causal. An output assigned to time
\(t\leq t_1\) is restricted to samples acquired at times \(u\leq t\), and the
same information boundary governs filter-state initialisation, detrending,
baseline correction, interpolation, artefact modelling and latency correction.
Zero-phase, forward--reverse and other acausal operations are confined to
exploratory robustness analyses; the confirmatory endpoint and comparator
exclude them. Online evaluation of the endpoint at \(t_1\) is optional. Any
later computation must reproduce the frozen rule using only samples available
by \(t_1\) and must remain blind to the assigned-delay labels.

For concreteness, the worked neuroscience anchor is a terminal CNV-like
pre-assignment slow-potential amplitude measured after a warning cue over a
fronto-central preparatory site and aligned to endpoint closure \(t_1\), before
the trial-specific delay to the imperative event is drawn. The planning anchor
is the baseline-corrected mean voltage over FCz and Cz during the final
\(250\,\mathrm{ms}\) before \(t_1\), with the sign reversed so that larger
\(\Apre\) denotes greater preparatory negativity. A confirmatory protocol may
adopt a different scientifically justified cluster or source summary, provided
that it names the exact channels or sources, window, baseline and aggregation
rule before delay-label access and requalifies the complete pipeline for that
specification.

The endpoint is selected because the CNV is a familiar anticipatory slow
potential with established sensitivity to foreperiod, temporal hazard,
expectation, preparation and task history
\citep{Walter1964,Macar2004,Trillenberg2000,Brunia2012}. The design assigns the
endpoint no prior anomalous status. The synthetic residual scale
\(1\,\mu\mathrm{V}\) serves simulation calibration only and carries no empirical
claim. In a neuroscience implementation, the label-blind single-trial residual
scale, retained-trial yield, participant heterogeneity, timing error and
delivery compliance are estimated in an independent calibration dataset or a
pilot whose endpoint and preprocessing are fixed before inspection of any pilot
endpoint-by-delay association. Those quantities determine the declared
resolution floor, participant target and trial target before the confirmatory
labels are opened.

\subsection{Forward-only comparator}
\label{sec:comparator}

The forward-only comparator is the prospectively specified high-capacity
past-adapted predictor used to absorb recorded anticipatory structure, improve
precision and quantify adjustment sensitivity. It does not supply the assignment
exclusion and need not be a complete causal model of anticipation; identification
instead derives from the operational scheduler, the past-adapted factorisation,
temporal integrity, and the retained-sample, frozen-object and route-specific
conditions required for the confirmatory analysis. Its covariate class may include elapsed foreperiod,
protocol-derived hazard, scheduler and block probabilities known before
\(t_1\), a disjoint-window CNV-like feature, arousal or vigilance proxies,
motor-preparation proxies, sequential trial history, reaction-time history,
time on task, preprocessing-state summaries, participant and session structure,
and prospectively declared interactions and nonlinear transforms. Previous
assigned delays may enter only as already realised history for later trials and
only under a route and fold construction that preserve the required
pre-assignment measurability. Every comparator input must be measurable with
respect to \(\F_{t_1}^{(j)}\). The current trial's assigned-delay label, realised
event latency, post-assignment delivery information and deterministic transforms
of those quantities are prohibited.

Let \(\mathbf{X}_j\) denote the locked covariate vector. Under the declared
squared-error working loss, the population comparator target is
\begin{equation}
\label{eq:comparator-target}
m_0(\mathbf{X}_j)
=
\mathbb{E}\!\left[\Apre^{(j)}\mid\mathbf{X}_j\right].
\end{equation}
The fitted family may be penalised regression, gradient boosting, random
forests or another prospectively specified supervised predictor
\citep{Bishop2006,Hastie2009}. The family, feature dictionary, standardisation,
imputation, tuning rule, stopping rule and loss are fixed before confirmatory
label access. Any alternative loss requires an explicitly stated population
target and residual construction. 

Cross-fitting follows route-specific partition rules. In the
assignment-isolation route, outer folds preserve whole randomisation blocks or
other assignment-relevant clusters. In the sequential route, folds are
prequential or participant-disjoint, restricting the nuisance prediction for
trial \(j\) to information that excludes endpoint-derived descendants of that
trial's assignment.
All data-adaptive preprocessing, feature selection and hyperparameter tuning
occur inside the relevant training partition. Each held-out trial or fold is
predicted by a nuisance fit constructed without information from that held-out
unit, following the separation principle used in machine-learning-assisted
covariate adjustment \citep{LuEtAl2025}.

For trial \(j\) in outer fold \(k(j)\), let
\(\widehat m_0^{\,(-k(j))}\) denote the nuisance predictor fitted without
that fold. The confirmatory residual is
\begin{equation}
\label{eq:residual}
\Apre^{\mathrm{resid},(j)}
=
\Apre^{(j)}
-
\widehat{m}_0^{\,(-k(j))}(\mathbf{X}_j).
\end{equation}
Held-out predictions and residuals are generated once and frozen before
confirmatory delay-label access. Route-specific inferential calibration uses
that frozen construction throughout: refitting, retuning, fold changes,
preprocessing changes and covariate reselection are prohibited. Residualisation
serves precision and robustness. Identification derives from the operational scheduler together with the
route-specific forward-only factorisation and retained-sample conditions. The
assignment-isolation route uses the admissible reassignment law; the sequential
route uses the declared current-trial conditional scheduler law
\citep{Freedman2008,Lin2013,ZhaoDing2021}.

Under the past-adapted factorisation, the qualified retained-sample and
frozen-object conditions, and the selected route's requirement that conditioning
on the frozen analysis object preserve the relevant scheduler law, a locked
label-blind comparator cannot systematically create an assigned-delay ordering.
Under these conditions, the conditional law of the assigned delay remains the
declared scheduler law after conditioning on the objects required by the
analysis. In particular,
\[
\mathbb{E}\!\left[
\tauL^{(pj)}
\mid
\mathcal{R}_{pj},
\mathcal{G}_{\mathrm{frz}},
S^{(j)}=1
\right]
=
\mathbb{E}_{P_{\mathrm{sch}}}
\!\left[
\tauL
\mid
\mathcal{R}_{pj}
\right]
\]
for the assignment-isolation conditioning set, with the corresponding
route-specific conditioning set used for the sequential construction.
The centred assigned delay is therefore orthogonal in population to a fixed
comparator error measurable with respect to the qualified pre-assignment
object. Such comparator error has zero conditional contribution to the
population slope and cannot systematically tilt that population target.

Conditional on the randomisation stratum, the frozen comparator object and
inclusion in the qualified retained sample, a fixed comparator error contributes
to the population slope through
\[
\mathbb{E}\!\left[
\left\{
\tauL^{(pj)}-
\mathbb{E}_{P_{\mathrm{sch}}}
\!\left[\tauL\mid\mathcal{R}_{pj}\right]
\right\}
\left\{
\widehat m_0^{\,(-k(j))}(\mathbf{X}_j)-m_0(\mathbf{X}_j)
\right\}
\middle|
\mathcal{R}_{pj},
\mathcal{G}_{\mathrm{frz}},
S^{(j)}=1
\right],
\]
which is zero under the qualified forward-only factorisation and the
route-specific scheduler-law invariance condition. Thus comparator
misspecification alone neither creates a population null departure nor
attenuates an endpoint-level injection constructed independently of the
comparator covariates. Finite-sample delay--covariate associations remain
possible and are handled by cross-fitting, freezing and route-specific
calibration.

This protection is class-relative. A non-factorising alternative may also
induce dependence between assigned delay and one or more comparator inputs or
the comparator prediction, in which case shared ordered structure can be
absorbed and population attenuation need not be zero. The false-adequacy
benchmark therefore qualifies the declared covariate-orthogonal endpoint-level
injection family, not every possible joint alternative. The confirmatory target
remains the frozen residual, while the unadjusted committed-endpoint slope is
reported as an adjustment-sensitivity diagnostic that can reveal substantial
attenuation or disagreement. Identification comes from the scheduler mechanism,
the factorisation and their route-specific qualifications, not from comparator
completeness.

\subsubsection*{Concrete covariate constructions}

Every declared construction below is \(\F_{t_1}^{(j)}\)-measurable and excludes
the current realised assigned-delay draw. The complete covariate class,
interactions, boundary conventions and fallback hierarchy are specified in
SI~\S3, Table~S2.

First, a CNV-like slow-potential covariate is obtained from a pre-endpoint window
that is disjoint from the confirmatory endpoint support. Let
\(V_c^{(j)}(t)\) be the causally filtered, baseline-corrected potential at the
prospectively declared fronto-central summary, and let
\(W_{\mathrm{cnv}}\subset[t_0,t_1)\) be the locked sample set. With \(\bar t\)
and \(\bar V_c^{(j)}\) denoting the corresponding sample means, define
\begin{equation}
\label{eq:cnv-cov}
x_{\mathrm{cnv}}^{(j)}
=
\frac{
\sum_{t\in W_{\mathrm{cnv}}}
(t-\bar t)\bigl(V_c^{(j)}(t)-\bar V_c^{(j)}\bigr)
}{
\sum_{t\in W_{\mathrm{cnv}}}(t-\bar t)^2
},
\end{equation}
provided the denominator is positive. This is an ordinary-least-squares summary
of slow-potential build-up before the terminal endpoint window
\citep{Walter1964,Trillenberg2000}. Disjoint support functions as a prospective
sensitivity safeguard by preventing the comparator from reproducing the
endpoint through reuse of the same samples. Identification derives from the
declared assignment and analysis conditions rather than from this safeguard.

Second, a protocol-derived hazard covariate represents the event-time
expectation available before the current delay draw. Let \(G\) be the declared
marginal distribution of cue-to-imperative-event time induced by the known
foreperiod schedule and the marginal post-endpoint scheduler law, and let \(g\)
be its density where one exists. Both quantities are constructed from the
declared marginal timing law without using the realised delay for trial \(j\).
At the elapsed cue time
\(e^{(j)}=t_1-t_{\mathrm{cue}}^{(j)}\), the instantaneous and integrated hazards
are
\begin{equation}
\label{eq:hazard-cov}
\lambda^{(j)}
=
\frac{g\!\left(e^{(j)}\right)}{1-G\!\left(e^{(j)}\right)},
\qquad
\Lambda^{(j)}
=
-\log\!\left(1-G\!\left(e^{(j)}\right)\right),
\end{equation}
on the prospectively defined region where
\(1-G(e^{(j)})>0\). For discrete or mixed timing laws, the protocol uses the
corresponding conditional event probability or cumulative hazard with a fixed
boundary convention. These quantities encode protocol-level event imminence;
the trial-specific post-endpoint draw contributes no information to their
construction \citep{Niemi1981,Nobre2010,Coull2009}.

Third, sequential foreperiod history is represented by the prospectively locked
pair
\begin{equation}
\label{eq:seq-cov}
x_{\mathrm{seq}}^{(j)}=e^{(j-1)},
\qquad
\Delta e^{(j)}=e^{(j)}-e^{(j-1)},
\end{equation}
with a fixed convention for the first trial in each participant, session and
block. Previous responses and previous assigned delays may enter the comparator
when they are already contained in the current pre-assignment history and are
permitted by the selected inferential route. Arousal, motor-preparation and
preprocessing-burden covariates enter through similarly locked pre-\(t_1\)
summaries. Identification remains grounded in the declared assignment and
analysis conditions. These covariates serve to protect the comparator against
ordinary explanations based on arousal, preparation, fatigue, timing and
session structure. Comparator instability, post-\(t_1\) inputs, failure of the frozen-object
invariance checks, or post-result retuning constitutes a protected-construction
failure and yields a diagnostic-failure outcome, with no additional evidential
weight assigned to the affected contrast.

\subsection{Participant-level estimand}
\label{sec:estimand}

The whole participant is the population replication and resampling unit. Trials
within a participant are treated as repeated observations from that participant,
with no status as independent biological replications. The primary analysis is
indexed by the assigned delay \(\tauL\). Measured delivered latency is reserved
for the delivery audit and compliance summaries; the confirmatory estimand
retains the randomised assigned-delay label.

The participant slope functional is fixed by the inferential route. In the
assignment-isolation route, within participant \(p\), trial \(j\), and
randomisation stratum \(\mathcal{R}_{pj}\), the stratum-centred working
regression is
\begin{equation}
\label{eq:participant-slope}
\Apre^{\mathrm{resid},(pj)}
=
a_{\mathcal{R}_{pj}}
+
\beta_{\tau,p}
\left\{
\tauL^{(pj)}
-
\mathbb{E}_{P_{\mathrm{sch}}}
\!\left[\tauL\mid\mathcal{R}_{pj}\right]
\right\}
+
\varepsilon_{pj}.
\end{equation}
Here \(\varepsilon_{pj}\) is the remainder from this
participant-specific, stratum-adjusted linear projection.
The strata and their intercepts are participant-specific; participant dependence
is implicit in \(a_{\mathcal{R}_{pj}}\) despite suppression of a separate
participant index. Equation~\eqref{eq:participant-slope} defines the
participant's stratum-adjusted linear projection: the intercepts and
\(\beta_{\tau,p}\) minimise the retained-sample squared error under the locked
eligibility rule. This projection permits a nonlinear conditional mean,
heteroskedastic errors and dependence among trial residuals.

In the sequential route, the participant slope is the conditional-score
functional specified in SI~\S4.3. It centres each assignment at its known
current-trial conditional mean and scales it by the corresponding conditional
assignment variance. The assignment-isolation and sequential quantities are
signed summaries of assigned-delay ordering. Under the structural exclusion,
the centred assignment score has zero conditional mean under the corresponding
route-specific validity conditions. For fixed-multiset assignment isolation,
this yields zero design expectation of the participant slope because the
retained leverage denominator is fixed by the multiset. For a ratio-form slope
with a random denominator, including the reported sequential slope, score
centring alone does not imply zero expectation of the finite-sample ratio;
inferential validity is instead supplied by the route-specific calibration.
Under independent within-stratum assignment and a constant linear slope, the two
functionals coincide. The route and its slope functional are fixed
prospectively, and the declared choice governs the analysis regardless of
comparative favourability.

For the declared route, the population estimand is the equal-participant mean
\begin{equation}
\label{eq:population-slope}
\beta_\tau
=
\mathbb{E}_p\!\left[\beta_{\tau,p}\right],
\end{equation}
where the expectation is taken over the prospectively defined
eligible-participant population. Let \(\widehat{\beta}_\tau\) denote the
corresponding equal-weight average of the estimable participant slopes. Equal
weighting assigns each participant the same contribution to the population
contrast, irrespective of the number of retained trials. The coefficient is a
signed linear summary over the declared assigned-delay support. Its
interpretation is confined to that support and does not require an exponential
shape. Because the endpoint is committed before assignment, the assigned delay
cannot act as a forward-time treatment on that earlier endpoint;
\(\beta_\tau\) is interpreted as an assignment-protected negative-control
slope.

Pre-assignment participant eligibility and post-assignment slope estimability
are separate qualifications. A participant may belong to the target population
while retaining insufficient assigned-delay support for slope estimation after
delivery, artefact rejection, preprocessing and inclusion. The protocol
therefore fixes minimum-leverage and non-estimability rules prospectively, with
the sign and magnitude of the participant slope excluded from those rules. It
reports the non-estimable fraction and its causes and compares estimable and
non-estimable participants using locked, label-blind summaries of the endpoint,
residual scale and retention. The participant-level estimability sensitivity
analysis of SI~\S8 is completed before any population-magnitude or
affirmative-null conclusion. A retained set containing fewer than
\(N_{\min}\) eligible and estimable participants is classified as
selection-limited or inconclusive. The same classification applies when the
declared bounds show that non-estimability could change the population
conclusion. These classifications preclude both support and a forward-only
adequate null.

The route-selection rule is fixed before confirmatory label access. The
assignment-isolation route applies when the experimental design and
prospectively declared reset or washout conditions justify invariance of the
frozen endpoint array under admissible reassignment. Potential carryover,
adaptation, fatigue or sequential dependence capable of transmitting one
assigned delay to later endpoints requires the sequential martingale/e-value
route. A prospectively specified diagnostic failure may activate a sequential fallback
only when that fallback, including its folds, e-value construction and operating
characteristics, has been fully declared and separately qualified in advance.
If assignment isolation is invalid and no such prospectively qualified
sequential fallback is available, no valid inferential route can be established
and the dataset is classified as inconclusive. After unblinding, a favourable
lagged-delay diagnostic carries no authority to replace the sequential route
with assignment isolation.

Under the sequential route, the current benchmark uses prospectively fixed,
participant-disjoint outer folds. Each participant's residuals are produced by
a comparator trained outside that participant's fold. One-step assignment
factors are multiplied and mixed over the fixed \(\lambda\)-grid within fold;
the fold e-values are then combined by a fixed convex mixture and are never
multiplied across folds. The e-value set is fixed by predictable eligibility,
while the later participant-estimability filter is used only for the reported
slope and magnitude bound. Assignment isolation is qualified only when, conditional on the locked retained
trial set, retained stratum pattern, frozen comparator object, complete frozen
residual array, and locked participant eligibility and estimability decisions,
the retained assigned-delay vector still follows the declared admissible
reassignment law under the sharp forward-only null. This joint retained-design
and endpoint-array invariance requirement is a route-specific statistical
condition, stronger than trial-local \(\mathrm{(R3^\star)}\), and is not
established by scheduler audits alone. The sequential route instead requires
predictable pre-assignment e-value eligibility, residual predictability under
the route's declared conditioning set, the declared current-trial conditional
scheduler law and the forward-only null. A dataset for which the selected route's
validity condition cannot be established is classified as inconclusive and cannot
yield either directional support or a clean forward-only adequate conclusion.
False-adequacy certification is performed separately for the route actually used.
Any loss of power from the conservative fold mixture is measured inside that
route-specific operating characteristic.

\section{Decision rule, audits and sensitivity analysis}
\label{sec:decision-audits-sensitivity}

The classifier asks four questions in a fixed order. Does the implementation
preserve the assignment boundary, is the retained analysis set qualified against
selection, can the retained design resolve the declared magnitude, and does the
slope meet the registered directional rule?

The confirmatory classifier brings together route-specific assignment
calibration, a population slope bound from whole-participant resampling,
resolution and participant-count requirements fixed in advance, operational
audits, trial-level selection qualification, an endpoint-by-delay collider
diagnostic, and participant-level estimability sensitivity. Every component, and
their order of precedence, is fixed before any confirmatory assigned-delay label
is analysed.

\subsection{Exclusion null, directional departure and inference}
\label{sec:inference}

Before selection, the forward-only no-ordering implication is sign-symmetric.
Under the operational scheduler in \(\mathrm{(R1)}\), temporal integrity in
\(\mathrm{(R2)}\) and the declared past-adapted factorisation, the assigned
delay is conditionally independent of every committed pre-selection statistic,
given the randomisation stratum. Carrying that exclusion into the
retained sample, for the committed endpoint and frozen residual, takes three
further ingredients: the retained-sample delay-neutrality condition
\(\mathrm{(R3^\star)}\), positivity on the declared delay support, and the
frozen-comparator independence condition of SI~\S1, Lemma~S1. This conditional
assignment exclusion is the structural null that the route-specific procedures
calibrate. The composite slope statement \(\beta_\tau\geq0\) does not exhaust
this structural null. The final classifier applies a separate magnitude
threshold to that null.

The population estimand is the equal-participant slope of the frozen residual
defined in \Cref{eq:population-slope}. The exclusion result covers the committed
endpoint as well. Any direct endpoint analysis is declared and reported
separately. Unless stated otherwise, \(\beta_\tau\),
\(\widehat{\beta}_\tau\), \(\UCB\) and \(\LCB\) in this section all refer to
the frozen-residual population estimand.

The registered directional departure is \(\beta_\tau<0\). Under the fixed sign
convention, larger residual values occur at shorter assigned delays and
attenuate as the assigned delay lengthens across the declared support.
Prospective fixation of the direction restricts the analysis to the registered
negative sign. It excludes a two-direction search and post hoc selection of the
favourable direction. The prediction specifies direction alone, without
mechanistic attribution. A qualified material positive slope is classified as
an opposite-direction diagnostic. It blocks both directional support and a
forward-only-adequate classification, but it is not a second confirmatory
level-\(\alpha\) class-rejection route. It remains ineligible as support for the
registered negative prediction and requires independent replication and
investigation.

Let \(\beta_{\min}\geq0\) be the resolution floor declared in advance, in the
endpoint's native units per second or on a fixed label-blind harmonised scale.
Let \(\UCB(\widehat{\beta}_\tau)\) be the one-sided \(95\%\) population upper
confidence bound from whole-participant resampling. The negative-direction
magnitude criterion is
\begin{equation}
\label{eq:materiality}
\UCB\!\left(\widehat{\beta}_\tau\right)<-\beta_{\min}.
\end{equation}
For the opposite-direction check, also declared in advance, let
\(\LCB(\widehat{\beta}_\tau)\) be the corresponding one-sided \(95\%\) lower confidence
bound. A positive departure is resolved only if the positive-tail route-specific
assignment calibration passes and
\begin{equation}
\label{eq:materiality-positive}
\LCB\!\left(\widehat{\beta}_\tau\right)>\beta_{\min}.
\end{equation}
The same resolution floor, participant-count requirement, audits and sensitivity
qualifications hold in both directions. The positive-direction rule defines a
prespecified opposite-direction diagnostic; it never defines directional support
or confirmatory class rejection.

No validated biological effect-size scale exists for an assigned-delay residual,
so \(\beta_{\min}\) is anchored to label-blind within-participant
resolvability. A fixed population significance level and a claimed threshold of
biological importance are absent from its definition. Applied to the
equal-participant population slope, \Cref{eq:materiality} also sets the minimum
magnitude that can be both resolved and reported. As participant-level
uncertainty decreases, the bound approaches the requirement that the population
slope itself lie below \(-\beta_{\min}\). Increasing the number of participants
reduces population-level uncertainty, while the resolution floor remains
determined by retained-trial yield, assigned-delay leverage and residual scale,
which together fix within-participant slope resolution. The locked resolution
rule sets this floor.
\begin{equation}
\label{eq:beta-min-main}
\beta_{\min}
=
\kappa\,
\frac{\sigma_{\mathrm{resid}}^{\mathrm{blind}}}
{\sigma_\tau\sqrt{\bar n_{\mathrm{ret}}}},
\end{equation}
where \(\sigma_{\mathrm{resid}}^{\mathrm{blind}}>0\) is the
prospectively locked label-blind residual-scale summary,
\(\sigma_\tau>0\) is the route-specific effective retained assignment scale,
\(\bar n_{\mathrm{ret}}>0\) is the arithmetic mean retained usable-trial yield
per estimable participant, and \(\kappa>0\) is a dimensionless multiplier fixed
prospectively. In the executable benchmark,
\(\sigma_\tau\sqrt{\bar n_{\mathrm{ret}}}\) is the square root of the mean
participant slope-denominator leverage under the selected route. SI~\S6.2 gives
the exact estimators, analysis-mask rule and contingency conventions. Planning
values come from the declared scheduler and conservative retention assumptions.
Qualification applies the locked construction to the realised retained data. No
input is drawn from the observed slope, endpoint values, injected departure or
an unblinded pilot.

Equation~\eqref{eq:beta-min-main} defines a non-compensatory resolution floor.
It is not an exact standard-error identity or a test size. Its purpose is to stop
support being declared when the retained design cannot resolve a slope within
participants; the unstandardised slope is reported separately. Throughout,
``material'' means resolved beyond the prospectively declared design-resolution
floor. It does not denote biological, clinical or theoretical importance.
SI~\S6.2 gives the full qualification and orientation diagnostics.

Inference uses two procedures, both selected in advance, with distinct validity
arguments. In the assignment-isolation route, the committed endpoints or frozen
residuals, together with the participant and stratum structure, are held fixed.
Fixed assignment multisets are permuted within declared blocks, or stochastic
assignments are redrawn from the declared scheduler law. The one-sided plus-one
randomisation value is the finite-replicate Monte Carlo rank value, taken from
the proportion of admissible reassignment statistics at least as negative as
the observed one, with the plus-one correction for finite-replicate validity
\citep{Harris2023,PhipsonSmyth2010}. The positive-direction check uses the
corresponding positive tail under the same admissible reassignment law.

The sequential route makes no assertion of a counterfactual reassignment
distribution for a fixed endpoint array. Under this route, current-trial
assignment increments are evaluated against their declared pre-assignment
conditional laws and combined through the martingale and e-value construction
of the SI. The positive-direction check uses the sign-reversed score declared in
advance. This route is used wherever carryover or other sequential dependence
makes frozen-array reassignment inexact.

The assignment-isolation route belongs to the conditional-randomisation family
\citep{Candes2018,Berrett2020}. Its scheduler and reassignment laws come from
the experiment and are not estimated from observational covariates. Its
finite-sample design validity rests on the admissible reassignment law,
frozen-array invariance and the retained-sample qualifications above. The
sequential route is different in kind, replacing frozen-array reassignment with
predictable current-trial increments calibrated under their declared
current-trial conditional scheduler laws and the forward-only null.

Population bounds come from a studentised participant bootstrap-\(t\),
resampling participants with all trials and frozen residuals intact. A
bias-corrected and accelerated bootstrap and a participant \(t\)-interval are
reported as sensitivity analyses. Let \(N\) be the number of analysable
participants. In executable benchmark package version~1.2.0, which was used
to generate the certified run, the minimum is \(N_{\min}=10\) eligible and
estimable participants. The canonical operating
characteristics use \(P=24\), so \(N_{\min}=10\) is a hard information floor,
not a claim that every design with ten participants has qualified
population-bound calibration. Below the floor, the assignment-calibrated result
may still be reported, but no population magnitude or adequacy conclusion is
drawn. Estimator and calibration details are in SI~\S5--\S6; affirmative-null
operating-characteristic qualification is in SI~\S9. 

\subsection{Decision rule}
\label{sec:decision}

The final classifier is an ordered, non-compensatory rule. Operational audit
failures are classified first. Retained-sample, collider and
participant-estimability qualifications are then applied according to their
declared scope, while failure to establish the selected route's validity is
routed to the inconclusive class. The inferential and magnitude components
determine whether a material departure is resolved, but no directional outcome
is assigned until the applicable qualifications have been evaluated. The scalar
trial-level selection-sensitivity gate is evaluated only for an already resolved
material departure; failure of that applicable gate yields a selection-limited
outcome. The ordering and outcome precedence are fixed in advance, so overlapping
signals cannot be sorted after the fact into the most favourable category.

A dataset earns negative-direction support only when all of the following hold:
\begin{enumerate}[leftmargin=2.2em,itemsep=2pt]
\item the selected route's validity condition and negative-tail assignment
calibration pass: joint retained-design and endpoint-array invariance for
assignment isolation, or predictable pre-assignment e-value eligibility,
residual predictability under the route's declared conditioning set, the
declared current-trial conditional scheduler law and the forward-only null for
the sequential route;
\item \(\UCB(\widehat{\beta}_\tau)<-\beta_{\min}\);
\item the analysable participant count satisfies \(N\geq N_{\min}\);
\item the randomisation, temporal-integrity, delivery and applicable
implementation-swap audits pass;
\item the retention audit and trial-level selection-sensitivity gate pass;
\item the endpoint-by-delay collider diagnostic does not fire; and
\item the participant-level estimability sensitivity gate passes.
\end{enumerate}

For an opposite-direction diagnostic, the positive-tail assignment calibration
must pass. The corresponding population materiality criterion is
\(\LCB(\widehat{\beta}_\tau)>\beta_{\min}\). The selected route's validity
condition and the same participant-count, audit, trial-selection, collider and
participant-estimability qualifications also apply. The result is reported
separately, blocks a forward-only-adequate classification, and remains outside
both directional support and confirmatory class rejection.

The committed endpoint carries the assignment exclusion directly. The
frozen-residual analysis is the declared confirmatory input to the classifier
because it is the precision-adjusted estimand that has been qualified in
advance. The same participant-level slope may also be reported descriptively on
the unadjusted committed endpoint, as an adjustment-sensitivity diagnostic. That
descriptive slope has no role as a support criterion or veto. If it disagrees
with the frozen-residual result, the executable outcome class remains unchanged.
It is reported to show how far the locked forward-only adjustment moves the
estimate, and so to guide interpretation of comparator sensitivity.

The frozen residual is confirmatory because the reported operating
characteristics, including the false-adequacy curve, qualify that residual-based
classifier. Using the unadjusted contrast as a support criterion or veto would
define a different classifier and require its own qualification. Every
empirical application therefore reports the unadjusted slope, its sign, and
whether it agrees with the frozen-residual slope. A supported residual
accompanied by a non-negative unadjusted slope retains its executable outcome
class and is flagged as adjustment-sensitive. It does not qualify as robust to
adjustment and must be interpreted and replicated under the same locked
comparator specification and residual-construction rule.

If the route-specific inferential component and the magnitude component disagree
in either direction, the result is ambiguous or inconclusive. A dataset becomes
eligible for a forward-only adequate classification only when neither
directional inferential component nor its matching magnitude component points to
a departure, all audits and affirmative-null-applicable qualifications pass,
\(N\geq N_{\min}\), retained assigned-delay leverage meets the registered
requirement, and the complete classifier satisfies the direction-specific
false-adequacy operating-characteristic criteria declared in advance. The
scalar trial-level selection-sensitivity gate is applicable only to an already
resolved material departure; it is not an affirmative-null gate. A
non-significant inferential value alone does not establish an affirmative null.

The affirmative null is a conditional certificate under
\(\mathrm{(R3^\star)}\), retained-support positivity and frozen-comparator
independence, within the audited sensitivity envelope. The retention audit,
endpoint-by-delay collider diagnostic and auxiliary negative controls qualify
reliance on \(\mathrm{(R3^\star)}\) for an affirmative-null interpretation
within that envelope. The scalar trial-level selection-sensitivity gate instead
addresses whether an already resolved material departure can be explained by
the declared marginal-selection pathway. The participant-level estimability
gate addresses a distinct aggregation problem: whether selection
into the estimable participant set can distort the equal-participant population
slope. None of these checks proves the relevant underlying independence
conditions globally. A failed trial-level selection or collider qualification
yields a selection-limited outcome. A failed participant-level estimability
qualification yields either a selection-limited or an inconclusive outcome,
according to whether the limiting feature is a plausible selection distortion
or a shortage of participant-level information.

Programme-level support requires independent replication as well. The
replication uses the same locked endpoint, comparator family, resolution floor,
inferential route, retained-sample qualification logic, participant estimability
rule and audit battery. A single qualified dataset supplies dataset-level
support only. \Cref{tab:decision} gives the complete outcome map.

The audit battery qualifies the operational scheduler and the
assignment-to-analysis implementation conditions and is reported whatever the
outcome. It does not verify the past-adapted factorisation that the confirmatory
contrast tests. The battery comprises a randomisation audit, checking that
realised assignments lie in the legal support of the declared scheduler law and
satisfy any prospectively imposed fixed-multiset, blocking or
restricted-randomisation constraints, together with seed and log integrity,
scheduler execution and concealment under \(\mathrm{(R1)}\); chance imbalance
under unrestricted stochastic assignment is reported descriptively and is not
by itself an audit failure; a temporal-integrity audit, which injects controlled
post-\(t_1\) impulses, steps, ringing patterns and event-locked transients and
requires no committed pre-event output beyond the tolerance fixed in advance
under \(\mathrm{(R2)}\); a delivery audit, bounding assigned-to-measured timing
non-compliance; a retention audit, recording stratum-specific delivery,
rejection, missingness and inclusion, with retained-versus-excluded endpoint
summaries available before confirmatory unblinding; and implementation-swap
checks, testing whether an association is peculiar to one generator, command
pathway, clock, trigger chain or hardware route. Audit definitions, pass
criteria and failure classifications are given in SI~\S7, Table~S3.

Negative controls are two-sided diagnostics. The assigned delay is the primary
negative-control exposure for the past-adapted factorisation. A material
association in either direction can reflect failure of the operational scheduler,
temporal leakage, delivery or implementation error, retained-sample
non-neutrality, or failure of the past-adapted statistical restriction itself
\citep{Lipsitch2010,Shi2020}. The audits and sensitivity gates are used to
separate the operational and selection explanations from the class-level
interpretation; they do not prove their absence globally. A qualified material
negative association may enter the confirmatory support rule. A qualified
positive association is the prespecified opposite-direction diagnostic: it
blocks a forward-only-adequate classification but carries no confirmatory
class-insufficiency claim. On an auxiliary negative control, an association of
either sign is a design alarm. A failed audit invalidates or limits the test and
supplies no evidence that the past-adapted class is sufficient.

The endpoint and label-blind comparator are fixed before the assigned-delay
label is available. After post-endpoint assignment, the population slope enters
confirmatory interpretation only if the timing, retention, collider and
participant-estimability qualifications are satisfied. A qualified material
negative slope supports conditional insufficiency; a qualified, adequately
sensitive result with neither directional departure yields the affirmative
null. Every other pattern is routed to a named diagnostic-failure,
selection-limited, opposite-direction or inconclusive outcome in
\Cref{tab:decision}.

\begingroup
\footnotesize
\setlength{\LTcapwidth}{\textwidth}
\setlength{\LTpre}{0.5\baselineskip}
\setlength{\LTpost}{0.5\baselineskip}

\begin{xltabular}{\textwidth}{@{}L{3.65cm}L{6.15cm}Y@{}}

\caption{Decision outcomes for the post-endpoint assignment-calibrated
Level~II-A test. The estimand is the equal-participant population slope
\(\beta_\tau\) defined in \Cref{eq:population-slope}. The structural null is
the endpoint-level no-ordering implication of the past-adapted factorisation,
conditional on the operational scheduler and the remaining design
qualifications; the registered support direction is negative. A forward-only
adequate outcome is stricter than non-rejection and additionally requires
two-directional, magnitude-indexed false-adequacy qualification within the
declared retained-sample envelope. Passing an audit, diagnostic or sensitivity
gate qualifies interpretation only within its declared envelope and does not
establish the corresponding independence condition globally.}
\label{tab:decision}
\\

\toprule
\textbf{Outcome} &
\textbf{Required statistical and qualification pattern} &
\textbf{Interpretation} \\
\midrule
\endfirsthead

\caption[]{Decision outcomes for the post-endpoint assignment-calibrated
Level~II-A test, continued.}
\\

\toprule
\textbf{Outcome} &
\textbf{Required statistical and qualification pattern} &
\textbf{Interpretation} \\
\midrule
\endhead

\midrule
\multicolumn{3}{r@{}}{\footnotesize\itshape Continued on next page}
\\
\endfoot

\bottomrule
\endlastfoot

Supported negative departure &
The selected route's validity condition and negative-tail assignment calibration
pass; \(\UCB(\widehat{\beta}_\tau)<-\beta_{\min}\);
\(N\geq N_{\min}\); all applicable audits, the trial-level selection gate,
the endpoint-by-delay collider diagnostic and the participant-level estimability
gate pass. &
Under the declared randomisation, temporal-integrity, frozen-residual and
retained-sample conditions, a qualified material negative ordering is
incompatible with the no-ordering implication of the declared past-adapted
class at the prespecified inferential calibration. The observed
assignment-calibrated departure therefore supports conditional insufficiency
of the class for the declared endpoint, linear contrast, delay support and
regime. It identifies no mechanism, and programme-level robustness still
requires independent replication.
\\
\addlinespace[0.7em]

Forward-only adequate (affirmative null) &
The selected route's validity condition holds; neither directional
route-specific inferential component nor its corresponding magnitude component
indicates a departure; all audits and affirmative-null-applicable
qualifications pass; \(N\geq N_{\min}\); assigned-delay leverage and the
route-specific false-adequacy qualification are adequate. The scalar
trial-level selection-sensitivity gate is not applicable in the absence of an
already resolved material departure. &
Within the declared retained-sample envelope and the evaluated additive
endpoint-level linear-injection family, inference route and nuisance grid, the
complete classifier is qualified against negative slopes of magnitude at least
\(\delta^\star_{r,-}\) and positive slopes of magnitude at least
\(\delta^\star_{r,+}\). This bounds the registered linear contrast, endpoint,
comparator class, delay support and regime; it is not a claim that residual
structure is globally absent.
\\
\addlinespace[0.7em]

Diagnostic failure &
With selected-route validity established, the assignment record is incompatible
with the declared law, concealment or log integrity fails, temporal leakage or
frozen-object non-invariance is detected, delivery integrity fails, an
implementation dependence is unresolved, or an auxiliary negative control
fails. &
The protected design-analysis boundary is not qualified. The dataset is neither
evidence for conditional insufficiency nor a clean affirmative null; the failure
must be resolved before confirmatory interpretation, and redesign or new data
may be required.
\\
\addlinespace[0.7em]

Selection-limited &
The retention audit or endpoint-by-delay collider diagnostic fires; the
trial-level selection gate fails for an already resolved material departure;
participant-level estimability sensitivity indicates a conclusion-changing
participant-level selection effect; or the affirmative-null false-adequacy
qualification is not met at the declared operating point. &
The retained trial or participant set, or the applicable affirmative-null
qualification, is insufficient for the claimed contrast within its declared
sensitivity envelope. The dataset is neither support nor a clean affirmative
null.
\\
\addlinespace[0.7em]

Opposite-direction departure &
Positive-tail route-specific assignment calibration passes;
\(\LCB(\widehat{\beta}_\tau)>\beta_{\min}\);
\(N\geq N_{\min}\); all applicable audits and sensitivity gates pass. &
The result runs counter to the registered directional prediction. It is a
prespecified opposite-direction diagnostic rather than a second confirmatory
class-rejection route. It blocks both directional support and a forward-only-
adequate conclusion, calls for independent replication and investigation, and
lends no support to the registered negative prediction.
\\
\addlinespace[0.7em]

Ambiguous or inconclusive &
The selected route's validity condition cannot be established, the inferential
and magnitude components disagree in either direction, \(N<N_{\min}\), or the
available participant-level information is insufficient to resolve a
population conclusion. &
The available route-specific inferential, magnitude or population information
cannot sustain either confirmatory class rejection or an affirmative null. All
components are reported descriptively.
\\

\end{xltabular}
\endgroup

\subsection{Omitted-pathway sensitivity analysis}
\label{sec:sensitivity}

\(\mathrm{(R3^\star)}\) cannot be settled by inspection. The trial-level
selection-sensitivity analysis therefore asks a narrower calibration question:
how large would a declared differential, endpoint-correlated retention pathway
have to be to reproduce the observed retained-sample slope under an otherwise
past-adapted account, and how large a pathway does the audit record still permit?
The gate maps an audited probability-scale imbalance onto the native
endpoint-slope scale, under a selection-model class declared in advance. It
draws on monotone sharp bounds \citep{Lee2009} and worst-case bounds without
monotonicity \citep{Manski1990}, within the wider sensitivity-analysis tradition
\citep{Rosenbaum2002,VanderWeeleDing2017,SmithVanderWeele2019}. The scale differs
from that of a risk-ratio E-value: the gate reports its result on the same slope
scale as the confirmatory estimand.

Let \(\Delta_{\mathrm{sel}}^{\mathrm{req}}\) be the smallest probability-scale
selection contrast that could reproduce the observed slope within the declared
model class, and \(\Delta_{\mathrm{sel}}^{\mathrm{aud}}\) the largest contrast
still compatible with the recorded delivery, retention, preprocessing,
inclusion and exclusion information. The bounds in this gate are the locked
selection-model bounds of SI~\S8; they are distinct from the whole-participant
resampling bounds applied to \(\widehat{\beta}_\tau\). The registered scalar
gate is
\begin{equation}
\label{eq:selection-gate-main}
\LCB\!\left(\Delta_{\mathrm{sel}}^{\mathrm{req}}\right)
>
\UCB\!\left(\Delta_{\mathrm{sel}}^{\mathrm{aud}}\right).
\end{equation}
The two one-sided bounds serve as a classifier gate fixed in advance. The
complete synthetic qualification assesses the gate's operational behaviour.
The bound comparison is not offered as a standalone \(95\%\) simultaneous
confidence statement for the difference. If
\Cref{eq:selection-gate-main} fails while a material slope is present, the
result is classified as selection-limited and support is blocked. Passing the
gate rules out only the declared marginal-imbalance pathway at the calibrated
operating point. It leaves \(\mathrm{(R3^\star)}\) unresolved in general.
SI~\S8 gives the full construction and a worked calculation.

Collider-stratification bias is a separate threat. Conditioning on inclusion
can induce an association in the retained sample even where the pre-selection
sample respects the randomisation boundary
\citep{Hernan2004,Greenland2003}. A pure endpoint-by-delay collider is the
sharpest case: inclusion depends on the joint configuration of the committed
endpoint and the assigned delay, with no main delay effect on retention, so a
retained-sample slope can be manufactured while marginal retention rates stay
roughly balanced. Because the committed endpoint exists for every assigned
trial, a diagnostic declared in advance can target that interaction directly. A
dataset that fires the diagnostic, or whose slope remains compatible with the
declared collider class outside the scalar marginal-retention summary, is
classified as selection-limited. In the benchmark, this diagnostic blocks the
manufactured balanced-collider slope; the scalar gate does not
(\Cref{sec:benchmark-role}; SI~\S8--\S9).

Participant-level estimability is qualified on its own terms. Trial-level
\(\mathrm{(R3^\star)}\) leaves unresolved whether the participants with
estimable slopes represent the eligible participant set behind the
equal-participant population estimand. The protocol therefore reports the
non-estimable fraction and its causes, and applies worst-case or monotone bounds,
declared in advance, to the slopes those participants could have contributed.
Support is blocked if those values could lift the population slope above
\(-\beta_{\min}\) or make its upper bound fail
\Cref{eq:materiality}; an affirmative null is blocked if they could conceal a
material departure in either direction. Details are in SI~\S8.

\section{Synthetic validation and practical implementation}
\label{sec:synthetic-validation}

\subsection{What the synthetic benchmarks establish}
\label{sec:benchmark-role}

The benchmarks qualify the complete analysis pipeline at specified operating
points against behaviour declared in advance. Before any empirical delay label is
analysed, the locked pipeline is run on simulated data with known generating
processes and must behave as declared across seven scenario families.
These benchmarks do not supply evidence that an assigned-delay residual exists
in human EEG. SI~\S9 reports the realised operating characteristics, produced by
the public benchmark package described in the Data accessibility statement.

Seven generators exercise the decision rule, each specified in SI~\S9. Under a
pure forward-only generator, the residual-slope distribution should centre on
zero and the complete classifier should control false support. Under an
endpoint-level injected departure, the pipeline should recover the registered
direction with the declared sensitivity and without excessive false adequacy.
Designated high-signal leakage, selection and collider generators should be routed
to their corresponding support-blocking outcomes at the rates required in advance.
An adversarial forward-only generator should preserve false-support control
despite a declared mixture of modelling stresses, and an opposite-direction
injection should be classified as an opposite-direction departure rather than as
directional support.

The adversarial generator includes nonlinear hazard, heavy-tailed heterogeneity
across participants, heteroskedasticity, autocorrelation, carryover and
comparator misspecification. Carryover makes the frozen endpoint array an inappropriate
sharp-null object under counterfactual label reshuffling. For this scenario, the
sequential martingale/e-value route consequently replaces assignment-isolation
reassignment. It uses the declared current-trial conditional scheduler law and
reports an e-value-derived inferential value; a plus-one frozen-array
randomisation value is not reported. Failure to meet any declared behaviour
leaves the pipeline operationally unqualified at that benchmark operating point,
and qualification extends only to the generators and design regimes actually
evaluated. The seven scenario families in SI~\S9 define that evaluated set and
form the pipeline's prespecified failure surface, with each generator targeting
a distinct way the classifier could mishandle known truth.

In the certified qualification run, run \LevelIIARunHash, each canonical
scenario contains \(M=\LevelIIAM\) datasets with \(P=24\) participants and 24
planned trials per assigned-delay bin. The clean forward-only anchor produced no
support (\(\LevelIIAAnchorSupportCountRate\)) and was classified as
forward-only adequate in \(\LevelIIAAnchorAdequateCountRate\), inconclusive in
\(\LevelIIAAnchorInconclusiveCountRate\), and selection-limited in
\(\LevelIIAAnchorSelectionLimitedCountRate\). Thus the clean-null benchmark
controlled false support but did not produce an affirmative conclusion in
\(127/1200=0.106\) datasets. This non-affirmation rate
is a conservative information and qualification cost, not a false-positive
rate; empirical planning must consider both false support and the probability
that a genuinely clean dataset remains inconclusive or selection-limited.
The injected \(-60\,\mu\mathrm{V\,s^{-1}}\) residual was supported in
\(\LevelIIAInjectedSupportCountRate\), with
\(\LevelIIAInjectedSelectionLimitedCountRate\) selection-limited outcomes and
\(\LevelIIAInjectedAdequateCountRate\) false-adequacy outcomes. The adversarial
forward-only null produced no support
(\(\LevelIIAAdversarialSupportCountRate\)) and yielded
\(\LevelIIAAdversarialAdequateCountRate\) forward-only adequate,
\(\LevelIIAAdversarialInconclusiveCountRate\) inconclusive,
\(\LevelIIAAdversarialSelectionLimitedCountRate\) selection-limited and
\(\LevelIIAAdversarialDiagnosticFailureCountRate\) diagnostic-failure outcomes
under the sequential route.

The leakage generator produced
\(\LevelIIALeakageDiagnosticFailureCountRate\) diagnostic failures. The
standard-selection generator produced
\(\LevelIIASelectionSelectionLimitedCountRate\) selection-limited and
\(\LevelIIASelectionDiagnosticFailureCountRate\) diagnostic-failure outcomes.
The balanced collider produced \(\LevelIIAColliderSelectionLimitedCountRate\)
selection-limited and \(\LevelIIAColliderDiagnosticFailureCountRate\)
diagnostic-failure outcomes; its clustered endpoint-by-delay interaction
diagnostic fired in every dataset. The positive injection was classified as an
opposite-direction departure in
\(\LevelIIAOppositeOppositeDirectionCountRate\), with no directional support
or false adequacy.

A route-matched auxiliary experiment separates route and generator effects. On
identical clean datasets, sequential adequacy exceeded assignment-isolation
adequacy by \(+0.101\), with \(95\%\) interval \([+0.083,+0.119]\). Under the
common sequential route, adversarial adequacy exceeded clean adequacy by
\(+0.007\), with interval \([-0.001,+0.015]\). The former clean-versus-
adversarial difference was therefore mainly a route effect. The adversarial
assignment-isolation cell is excluded because carryover violates endpoint-array
invariance.

False adequacy is indexed by route, direction and injected magnitude. The
certification family contains 40 cells: two routes, two directions and the grid
\(\{5,10,15,20,30,40,50,60,75,90\}\,\mu\mathrm{V\,s^{-1}}\). Pointwise Wilson
intervals are descriptive. Certification uses a one-sided,
Bonferroni-adjusted Clopper--Pearson familywise upper-bound envelope at the
candidate magnitude and all larger evaluated magnitudes. This deliberately
conservative construction assumes no monotonicity of the realised
false-adequacy curve; the maximum over larger magnitudes protects against
empirical reversals. A shape-constrained alternative would define a different
certificate and require separate qualification. Assignment isolation
first passes in both directions at
\(15\,\mu\mathrm{V\,s^{-1}}\). The sequential e-value route first passes in
both directions at \(30\,\mu\mathrm{V\,s^{-1}}\). At the sequential
\(20\,\mu\mathrm{V\,s^{-1}}\) cells, false adequacy remains \(0.318\) and
\(0.295\), so a single route-neutral certificate would be misleading. These
values are false-adequacy resolution boundaries for the complete classifier
under the evaluated additive endpoint-level linear-injection family, inference
routes, nuisance grid and retained-sample envelope. They are distinct from
biological materiality and directional-support power.
At the displayed \(\pm60\,\mu\mathrm{V\,s^{-1}}\) points, false adequacy is
\(0/1200\) under both routes and directions, with simultaneous upper bound
\(0.0056\).

Operating rates are realised Monte Carlo rates from \(M=1200\) datasets and
retain binomial Monte Carlo uncertainty. In particular, \(0/1200\) support
means that no supported outcome arose in the realised run; it does not set the
underlying false-support probability to zero. Pointwise intervals for selected
canonical rates, full outcome decompositions, route-matched contrasts, collider
sweeps and route-specific adequacy tables are reported in SI~\S9 and in the
archived machine-readable outputs \citep{Wilson1927,BrownCaiDasGupta2001}.

\begin{figure}[t]
\centering
\includegraphics[width=\textwidth]{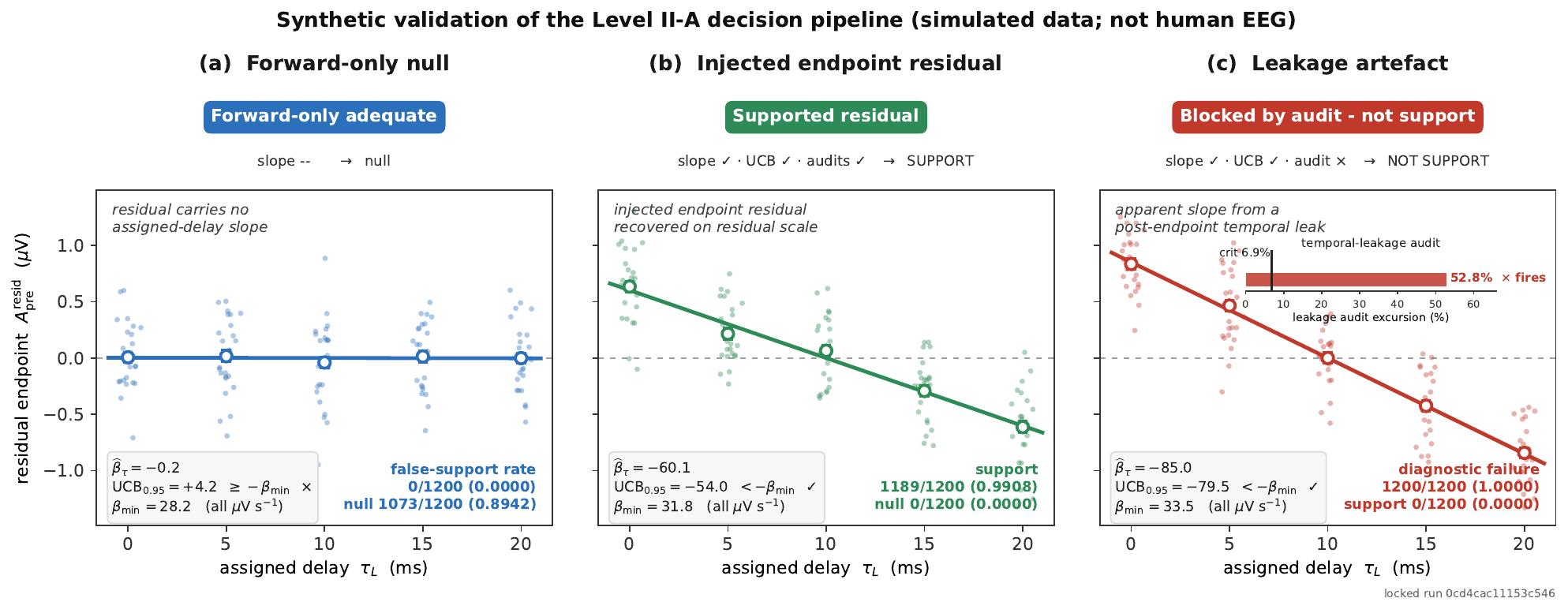}
\caption{Synthetic validation of the Level~II-A pipeline using simulated data
only. The reference grid has \(T_0=20\,\mathrm{ms}\), five origin-coded delay
bins, 24 participants and 24 planned trials per bin. Annotations combine
representative-dataset quantities (the slope, one-sided
upper bound and dataset-specific resolution floor) with selected operating
rates across the 1200 simulated datasets. (a) The forward-only anchor yields no support and
\(\LevelIIAAnchorAdequateCountRate\) forward-only-adequate outcomes. (b) The
endpoint-level \(-60\,\mu\mathrm{V\,s^{-1}}\) injection is supported in
\(\LevelIIAInjectedSupportCountRate\), with
\(\LevelIIAInjectedAdequateCountRate\) false-adequacy outcomes at that
magnitude.
The separate adequacy sweep gives certified false-adequacy boundaries of
\(15\,\mu\mathrm{V\,s^{-1}}\) for assignment isolation and
\(30\,\mu\mathrm{V\,s^{-1}}\) for the sequential e-value route, in both
directions. (c) A post-endpoint temporal
leak creates an apparent resolved slope, while the temporal-integrity audit
fires in every dataset and blocks support. Run \LevelIIARunHash; full operating
characteristics are reported in SI~\S9.}
\label{fig:synthetic}
\end{figure}

\subsection{Synthetic anchor and planning power}
\label{sec:benchmark-anchor}

The reference benchmark uses a deliberately short synthetic support
\(T_0=20\,\mathrm{ms}\), with five assigned-delay bins
\(\{0,5,10,15,20\}\,\mathrm{ms}\) and, where a departure is injected, an additive
slope at the endpoint-generating level. The \(0\)-ms bin is an origin-coded
synthetic analysis label.

In an empirical implementation, the assigned-delay support must begin at a
minimum interval after the label has been generated and concealed up to
\(t_1\). The interval must be strictly positive, logged and auditable, and must
be determined from the randomisation service, command path, display or auditory
delivery path, trigger chain, timestamping precision and concealment barrier. If
the empirical grid is plotted after this minimum has been subtracted, the axis
must be labelled as an offset from the empirical minimum; it no longer
represents an absolute post-endpoint latency.

The synthetic grid is a short-support stress test of the estimator and
classifier. Any systematic non-null operating pattern must trace back to the
injected departure or to a deliberately injected artefact. Finite-sample slopes
under the forward-only generator are handled by the declared inferential
calibration.

The injected departure is added to the committed endpoint before comparator
fitting and residualisation. It is not added to the post-comparator residual.
Endpoint-level and residual-scale injections are similar in the clean anchor
where the past-adapted covariates are independent of the assigned delay by
construction and the comparator is label blind. The endpoint-level injection
remains the primary end-to-end benchmark because it exercises the comparator,
cross-fitting, freezing, residualisation, route-specific inferential
calibration, participant-level bound, resolution floor and decision rule as one
coupled pipeline.

In the retained-sample synthetic benchmark, the estimator meets the planning
target only when the injected slope is large enough relative to residual noise
and the retained bins are sufficiently balanced. As synthetic estimator
operating characteristics, the favourable grid reaches that target with roughly
five retained trials per bin, and the higher-noise grid reaches it with roughly
twenty.

The scaling behind them carries the feasibility caveat. By
\eqref{eq:beta-min-main},
\[
\beta_{\min}
\propto
\frac{\sigma_{\mathrm{resid}}^{\mathrm{blind}}}
     {\sigma_\tau\sqrt{\bar n_{\mathrm{ret}}}},
\]
with the normal-approximation orientation diagnostic confined to SI~\S6.2. In a
balanced assignment-isolation design, the trial budget needed to reach a fixed
within-participant resolution scale grows with the square of the label-blind
single-trial residual scale and falls with the square of the assigned-delay
spread. More generally, the relevant quantity is the route-specific effective
retained assignment leverage. Loss of retained assignment support or
route-specific denominator leverage can therefore raise the floor or render the
slope non-estimable even when the planned support was wide.

The synthetic anchor validates the complete pipeline at a declared operating
point. An empirical protocol must set the per-participant retained-trial target
in advance, using label-blind estimates of residual variance, participant
heterogeneity, retention yield, timing precision, delivery compliance and the
declared resolution floor.

An illustrative plug-in calculation shows the resulting empirical burden.
Suppose the frozen forward-only comparator leaves a label-blind single-trial
residual scale
\(\sigma_{\mathrm{resid}}^{\mathrm{blind}}\approx4\,\mu\mathrm{V}\).
For a balanced assignment-isolation design with approximately uniform empirical
support \([\tau_{\min},\tau_{\min}+600\,\mathrm{ms}]\), where
\(\tau_{\min}>0\) is the auditable minimum assigned delay,
\(\sigma_\tau\approx0.6/\sqrt{12}\approx0.17\,\mathrm{s}\).
With \(\bar n_{\mathrm{ret}}\approx300\) retained trials per participant and
\(\kappa=2\), \Cref{eq:beta-min-main} gives
\(\beta_{\min}\approx2.7\,\mu\mathrm{V\,s^{-1}}\). On the same
illustrative scale, this is a \(0.27\,\mu\mathrm{V}\) change per
\(100\,\mathrm{ms}\) of assigned delay, about \(7\%\) of the assumed
single-trial residual scale; across the full \(600\,\mathrm{ms}\) support,
the corresponding change is \(1.62\,\mu\mathrm{V}\), about \(40\%\) of
that scale.

This value is the within-participant resolution scale that enters the registered
magnitude rule. It is not a predicted biological effect size and, on its own,
does not guarantee population-level power. Population-level performance depends
further on participant-level slope dispersion, the analysable participant count
and the complete simulated operating characteristics. Those quantities are
estimated in the label-blind calibration stage and used to set \(N_{\min}\). An
unresolvable declared floor, whether from residual scale, retained yield,
assigned-delay spread, participant-level dispersion or audit tolerances, routes
the dataset to the predeclared inconclusive outcome.

This arithmetic is an inherent feature of the exclusion boundary. It identifies
which endpoint classes can support an informative affirmative null. Late-CNV is
a conservative neuroscience anchor with demanding single-trial noise properties
and is used here as a worked case, not as a claim that it is the most efficient
first empirical endpoint.
Decoder outputs, adaptive-control signals, physiological-feedback
summaries and other high-throughput committed statistics may offer more
favourable regimes, provided the same post-endpoint assignment and
assignment-to-analysis integrity conditions can be audited.

The admissible delay support is itself a design variable with opposing
consequences. A wider support raises assigned-delay variance, and so statistical
leverage. It also raises post-endpoint waiting-time heterogeneity, the delivery
and timestamping burden, the opportunities for post-assignment attrition, and
cross-trial carryover through trial duration, adaptation and fatigue. Those
costs may call for stronger reset or washout procedures, richer recorded
pre-assignment histories, or the sequential calibration route.

Candidate delay supports are therefore compared using the design-known scheduler
law, label-blind estimates of residual scale and retention, and
operating-characteristic simulations declared in advance. The chosen support is
frozen before confirmatory label access, and is never revised in response to an
observed endpoint-by-delay association.

The reported anchor leaves an empirical implementation unqualified on its own.
Before confirmatory use, a planned design should be evaluated over a label-blind
operating-characteristic surface spanning residual scale, retained-trial yield,
participant-level slope dispersion, assigned-delay support, participant count,
timing error, retention imbalance and comparator family. A planned
implementation is adequately qualified only where forward-only generators
control false support, injected material departures are recovered with the
declared sensitivity, leakage and selection artefacts are blocked, and the
affirmative-null classifier stays informative rather than mostly inconclusive.
Benchmark specifications and the planning grid are given in SI~\S9.

\subsection{Practical implementation sequence}
\label{sec:implementation}

The most economical programme is staged. First, qualify the hardware, scheduler,
concealment barrier, delivery logging, retention rules and analysis
implementation on synthetic signals and hardware loopback data. An independent,
non-confirmatory pilot then estimates feasibility quantities using an endpoint,
comparator family, fold structure and causal preprocessing rules that are all
fixed before any pilot delay association is inspected.

The pilot may estimate achieved retention yield and the label-blind residual
scale at the single-trial level. It may also estimate heterogeneity across
participants, timing error and delivery compliance. It may not select or revise
the endpoint, comparator, preprocessing, exclusion rules or analysis window in
response to an observed endpoint-by-delay pattern. If the pilot motivates a
design change, the revised design must be frozen and, where necessary,
recalibrated in a separate label-blind qualification stage before confirmatory
use.

Second, turn the calibrated design into a complete empirical protocol,
preferably for Stage~1 Registered Report review
\citep{Chambers2013,Nosek2018}. The protocol fixes the scheduler, committed
endpoint, comparator, folds, resolution floor, participant and trial targets,
exclusions, audits, selection-sensitivity analysis, inferential route and
non-compensatory decision rule at the level of detail prospective execution
requires.

\begin{figure}[t]
\centering
\begin{tikzpicture}[
>=Stealth,
font=\footnotesize,
node distance=5mm,
stage/.style={
    draw,
    rounded corners=2pt,
    fill=gray!8,
    align=center,
    text width=2.75cm,
    minimum height=0.86cm,
    inner sep=3pt
},
stagewide/.style={
    draw,
    rounded corners=2pt,
    fill=gray!8,
    align=center,
    text width=3.45cm,
    minimum height=0.78cm,
    inner sep=3pt
},
barrierwide/.style={
    draw,
    dashed,
    rounded corners=2pt,
    fill=red!6,
    align=center,
    text width=3.35cm,
    minimum height=0.78cm,
    inner sep=3pt,
    font=\scriptsize
}
]

\node[stage] (lock)
{Lock scheduler, endpoint, comparator, folds,
\(\beta_{\min}\), \(N_{\min}\)};

\node[stagewide,right=7mm of lock] (fit)
{Fit label-blind comparator and freeze confirmatory objects};

\node[barrierwide,right=7mm of fit] (audit)
{Randomisation, temporal, delivery, retention and implementation audits};

\node[
above=0pt of audit,
font=\scriptsize\itshape,
text=red!60!black
]
{operational and assignment-to-analysis integrity};

\node[stagewide,below=6mm of audit] (infer)
{Route validity, route-specific inference and participant upper bound};

\node[stagewide,left=7mm of infer] (qual)
{Applicable selection, collider and participant-estimability qualifications};

\node[stagewide,left=7mm of qual] (dec)
{Non-compensatory decision rule (\Cref{tab:decision})};

\draw[->] (lock) -- (fit);
\draw[->] (fit) -- (audit);
\draw[->] (audit) -- (infer);
\draw[->] (infer) -- (qual);
\draw[->] (qual) -- (dec);

\end{tikzpicture}

\caption{Analysis-access schematic, not within-trial event chronology or
executable classifier precedence. The scheduler, endpoint, comparator, folds,
resolution floor and minimum participant count are fixed before confirmatory
label access. Label-blind comparator fitting and freezing are followed by the
audit battery and selected-route validity assessment, route-specific
inferential calibration and the participant-level upper bound, and the
applicable selection, collider and participant-estimability qualifications
before final non-compensatory classification. The scalar trial-level
selection-sensitivity gate is applicable only to an already resolved material
departure.}
\label{fig:pipeline}
\end{figure}
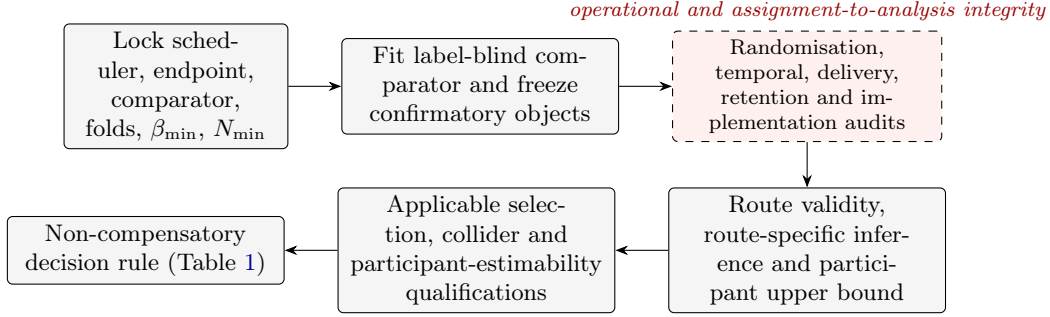

Third, run the dataset-level test and report the complete assignment,
implementation, audit, retention, feasibility and decision record, whatever the
outcome. Fourth, repeat the same locked test in an independent sample. This
sequence narrows analytic flexibility \citep{Simmons2011}, makes a surprising
result hard to manufacture through post hoc analysis choices, and leaves that
result readily open to challenge through replication and implementation audit.
\Cref{fig:pipeline} summarises the analysis-access sequence and the integrity
checks that separate label-blind construction from confirmatory inference.

\section{Discussion}
\label{sec:discussion}

This article puts a standing assumption of anticipatory neuroscience to a test it
could fail. The assumption is that the systematic part of a pre-event EEG
endpoint is exhausted by past-adapted information: foreperiod, hazard, temporal
expectation, vigilance, motor preparation, trial history and session structure,
all of it available before the endpoint window closes. Level~II-A sets prediction
aside and asks a different question. Once the endpoint and comparator are locked,
can the registered confirmatory statistic still be ordered by a delay label
assigned only after that statistic was committed and frozen?

The primary positive output is a route-qualified affirmative null: a clean,
adequately sensitive result stating that the declared account leaves no
material linear assigned-delay ordering within the certified range for the
inference route used. The present benchmark qualifies assignment isolation from
\(15\,\mu\mathrm{V\,s^{-1}}\) and the sequential e-value route from
\(30\,\mu\mathrm{V\,s^{-1}}\), in both directions. These bounds are indexed to
the declared synthetic generators, comparator, audit envelope, magnitude grid
and familywise Monte Carlo convention. A route whose operating characteristic
exceeds the false-adequacy bound at the scientifically relevant magnitude
remains unqualified for an affirmative interpretation.

The design supports a class-level conclusion in the registered negative
direction. The past-adapted factorisation, not the fit of any particular
comparator, defines the scientific class established in \Cref{sec:conditions}. Under the
factorisation, \Cref{prop:exclusion} and its retained-sample counterpart rule out
systematic assigned-delay ordering of either sign for every member of the class.
The confirmatory level-\(\alpha\) claim is reserved for a qualified material
negative departure that clears the non-compensatory rule; such a result supports
conditional insufficiency of the class for the declared endpoint and contrast.
Any undetected scheduler, temporal, frozen-object or selection failure would
constitute a breach of an operational or identifying qualification, just as an
unrecognised premise violation can affect any randomisation or causal-design
argument. Conditional on those qualifications, the negative departure is
incompatible with the past-adapted factorisation. For temporal-preparation
research, that outcome would show that foreperiod, hazard, CNV, arousal,
preparation and trial history do not exhaust the registered endpoint after
flexible label-blind adjustment. The result remains
mechanism-neutral, and programme-level robustness requires independent replication.
At the dataset level, the negative-direction inference is a genuine conditional
class rejection rather than the designation of an unexplained anomaly. A qualified
positive association remains a prespecified diagnostic and carries no parallel
confirmatory class-rejection claim.

The contribution is the boundary and the qualified adequacy statements it
supports. The operational scheduler creates the post-commitment probe. Under the
past-adapted factorisation, the probe is conditionally independent of any
integrable pre-assignment statistic before selection. Under retained-sample
delay-neutrality, retained-support positivity and frozen-comparator independence
as stated in SI~\S1, the retained-sample analogue extends the trial-local
no-ordering implication to the committed endpoint and frozen residual.
Participant-level slope inference additionally requires the selected route's
joint-design or predictable-score conditions and the participant-estimability
qualification. The forward-only comparator can sharpen precision and express
the contrast after adjustment for registered anticipatory structure.
Identification derives from the operational scheduler, the past-adapted
factorisation, temporal integrity, frozen-object construction, retained-sample
conditions and route-specific inferential qualification.

Comparator sensitivity (\Cref{sec:comparator}) is quantified prospectively for
the evaluated endpoint-level linear-injection family: injected material
departures are passed through the full frozen-comparator pipeline, and the
resulting false-adequacy operating characteristic is reported (SI~\S9).
A failed operational audit yields a diagnostic-failure outcome, while a failed
retention qualification or collider diagnostic yields a selection-limited
outcome. For an already resolved material departure, failure of the applicable
scalar selection-sensitivity gate also yields a selection-limited outcome.
These classifications block directional support. Operational, retention and
collider qualifications additionally govern eligibility for a clean affirmative
null; the scalar selection-sensitivity gate is not itself an affirmative-null
gate.

This construction closes a specific methodological gap for EEG and MEG.
Goodness of fit measures approximation and therefore leaves unresolved whether
a past-adapted class exhausts a committed endpoint. Level~II-A introduces a
post-commitment randomised probe and, under the declared past-adapted
factorisation and its analysis qualifications, tests the resulting no-ordering
implication for one registered linear contrast. The resulting inference is
bounded to the tested endpoint, declared delay support, retained-sample envelope
and simulation-qualified magnitude range. Adequacy claims for nonlinear,
temporal or subgroup-specific features of anticipatory activity remain outside
this boundary.

The construction is related to equivalence testing and severity assessment,
which interpret a sufficiently informative non-rejection as bounded positive
evidence \citep{AltmanBland1995,Lakens2017,MayoSpanos2006}. Bayes-factor null
assessment makes a comparable move in neuroscience \citep{Keysers2020}. The
present construction differs in two respects. First, the design generates the
probe after endpoint commitment. Second, admission of the affirmative conclusion
depends on both a prospectively fixed margin and the false-adequacy operating
characteristic of the complete classifier.
The qualification is reported on the native assigned-delay slope scale as
route- and direction-indexed \(\delta^\star_{r,d}\). False adequacy is the
simulation-estimated probability of classifying a genuine linear departure as
forward-only adequate. Under the current declared generators, grid and Monte
Carlo convention, the route-specific boundaries reported above therefore mark
the smallest evaluated departure magnitudes for which the false-adequacy
criterion is satisfied. The larger boundary for the sequential fold-mixture
classifier records the resolution cost of the conservative valid fold mixture
rather than a defect in its calibration. Because the label is assigned only
after commitment, a past-adapted comparator has no legitimate access to the
probe during construction or tuning.

\subsection{Relation to broader theoretical interpretations}
\label{sec:broader}

A qualified, independently replicated departure would raise a theoretical question
without answering it. Level~II-A supplies the exclusion boundary, its identifying
conditions, its qualification checks and its falsification rule. It does not supply a
mechanism. The required restraint is correspondingly precise: interpretation must
not slide from a surviving association to a mechanism, from a mechanism to a
substrate, or from a failed replication to a philosophical dispute. The first task
is to determine whether the declared past-adapted class is adequate for the
declared endpoint and contrast. Explanatory attributions belong downstream.

Level~II-A is domain-general at the level of design rather than subject matter.
It can transfer wherever a scientifically or operationally meaningful quantity
can be irrevocably committed before an exogenous label is generated, with
label-blind construction and qualified post-assignment selection. Applications
include anticipatory neurophysiology and behavioural experiments, randomised
clinical and social interventions, forecasting and machine-learning pipelines,
algorithmic decision systems, and engineering or operational systems in which
predictions or state estimates can be frozen before randomised audit, routing or
deployment decisions.

Transfer beyond EEG nevertheless takes two distinct inferential forms. An
exogenous post-commitment label can act as a pipeline-integrity negative control,
testing whether a committed quantity or frozen residual stays independent of
information the generating pipeline could not have used
\citep{Lipsitch2010,Shi2020}. It becomes a substantive sufficiency probe only
when a declared alternative predicts ordering by that particular label. An
arbitrary queue token therefore can audit integrity, but it cannot by itself
convert predictive adequacy into a scientific model comparison. The EEG worked
case meets the substantive criterion: \Cref{sec:exclusion} declares a
non-factorising joint-history alternative that predicts ordering by the assigned
delay while preserving the identifying conditions. The affirmative null
therefore bounds the asserted substantive ordering and its corresponding
class-level adequacy implication.

In forecasting or machine learning, a score may be locked and timestamped at time
\(t\), with any residual built only from information admissible at \(t\), before a
randomised audit arm, release order or deployment route is assigned. Association
with that later label diagnoses a failure of assignment, processing, residual
construction or selection. A clean, adequately sensitive null supports a bounded
integrity statement and leaves sufficiency for the actual prediction target
unresolved. A withheld observed target is not automatically exogenous, since the
same latent state may drive both forecast and target. Allocation triggered
deterministically by the committed score likewise lacks the required exogeneity.
Score-stratified randomisation, declared in advance, remains admissible only if
the score enters the stratum and inference uses the corresponding conditional
scheduler law together with the relevant forward-only null restriction. These
safeguards complement, rather than replace, the ordinary protections against
circular analysis and train--test leakage
\citep{Kriegeskorte2009,VarmaSimon2006}.

A familiar substantive candidate is a risk score locked before randomisation to a
treatment arm. The arm functions initially as an integrity probe. It becomes a substantive
sufficiency probe only if a declared non-past-adapted alternative predicts
ordering of the locked score or residual by that future arm; outcome relevance
alone leaves this criterion unmet. In either form, transfer requires irrevocable commitment, a
known randomisation law or a defensible exogeneity argument, label-blind endpoint
and residual construction, qualified post-assignment selection, and prospective
evaluation of the declared departures that the pipeline might miss. Under the integrity
formulation, a surviving association diagnoses pipeline failure. Under the substantive
formulation, and conditional on the identifying conditions, it rejects the declared
past-adapted class.

A downstream Level~II-B programme would begin only after independent
replication of a Level~II-A class-level departure. Its task would be to compare
restricted non-past-adapted joint-history models, with the Level~II-A exclusion
retained as the programme's entry condition. One natural route is to freeze the
Level~II-A past-adapted law as a reference and add
a low-dimensional reciprocal dependence, one that preserves the declared scheduler
marginal while predicting held-out delay values, the complete pre-endpoint
temporal profile, cross-window dependence and transport across scheduler or
implementation changes. Building that model lies outside the present Perspective.
Level~II-A asks whether the declared past-adapted class is adequate; Level~II-B
would ask which restricted joint-history extension, if any, explains a replicated
failure of that class.

\subsection{Conclusion}
\label{sec:discussion-conclusion}

Anticipation is usually analysed as a forward process, and that is the right
default: brains learn regularities, track hazards, prepare actions and tune gain
from information already in hand. But a default is not a measured boundary.
Level~II-A turns the default into a measurement, fixing what must be locked,
hidden and audited, what counts as a qualified departure, what counts as a clean
and adequately sensitive null, and what must be set aside as unqualified, so that
no outcome is wasted. It reports no empirical decision; it fixes the rule by
which one could be made, and shows on synthetic data that the rule behaves as
declared. What is at stake is the line between what the past explains and what,
if anything, remains ordered by a label assigned only after the measurement
window has closed. Level~II-A does not assume where that line falls. It builds
the apparatus to locate it, commits in advance to the results that would reveal
it, and states where interpretation must stop.

\section*{Ethics}

This article reports methodological design and synthetic benchmark analyses
only. No human participants, animal subjects, human tissue, identifiable
personal data, or clinical data were collected or analysed. No ethics approval
or informed consent was required for the work reported here.

\section*{Data accessibility}
\label{sec:data-access}

All code, configurations, synthetic-data generators, tests and certified
benchmark artefacts supporting the synthetic benchmark results are publicly
archived as version~1.2.1 of the Level~II-A post-endpoint randomisation
benchmark pipeline~\cite{Sopasakis2026LevelIIA}. The version-specific archive
is available at
\url{https://doi.org/10.5281/zenodo.21887583}, and the corresponding public
source repository is maintained at
\url{https://github.com/geosop/LEVEL-IIA}. Release~v1.2.1 preserves certified
benchmark run \texttt{0cd4cac11153c546} and its numerical results; the
executable benchmark package version used to generate that certified run
remains version~1.2.0, and no Monte Carlo benchmark was rerun.

\section*{Use of artificial intelligence and AI-assisted technologies}

ChatGPT (OpenAI; GPT-5.6 Sol) was used for language and readability editing,
LaTeX and bibliographic formatting, and terminology, clarity and consistency checks.
All scientific content, analyses, interpretations and conclusions were reviewed
and verified by the authors, who retain full responsibility for the integrity
and accuracy of the article.

\section*{Author contributions}

George Sopasakis led the conceptual development, methodological and inferential
design, benchmark specification, and manuscript preparation. Alexandros
Sopasakis contributed to the methodology, inferential development, benchmark
specification, and critical revision of the manuscript. Both authors reviewed
and approved the final manuscript.

\section*{Competing interests}

The authors declare no competing interests.

\section*{Funding}

The authors received no specific funding for this work.


\begingroup
\small
\setlength{\bibsep}{2pt}
\bibliographystyle{unsrtnat}
\bibliography{LeveLIIArefs}

@article{Walter1964,
  author  = {Walter, W. G. and Cooper, R. and Aldridge, V. J. and McCallum, W. C. and Winter, A. L.},
  year    = {1964},
  title   = {Contingent negative variation: an electric sign of sensori-motor association and expectancy in the human brain},
  journal = {Nature},
  volume  = {203},
  pages   = {380--384},
  doi     = {10.1038/203380a0}
}

@article{Macar2004,
  author  = {Macar, F. and Vidal, F.},
  year    = {2004},
  title   = {Event-related potentials as indices of time processing: a review},
  journal = {Journal of Psychophysiology},
  volume  = {18},
  pages   = {89--104},
  doi        = {10.1027/0269-8803.18.23.89}
}

@article{Trillenberg2000,
  author  = {Trillenberg, P. and Verleger, R. and Wascher, E. and Wauschkuhn, B. and Wessel, K.},
  year    = {2000},
  title   = {{CNV} and temporal uncertainty with ageing and non-ageing {S1--S2} intervals},
  journal = {Clinical Neurophysiology},
  volume  = {111},
  pages   = {1216--1226},
  doi     = {10.1016/S1388-2457(00)00274-1}
}

@article{Niemi1981,
  author  = {Niemi, P. and N{\"a}{\"a}t{\"a}nen, R.},
  year    = {1981},
  title   = {Foreperiod and simple reaction time},
  journal = {Psychological Bulletin},
  volume  = {89},
  pages   = {133--162},
  doi     = {10.1037/0033-2909.89.1.133}
}

@article{Los2001,
  author  = {Los, S. A. and Van den Heuvel, C. E.},
  year    = {2001},
  title   = {Intentional and unintentional contributions to nonspecific preparation during reaction time foreperiods},
  journal = {Journal of Experimental Psychology: Human Perception and Performance},
  volume  = {27},
  pages   = {370--386},
  doi     = {10.1037/0096-1523.27.2.370}
}

@book{Nobre2010,
  editor    = {Nobre, A. C. and Coull, J. T.},
  year      = {2010},
  title     = {Attention and Time},
  publisher = {Oxford University Press},
  address   = {Oxford, UK},
  doi        = {10.1093/acprof:oso/9780199563456.001.0001}
}

@article{Coull2009,
  author  = {Coull, J. T.},
  year    = {2009},
  title   = {Neural substrates of mounting temporal expectation},
  journal = {PLoS Biology},
  volume  = {7},
  pages   = {e1000166},
  doi     = {10.1371/journal.pbio.1000166}
}

@article{Miniussi1999,
  author  = {Miniussi, C. and Wilding, E. L. and Coull, J. T. and Nobre, A. C.},
  year    = {1999},
  title   = {Orienting attention in time: Modulation of brain potentials},
  journal = {Brain},
  volume  = {122},
  pages   = {1507--1518},
  doi     = {10.1093/brain/122.8.1507}
}

@article{Vanrullen2013,
  author  = {VanRullen, R.},
  year    = {2013},
  title   = {Visual attention: a rhythmic process?},
  journal = {Current Biology},
  volume  = {23},
  pages   = {R1110--R1112},
  doi     = {10.1016/j.cub.2013.11.006}
}

@article{Steinborn2008,
  author  = {Steinborn, Michael B. and Rolke, Bettina and
             Bratzke, Daniel and Ulrich, Rolf},
  year    = {2008},
  title   = {Sequential effects within a short foreperiod context:
             Evidence for the conditioning account of temporal preparation},
  journal = {Acta Psychologica},
  volume  = {129},
  number  = {2},
  pages   = {297--307},
  doi     = {10.1016/j.actpsy.2008.08.005}
}

@article{Friston2010,
  author  = {Friston, K.},
  year    = {2010},
  title   = {The free-energy principle: a unified brain theory?},
  journal = {Nature Reviews Neuroscience},
  volume  = {11},
  pages   = {127--138},
  doi     = {10.1038/nrn2787}
}

@article{Clark2013,
  author  = {Clark, A.},
  year    = {2013},
  title   = {Whatever next? {Predictive} brains, situated agents, and the future of cognitive science},
  journal = {Behavioral and Brain Sciences},
  volume  = {36},
  pages   = {181--204},
  doi     = {10.1017/S0140525X12000477}
}

@book{Hohwy2013,
  author    = {Hohwy, J.},
  year      = {2013},
  title     = {The Predictive Mind},
  publisher = {Oxford University Press},
  address   = {Oxford, UK},
  doi       = {10.1093/acprof:oso/9780199682737.001.0001}
}

@misc{Millidge2021,
  author        = {Millidge, B. and Seth, A. K. and Buckley, C. L.},
  year          = {2021},
  title         = {Predictive coding: a theoretical and experimental review},
  eprint        = {2107.12979},
  archivePrefix = {arXiv},
  primaryClass  = {q-bio.NC},
  url        = {https://arxiv.org/abs/2107.12979}
}

@article{FristonKiebel2009,
  author  = {Friston, K. and Kiebel, S. J.},
  year    = {2009},
  title   = {Predictive coding under the free-energy principle},
  journal = {Philosophical Transactions of the Royal Society B},
  volume  = {364},
  pages   = {1211--1221},
  doi     = {10.1098/rstb.2008.0300}
}

@book{Bishop2006,
  author    = {Bishop, C. M.},
  year      = {2006},
  title     = {Pattern Recognition and Machine Learning},
  series    = {Information Science and Statistics},
  publisher = {Springer},
  address   = {New York, NY}
}

@article{Freedman2008,
  author  = {Freedman, D. A.},
  year    = {2008},
  title   = {On regression adjustments in experiments with several treatments},
  journal = {Annals of Applied Statistics},
  volume  = {2},
  pages   = {176--196},
  doi     = {10.1214/07-AOAS143}
}

@article{Kriegeskorte2009,
  author  = {Kriegeskorte, N. and Simmons, W. K. and Bellgowan, P. S. F. and Baker, C. I.},
  year    = {2009},
  title   = {Circular analysis in systems neuroscience: the dangers of double dipping},
  journal = {Nature Neuroscience},
  volume  = {12},
  pages   = {535--540},
  doi     = {10.1038/nn.2303}
}

@article{VarmaSimon2006,
  author  = {Varma, S. and Simon, R.},
  year    = {2006},
  title   = {Bias in error estimation when using cross-validation for model selection},
  journal = {BMC Bioinformatics},
  volume  = {7},
  pages   = {91},
  doi     = {10.1186/1471-2105-7-91}
}

@article{Lin2013,
  author  = {Lin, W.},
  year    = {2013},
  title   = {Agnostic notes on regression adjustments to experimental data: reexamining Freedman's critique},
  journal = {Annals of Applied Statistics},
  volume  = {7},
  pages   = {295--318},
  doi     = {10.1214/12-AOAS583}
}

@incollection{Brunia2012,
  author    = {Brunia, C. H. M. and van Boxtel, G. J. M. and B{\"o}cker, K. B. E.},
  year      = {2012},
  title     = {Negative slow waves as indices of anticipation: the {Bereitschaftspotential}, the contingent negative variation, and the stimulus-preceding negativity},
  booktitle = {The Oxford Handbook of Event-Related Potential Components},
  editor    = {Kappenman, E. S. and Luck, S. J.},
  pages     = {189--207},
  publisher = {Oxford University Press},
  address   = {Oxford, UK},
  doi       = {10.1093/oxfordhb/9780195374148.013.0108}
}

@book{Luck2014,
  author    = {Luck, S. J.},
  year      = {2014},
  title     = {An Introduction to the Event-Related Potential Technique},
  edition   = {2},
  publisher = {MIT Press},
  address   = {Cambridge, MA}
}

@article{Blankertz2011,
  author  = {Blankertz, B. and Lemm, S. and Treder, M. and Haufe, S. and M{\"u}ller, K. R.},
  year    = {2011},
  title   = {Single-trial analysis and classification of {ERP} components - a tutorial},
  journal = {NeuroImage},
  volume  = {56},
  pages   = {814--825},
  doi     = {10.1016/j.neuroimage.2010.06.048}
}

@article{Schwarzkopf2014,
  author  = {Schwarzkopf, D. S.},
  year    = {2014},
  title   = {We should have seen this coming},
  journal = {Frontiers in Human Neuroscience},
  volume  = {8},
  pages   = {332},
  doi     = {10.3389/fnhum.2014.00332}
}

@article{Chambers2013,
  author  = {Chambers, C. D.},
  year    = {2013},
  title   = {Registered reports: a new publishing initiative at Cortex},
  journal = {Cortex},
  volume  = {49},
  pages   = {609--610},
  doi     = {10.1016/j.cortex.2012.12.016}
}

@article{Nosek2018,
  author  = {Nosek, B. A. and Ebersole, C. R. and DeHaven, A. C. and Mellor, D. T.},
  year    = {2018},
  title   = {The preregistration revolution},
  journal = {Proceedings of the National Academy of Sciences USA},
  volume  = {115},
  pages   = {2600--2606},
  doi     = {10.1073/pnas.1708274114}
}

@article{Simmons2011,
  author  = {Simmons, J. P. and Nelson, L. D. and Simonsohn, U.},
  year    = {2011},
  title   = {False-positive psychology: undisclosed flexibility in data collection and analysis allows presenting anything as significant},
  journal = {Psychological Science},
  volume  = {22},
  pages   = {1359--1366},
  doi     = {10.1177/0956797611417632}
}

@article{Wilson1927,
  author  = {Wilson, E. B.},
  year    = {1927},
  title   = {Probable inference, the law of succession, and statistical inference},
  journal = {Journal of the American Statistical Association},
  volume  = {22},
  pages   = {209--212},
  doi     = {10.1080/01621459.1927.10502953}
}

@article{BrownCaiDasGupta2001,
  author  = {Brown, L. D. and Cai, T. T. and DasGupta, A.},
  year    = {2001},
  title   = {Interval estimation for a binomial proportion},
  journal = {Statistical Science},
  volume  = {16},
  pages   = {101--133},
  doi     = {10.1214/ss/1009213286}
}

@article{PhipsonSmyth2010,
  author  = {Phipson, B. and Smyth, G. K.},
  year    = {2010},
  title   = {Permutation {$p$}-values should never be zero: calculating exact {$p$}-values when permutations are randomly drawn},
  journal = {Statistical Applications in Genetics and Molecular Biology},
  volume  = {9},
  pages   = {Article 39},
  doi     = {10.2202/1544-6115.1585}
}

@article{AltmanBland1995,
  author  = {Altman, D. G. and Bland, J. M.},
  year    = {1995},
  title   = {Statistics notes: Absence of evidence is not evidence of absence},
  journal = {BMJ},
  volume  = {311},
  pages   = {485},
  doi     = {10.1136/bmj.311.7003.485}
}

@article{Lakens2017,
  author  = {Lakens, D.},
  year    = {2017},
  title   = {Equivalence tests: a practical primer for {$t$} tests, correlations, and meta-analyses},
  journal = {Social Psychological and Personality Science},
  volume  = {8},
  pages   = {355--362},
  doi     = {10.1177/1948550617697177}
}

@article{MayoSpanos2006,
  author  = {Mayo, D. G. and Spanos, A.},
  year    = {2006},
  title   = {Severe testing as a basic concept in a {Neyman--Pearson} philosophy of induction},
  journal = {British Journal for the Philosophy of Science},
  volume  = {57},
  pages   = {323--357},
  doi     = {10.1093/bjps/axl003}
}

@article{Keysers2020,
  author  = {Keysers, C. and Gazzola, V. and Wagenmakers, E.-J.},
  year    = {2020},
  title   = {Using {Bayes} factor hypothesis testing in neuroscience to establish evidence of absence},
  journal = {Nature Neuroscience},
  volume  = {23},
  pages   = {788--799},
  doi     = {10.1038/s41593-020-0660-4}
}

@book{Hastie2009,
  author    = {Hastie, T. and Tibshirani, R. and Friedman, J.},
  year      = {2009},
  title     = {The Elements of Statistical Learning},
  series    = { Springer Series in Statistics},
  edition   = {2},
  publisher = {Springer},
  address   = {New York, NY},
  doi        = {10.1007/978-0-387-84858-7}
}

@article{ZhaoDing2021,
  author  = {Zhao, A. and Ding, P.},
  year    = {2021},
  title   = {Covariate-adjusted {Fisher} randomization tests for the average treatment effect},
  journal = {Journal of Econometrics},
  volume  = {225},
  pages   = {278--294},
  doi     = {10.1016/j.jeconom.2021.04.007}
}

@misc{LuEtAl2025,
  author        = {Lu, X. and Shi, L. and Liu, H. and Ding, P.},
  year          = {2025},
  title         = {Conditional cross-fitting for unbiased machine-learning-assisted covariate adjustment in randomized experiments},
  eprint        = {2508.15664},
  archivePrefix = {arXiv},
  url        = {https://arxiv.org/abs/2508.15664}
}

@misc{Harris2023,
  author        = {Harris, K. D. and Miller, K. J.},
  year          = {2023},
  title         = {Conditional randomization tests for behavioral and neural time series},
  eprint        = {2311.03554},
  archivePrefix = {arXiv},
  url        = {https://arxiv.org/abs/2311.03554}
}

@article{Candes2018,
  author  = {Cand{\`e}s, E. and Fan, Y. and Janson, L. and Lv, J.},
  year    = {2018},
  title   = {Panning for gold: model-{X} knockoffs for high-dimensional controlled variable selection},
  journal = {Journal of the Royal Statistical Society: Series B},
  volume  = {80},
  pages   = {551--577},
  doi     = {10.1111/rssb.12265}
}

@article{Berrett2020,
  author  = {Berrett, T. B. and Wang, Y. and Barber, R. F. and Samworth, R. J.},
  year    = {2020},
  title   = {The conditional permutation test for independence while controlling for confounders},
  journal = {Journal of the Royal Statistical Society: Series B},
  volume  = {82},
  pages   = {175--197},
  doi     = {10.1111/rssb.12340}
}

@article{Lipsitch2010,
  author  = {Lipsitch, M. and Tchetgen Tchetgen, E. and Cohen, T.},
  year    = {2010},
  title   = {Negative controls: a tool for detecting confounding and bias in observational studies},
  journal = {Epidemiology},
  volume  = {21},
  pages   = {383--388},
  doi        = {10.1097/EDE.0b013e3181d61eeb}
}

@article{Shi2020,
  author  = {Shi, X. and Miao, W. and Tchetgen Tchetgen, E.},
  year    = {2020},
  title   = {A selective review of negative control methods in epidemiology},
  journal = {Current Epidemiology Reports},
  volume  = {7},
  pages   = {190--202},
  doi        = {10.1007/s40471-020-00243-4}
}

@article{Rousselet2012,
  author  = {Rousselet, G. A.},
  year    = {2012},
  title   = {Does filtering preclude us from studying {ERP} time-courses?},
  journal = {Frontiers in Psychology},
  volume  = {3},
  pages   = {131},
  doi     = {10.3389/fpsyg.2012.00131}
}

@article{Widmann2015,
  author  = {Widmann, A. and Schr{\"o}ger, E. and Maess, B.},
  year    = {2015},
  title   = {Digital filter design for electrophysiological data - a practical approach},
  journal = {Journal of Neuroscience Methods},
  volume  = {250},
  pages   = {34--46},
  doi     = {10.1016/j.jneumeth.2014.08.002}
}

@article{deCheveigne2019,
  author  = {de Cheveign{\'e}, A. and Nelken, I.},
  year    = {2019},
  title   = {Filters: when, why, and how (not) to use them},
  journal = {Neuron},
  volume  = {102},
  pages   = {280--293},
  doi     = {10.1016/j.neuron.2019.02.039}
}

@article{Lee2009,
  author  = {Lee, D. S.},
  year    = {2009},
  title   = {Training, wages, and sample selection: estimating sharp bounds on treatment effects},
  journal = {Review of Economic Studies},
  volume  = {76},
  pages   = {1071--1102},
  doi        = {10.1111/j.1467-937X.2009.00536.x}
}

@article{Manski1990,
  author  = {Manski, C. F.},
  year    = {1990},
  title   = {Nonparametric bounds on treatment effects},
  journal = {American Economic Review},
  volume  = {80},
  pages   = {319--323}
}

@article{FrangakisRubin2002,
  author  = {Frangakis, C. E. and Rubin, D. B.},
  year    = {2002},
  title   = {Principal stratification in causal inference},
  journal = {Biometrics},
  volume  = {58},
  number  = {1},
  pages   = {21--29},
  doi     = {10.1111/j.0006-341X.2002.00021.x}
}

@article{VanderWeeleDing2017,
  author  = {VanderWeele, T. J. and Ding, P.},
  year    = {2017},
  title   = {Sensitivity analysis in observational research: introducing the {E-value}},
  journal = {Annals of Internal Medicine},
  volume  = {167},
  pages   = {268--274},
  doi     = {10.7326/M16-2607}
}

@article{SmithVanderWeele2019,
  author  = {Smith, L. H. and VanderWeele, T. J.},
  year    = {2019},
  title   = {Bounding bias due to selection},
  journal = {Epidemiology},
  volume  = {30},
  pages   = {509--516},
  doi     = {10.1097/EDE.0000000000001032}
}

@book{Rosenbaum2002,
  author    = {Rosenbaum, P. R.},
  year      = {2002},
  title     = {Observational Studies},
  edition   = {2},
  publisher = {Springer},
  address   = {New York, NY},
  doi        = {10.1007/978-1-4757-3692-2}
}

@article{Hernan2004,
  author  = {Hern{\'a}n, M. A. and Hern{\'a}ndez-D{\'i}az, S. and Robins, J. M.},
  year    = {2004},
  title   = {A structural approach to selection bias},
  journal = {Epidemiology},
  volume  = {15},
  pages   = {615--625},
  doi     = {10.1097/01.ede.0000135174.63482.43}
}

@article{Greenland2003,
  author  = {Greenland, S.},
  year    = {2003},
  title   = {Quantifying biases in causal models: classical confounding versus collider-stratification bias},
  journal = {Epidemiology},
  volume  = {14},
  pages   = {300--306},
  doi     = {10.1097/01.EDE.0000042804.12056.6C}
}

@article{Kekecs2023,
  author  = {Kekecs, Z. and others},
  year    = {2023},
  title   = {Raising the value of research studies in psychological science by increasing the credibility of research reports: the Transparent Psi project},
  journal = {Royal Society Open Science},
  volume  = {10},
  pages   = {191375},
  doi        = {10.1098/rsos.191375}
}

@article{Mossbridge2012,
  author  = {Mossbridge, J. and Tressoldi, P. and Utts, J.},
  year    = {2012},
  title   = {Predictive physiological anticipation preceding seemingly unpredictable stimuli: a meta-analysis},
  journal = {Frontiers in Psychology},
  volume  = {3},
  pages   = {390},
  doi     = {10.3389/fpsyg.2012.00390}
}

@article{NobreVanEde2018,
  author  = {Nobre, A. C. and van Ede, F.},
  year    = {2018},
  title   = {Anticipated moments: temporal structure in attention},
  journal = {Nature Reviews Neuroscience},
  volume  = {19},
  pages   = {34--48},
  doi     = {10.1038/nrn.2017.141}
}

@article{Ramdas2023,
  author  = {Ramdas, A. and Gr{\"u}nwald, P. and Vovk, V. and Shafer, G.},
  year    = {2023},
  title   = {Game-theoretic statistics and safe anytime-valid inference},
  journal = {Statistical Science},
  volume  = {38},
  pages   = {576--601},
  doi     = {10.1214/23-STS894}
}

@article{GrunwalddeHeideKoolen2024,
  author  = {Gr{\"u}nwald, P. and de Heide, R. and Koolen, W. M.},
  year    = {2024},
  title   = {Safe testing},
  journal = {Journal of the Royal Statistical Society: Series B},
  volume  = {86},
  pages   = {1091--1128},
  doi     = {10.1093/jrsssb/qkae011}
}

@article{Shafer2021,
  author  = {Shafer, G.},
  year    = {2021},
  title   = {Testing by betting: a strategy for statistical and scientific communication},
  journal = {Journal of the Royal Statistical Society: Series A},
  volume  = {184},
  pages   = {407--431},
  doi     = {10.1111/rssa.12647}
}

@article{VovkWang2021,
  author  = {Vovk, V. and Wang, R.},
  year    = {2021},
  title   = {E-values: calibration, combination and applications},
  journal = {Annals of Statistics},
  volume  = {49},
  pages   = {1736--1754},
  doi     = {10.1214/20-AOS2020}
}

@book{Hall1992,
  author    = {Hall, P.},
  year      = {1992},
  title     = {The Bootstrap and Edgeworth Expansion},
  publisher = {Springer},
  address   = {New York, NY},
  doi        = {10.1007/978-1-4612-4384-7}
}

@article{DiCiccioEfron1996,
  author  = {DiCiccio, T. J. and Efron, B.},
  year    = {1996},
  title   = {Bootstrap confidence intervals},
  journal = {Statistical Science},
  volume  = {11},
  pages   = {189--228},
  doi     = {10.1214/ss/1032280214}
}

@book{EfronTibshirani1993,
  author    = {Efron, B. and Tibshirani, R. J.},
  year      = {1994},
  title     = {An Introduction to the Bootstrap},
  publisher = {Chapman \& Hall},
  address   = {New York, NY},
  doi       = {10.1201/9780429246593}
}

@misc{Sopasakis2026LevelIIA,
  author       = {Sopasakis, George and Sopasakis, Alexandros},
  title        = {{Level II-A} post-endpoint randomisation benchmark pipeline},
  year         = {2026},
  howpublished = {Zenodo, version 1.2.1},
  doi          = {10.5281/zenodo.21887583},
  note         = {\href{https://doi.org/10.5281/zenodo.21887583}
                  {doi:10.5281/zenodo.21887583}}
}
\endgroup

\end{document}


\maketitle

\noindent
This supplement gives the formal and operational detail behind the Level~II-A
test with post-endpoint assignment calibration described in the main text.
It supports the Level~II-A design only. It contains no mechanistic, dynamical,
or physical model of any residual; the construction and evaluation of such
models are separate, prospectively specified work. Section numbering uses an \textbf{S} prefix and is
self-contained. All references to ``the main text'' point to the accompanying
Perspective.

\tableofcontents
\bigskip


\section{Formal randomisation boundary}
\label{sec:si-randomisation-boundary}

This section proves the conditional-exclusion result stated as
Proposition~1 in the main text. The result is elementary but central to the
Level~II-A design: if an integrable statistic is measurable with respect to the
pre-assignment filtration and the declared past-adapted factorisation holds
conditional on the randomisation stratum, then the statistic cannot be
systematically ordered by the assigned delay before post-assignment selection.

The argument has two layers. The first layer is a pre-selection theorem:
post-endpoint randomisation excludes assigned-delay ordering for any integrable
past-adapted statistic before post-assignment selection is imposed. The second
layer concerns the retained analysis sample. The retained sample is formed after
assignment, through delivery, trigger timing, artefact rejection, missingness,
preprocessing success, exclusion, and analysis inclusion. Conditioning on that
sample can create endpoint-by-delay dependence even when the pre-selection
randomisation theorem is true. Retained-sample exclusion therefore requires an
additional structural condition, stated below as retained-sample
delay-neutrality. The audits and sensitivity analyses do not prove that
condition globally; they qualify it within the declared operational and
sensitivity envelope.

The proof is included to make the inferential target precise. A supported
Level~II-A residual is not defined as any association between a pre-event
endpoint and a later label. It is defined as an association that remains after
the endpoint has been locked, the forward-only comparator and residual array
have been frozen, the assigned-delay label has been generated after endpoint
closure, and the randomisation, leakage, delivery, retention, implementation,
and selection conditions have not blocked the retained-sample contrast.

\subsection{Setup and filtration}
\label{sec:si-filtration}

The notation below separates three objects that must not be conflated: the
operational scheduler that generates and conceals the current assigned delay,
the class-level factorisation implied by a past-adapted account, and the
testable no-ordering consequence for the committed endpoint. The temporal
barrier makes the scheduler mechanism auditable. It does not by itself establish
statistical independence between the current draw and the entire pre-assignment
record under every hypothesis.

Fix a trial \(j\) with randomisation stratum \(\mathcal{R}_j\) and committed
endpoint window \(I_{\mathrm{pre}}=[t_0,t_1]\). Let
\((\Omega,\mathcal{A},\mathbb{P})\) carry the trial process. Let
\(\{\F_t^{(j)}\}_{t\le t_1}\) be the filtration generated by all information
available before the assigned-delay draw for trial \(j\): neural and
physiological history, protocol state, elapsed foreperiod, conditional hazard,
arousal and vigilance proxies, motor-preparation proxies, block and session
state, time on task, previous-trial variables, and all preprocessing quantities
permitted by the endpoint lock. The stratum label \(\mathcal{R}_j\) is
\(\F_{t_1}^{(j)}\)-measurable.

The assigned delay \(\tauL^{(j)}\in[0,T_0]\) is selected after \(t_1\) by the
registered randomisation service with declared conditional law
\[
P_{\mathrm{sch}}
\!\left(
\mathrm{d}\tauL
\mid
\mathcal{R}_j
\right).
\]
Here \([0,T_0]\) may denote an origin-coded analysis coordinate. In an
empirical implementation, the corresponding physical assigned-delay support is
translated by the strictly positive auditable minimum specified in
\Cref{sec:si-anchor}.
The realised assigned-delay label is therefore not an element of
\(\F_{t_1}^{(j)}\). Scheduler information that is available before \(t_1\), such
as the stratum, block probabilities, or conditional scheduler law, may belong
to \(\F_{t_1}^{(j)}\). The specific post-endpoint draw may not.

\begin{definition}[Past-adapted statistic]
\label{def:past-adapted}
A real-valued statistic \(Y^{(j)}\) is temporally admissible before assignment
when it is \(\F_{t_1}^{(j)}\)-measurable, meaning that it is computable from
information available at or before \(t_1\). It belongs to the declared
past-adapted class only under an account whose joint law with the current
scheduler draw satisfies the factorisation in \eqref{eq:si-exogeneity}.
Measurability alone does not establish that factorisation.
\end{definition}

The committed endpoint \(\Apre^{(j)}\), any locked comparator covariate
\(\mathbf{X}_j\), and any prospectively fixed transform of the committed
endpoint are intended to be temporally admissible. Temporal admissibility is
lost if preprocessing uses post-\(t_1\) samples or if fold construction or
nuisance fitting uses current assigned-delay labels. Membership in the
past-adapted null class additionally requires the statistical factorisation
below. Post-assignment inclusion is handled separately because conditioning can
open an endpoint-by-delay selection path.

\begin{definition}[Operational post-endpoint scheduler]
\label{def:exogenous}
Condition~(R1) holds operationally when the registered service selects the
current assigned delay after endpoint commitment from
\(P_{\mathrm{sch}}(\cdot\mid\mathcal{R}_j)\), using only the declared
scheduler inputs, and conceals the realised label from every operation that can
revise the endpoint or frozen confirmatory objects. Scheduler, seed, timestamp,
logging, concealment and implementation checks qualify conformance to this
mechanism.
\end{definition}

Under a past-adapted account, the operational scheduler additionally entails the
class-level forward-only factorisation
\begin{equation}
\label{eq:si-exogeneity}
\tauL^{(j)}
\ \perp\!\!\!\perp\
\F_{t_1}^{(j)}
\ \mid\
\mathcal{R}_j .
\end{equation}
Equation~\eqref{eq:si-exogeneity} is the statistical null restriction for the
complete pre-assignment filtration, not the operational content of (R1) itself.
The scheduler audits qualify generation and concealment; the confirmatory test
interrogates an endpoint-level consequence of the factorisation. The distinction
allows a substantive alternative to preserve the scheduler mechanism while
violating \eqref{eq:si-exogeneity}. Equation~\eqref{eq:si-exogeneity} is also a
pre-selection restriction: even when it holds, conditioning on delivery,
retention, preprocessing success, missingness, exclusion or analysis inclusion
can break the corresponding relation in the analysed sample. Those steps are
handled by the retained-sample condition below and by the audits and sensitivity
analyses in \Cref{sec:si-leakage-controls} and
\Cref{sec:si-sensitivity-analysis}.

\begin{definition}[Frozen comparator object]
\label{def:si-frozen-comparator-object}
Let \(\mathcal{G}_{\mathrm{frz}}\) denote the frozen comparator object: the
locked fold structure, nuisance-fitting rule, tuning choices, fitted nuisance
functions, and residual-construction objects used to produce the confirmatory
held-out residuals. This object is admissible for confirmatory inference only
when it is constructed without assigned-delay labels, without delay-labelled
residual inspection, and without assignment descendants that would make the
endpoint, comparator, folds, nuisance fits, or residual array change under the
admissible assignment scheme.
\end{definition}

The frozen comparator object is included explicitly because the confirmatory
residual for trial \(j\),
\[
\Apre^{\mathrm{resid},(j)}
=
\Apre^{(j)}
-
\widehat m_0^{(-k(j))}(\mathbf{X}_j),
\]
is not a function of \((\Apre^{(j)},\mathbf{X}_j)\) alone before the comparator has been
locked and frozen. It is a prospectively fixed function of the trial-level
object and \(\mathcal{G}_{\mathrm{frz}}\) after the locked cross-fitting and
residual-freezing protocol has been completed. The randomisation argument for the frozen residual therefore conditions on
\(\mathcal{G}_{\mathrm{frz}}\). Under the forward-only null it requires the
class-level factorisation to remain valid for the locked trial object given the
stratum and frozen object; after selection it additionally requires the
retained-sample conditions below.

\subsection{Conditional exclusion before retained-sample selection}
\label{sec:si-exclusion-proof}

The following proposition states the endpoint-level implication of the
past-adapted factorisation before post-assignment selection. The operational
scheduler supplies the probe; the statistical no-ordering conclusion requires
the forward-only restriction in \eqref{eq:si-exogeneity}.

\begin{proposition}[Endpoint exclusion under the past-adapted factorisation]
\label{prop:si-exclusion}
Let \(Y^{(j)}\) be integrable, with
\(\mathbb{E}\lvert Y^{(j)}\rvert<\infty\).
\begin{enumerate}[leftmargin=2.0em,itemsep=2pt]
\item If \(Y^{(j)}\) is \(\F_{t_1}^{(j)}\)-measurable and the past-adapted
factorisation \eqref{eq:si-exogeneity} holds, then
\begin{equation}
\label{eq:si-fwd-null}
\mathbb{E}
\!\left[
Y^{(j)}
\mid
\sigma(\tauL^{(j)}),\mathcal{R}_j
\right]
=
\mathbb{E}
\!\left[
Y^{(j)}
\mid
\mathcal{R}_j
\right]
\quad\text{a.s.}
\end{equation}
Equivalently, for every Borel set \(D\subseteq[0,T_0]\) with
\(P_{\mathrm{sch}}(D\mid\mathcal{R}_j)>0\),
\[
\mathbb{E}
\!\left[
Y^{(j)}
\mid
\tauL^{(j)}\in D,\mathcal{R}_j
\right]
=
\mathbb{E}
\!\left[
Y^{(j)}
\mid
\mathcal{R}_j
\right].
\]

\item Equation~\eqref{eq:si-fwd-null} applies with
\(Y^{(j)}=\Apre^{(j)}\), and with any prospectively fixed transform
\(g(\Apre^{(j)})\) whose definition and parameters do not depend on the
assigned-delay vector, provided \(\Apre^{(j)}\) is temporally admissible. This
is the endpoint-level consequence of the operational scheduler in (R1),
temporal integrity in (R2), and past-adapted sufficiency.

\item Equation~\eqref{eq:si-fwd-null} is a pre-selection statement. It need not
hold in a sample selected after assignment. If inclusion depends on the assigned
delay given the endpoint, locked covariates, stratum, or frozen comparator
object, conditioning on \(S^{(j)}=1\) can induce delay dependence in the retained
sample even when \eqref{eq:si-fwd-null} holds before selection. Retained-sample exclusion is therefore not a consequence of the operational
scheduler, temporal integrity and pre-selection factorisation alone. It requires
a separate
retained-sample condition, stated in \Cref{def:r3star}, and holds for the
confirmatory statistic under that condition by \Cref{lem:retained-exclusion}.
\end{enumerate}
\end{proposition}

\begin{proof}
For part~1, \(Y^{(j)}\) is \(\F_{t_1}^{(j)}\)-measurable by assumption.
By the past-adapted factorisation \eqref{eq:si-exogeneity},
\(\tauL^{(j)}\) is conditionally independent of \(\F_{t_1}^{(j)}\) given
\(\mathcal{R}_j\). Therefore \(Y^{(j)}\) and
\(\tauL^{(j)}\) are conditionally independent given \(\mathcal{R}_j\).

For any bounded measurable function \(\psi\),
\[
\mathbb{E}
\!\left[
Y^{(j)}\psi(\tauL^{(j)})
\mid
\mathcal{R}_j
\right]
=
\mathbb{E}
\!\left[
Y^{(j)}
\mid
\mathcal{R}_j
\right]
\mathbb{E}
\!\left[
\psi(\tauL^{(j)})
\mid
\mathcal{R}_j
\right].
\]
Taking \(\psi=\mathbf{1}_D\) gives
\[
\mathbb{E}
\!\left[
Y^{(j)}\mathbf{1}_{\{\tauL^{(j)}\in D\}}
\mid
\mathcal{R}_j
\right]
=
\mathbb{E}
\!\left[
Y^{(j)}
\mid
\mathcal{R}_j
\right]
P_{\mathrm{sch}}(D\mid\mathcal{R}_j).
\]
On the event where \(P_{\mathrm{sch}}(D\mid\mathcal{R}_j)>0\), division by this
conditional probability yields the stated equality conditional on
\(\{\tauL^{(j)}\in D\}\) and \(\mathcal{R}_j\). Borel sets with zero conditional
assignment probability do not define an estimable conditional contrast under
the declared scheduler. The conditional-expectation form
in \eqref{eq:si-fwd-null} follows from the defining property of conditional
expectation with respect to
\(\sigma(\tauL^{(j)})\vee\sigma(\mathcal{R}_j)\), using a monotone-class
argument to extend from indicator functions to all integrable \(Y^{(j)}\).

Part~2 is the special case \(Y^{(j)}=\Apre^{(j)}\), or
\(Y^{(j)}=g(\Apre^{(j)})\). Under condition~(R2), confirmatory preprocessing is
strictly causal and label-blind, so the committed endpoint and any
prospectively fixed transform of it are \(\F_{t_1}^{(j)}\)-measurable.

Part~3 is not an additional positive theorem. It records the limit of the
pre-selection result. Let \(S^{(j)}\) be the inclusion indicator for the
confirmatory retained sample. Even when \eqref{eq:si-fwd-null} holds before
selection, conditioning on \(S^{(j)}=1\) can induce dependence between
\(Y^{(j)}\) and \(\tauL^{(j)}\) if inclusion depends on the endpoint, the
assigned delay, delivery success, trigger timing, preprocessing success,
missingness, exclusion, analysis inclusion, or variables associated with them.
The retained sample therefore inherits no exclusion result from
\eqref{eq:si-exogeneity} alone. A sufficient retained-sample condition for the
confirmatory statistic is stated next.
\end{proof}

The proposition separates mechanism from scientific null. Condition~(R1) and its
audits qualify the scheduler's inputs, timing, concealment and implementation.
They do not establish \eqref{eq:si-exogeneity} against an alternative in which a
committed quantity anticipates the current draw. Proposition~\ref{prop:si-exclusion}
states what the past-adapted factorisation implies for any temporally admissible
statistic. A qualified ordering negates that class-level implication while
leaving the operational scheduler intact, subject to the remaining temporal,
frozen-object and retained-sample qualifications.

The exclusion is class-wide. Every member of the declared past-adapted class
satisfies both temporal admissibility and the factorisation in
\eqref{eq:si-exogeneity}. The pre-selection no-ordering implication follows from
that statistical restriction under the operational scheduler, not from temporal
order alone and not from the functional form, capacity or goodness of fit of any
one comparator. Once frozen-object invariance and the retained-sample
qualifications below are also satisfied, a material surviving association bears
conditionally on the declared class rather than only on the particular fitted
comparator. Additional predictive flexibility can improve fit or precision
within the class, but it cannot remove the no-ordering implication without
leaving the class or violating an identifying condition.

\subsection{Retained-sample delay-neutrality}
\label{sec:si-retained-neutrality}

The equations below state the additional condition required after artefact
rejection and other exclusions: among trials with the same committed endpoint,
locked covariates, randomisation stratum and frozen comparator, the probability
of retention must not vary with assigned delay.

Let
\[
\mathcal{Z}_j
:=
\big(\Apre^{(j)},\mathbf{X}_j\big),
\]
where \(\Apre^{(j)}\) is the committed endpoint and \(\mathbf{X}_j\) is the locked
past-adapted comparator-covariate vector. The factorisation needed for a
retained-sample theorem closes only for statistics that are measurable with
respect to the conditioning object used in the retained-sample condition. In the
confirmatory Level~II-A analysis this object is \(\mathcal{Z}_j\), the stratum, and the
frozen comparator object \(\mathcal{G}_{\mathrm{frz}}\). The theorem below is
therefore deliberately scoped to the committed endpoint and the frozen residual,
not to arbitrary past-adapted statistics in the full filtration.

\begin{definition}[Retained-sample delay-neutrality]
\label{def:r3star}
The retained-sample delay-neutrality condition is
\begin{equation}
\label{eq:si-r3star}
\mathrm{(R3^\star)}
\qquad
S^{(j)}
\ \perp\!\!\!\perp\
\tauL^{(j)}
\ \big|\ 
\big(
\mathcal{Z}_j,\mathcal{R}_j,\mathcal{G}_{\mathrm{frz}}
\big),
\end{equation}
with
\begin{equation}
\label{eq:si-retained-positivity}
\mathbb{P}
\!\left\{
S^{(j)}=1
\mid
\mathcal{Z}_j,\mathcal{R}_j,\mathcal{G}_{\mathrm{frz}}
\right\}
>0
\end{equation}
on the retained-support region used for the confirmatory estimand.

Equivalently, after conditioning on the committed endpoint, the locked
past-adapted covariates, the randomisation stratum, and the frozen comparator
object, inclusion in the confirmatory retained sample must not depend on the
assigned-delay label. This is a sufficient structural condition for
retained-sample exclusion of the committed endpoint and frozen residual. It is
not implied by randomisation. Where positivity fails, or where overlap is only
partial, the retained-sample estimand is restricted to the overlap region or the
dataset is routed to the declared selection-sensitivity and selection-limited
procedures.
\end{definition}

Condition~\eqref{eq:si-r3star} is intentionally a condition on the selection
mechanism, not a restatement of the desired endpoint-delay conclusion. It rules
out a retained-sample collider in which the assigned delay influences inclusion
after the endpoint and covariates are known. It also makes clear why marginal
retention balance is not enough: a selection rule may have no marginal
delay effect while still depending on the endpoint-by-delay configuration and
therefore reweighting the retained endpoint distribution across assigned-delay
bins.

This problem is related to, but distinct from, principal stratification and
other causal frameworks for post-assignment variables and missing outcomes
\citep{FrangakisRubin2002}. Those frameworks emphasise that randomisation of an
assigned variable does not by itself validate a contrast after analysis
inclusion has become an assignment descendant. Level~II-A does not define a
treatment-effect estimand or introduce latent principal strata. Instead, it
states a sufficient conditional delay-neutrality condition for the committed
endpoint and frozen residual and routes datasets to declared sensitivity or
selection-limited procedures when that condition cannot be qualified.

\begin{lemma}[Retained-sample exclusion for the confirmatory statistic]
\label{lem:retained-exclusion}
Assume the operational scheduler in (R1), temporal integrity in (R2), and the
pre-selection past-adapted factorisation \eqref{eq:si-exogeneity}. Let
\(\mathcal{Z}_j=(\Apre^{(j)},\mathbf{X}_j)\), and suppose
\(\mathrm{(R3^\star)}\) holds. Assume also that, conditional on the frozen
comparator object, the assigned-delay label remains independent of the locked
pre-assignment trial object:
\begin{equation}
\label{eq:si-frozen-independence}
\tauL^{(j)}
\ \perp\!\!\!\perp\
\mathcal{Z}_j
\ \big|\ 
\big(
\mathcal{R}_j,\mathcal{G}_{\mathrm{frz}}
\big).
\end{equation}
Condition~\eqref{eq:si-frozen-independence} imposes an additional
conditional-independence restriction that does not follow from the
pre-selection factorisation in \eqref{eq:si-exogeneity}. It conditions on the fitted object
\(\mathcal{G}_{\mathrm{frz}}\), and such conditioning can reintroduce dependence
between \(\tauL^{(j)}\) and \(\mathcal{Z}_j\) even when the pre-selection
factorisation holds. It is therefore a route-specific statistical condition
qualified by the frozen construction and its diagnostics, not an operational
scheduler condition or a corollary of randomisation. The pathways by which it
can fail, and the checks used to qualify it, are stated in
\Cref{sec:si-frozen-operational}.
Let \(h\) be any integrable prospectively fixed function of
\((\mathcal{Z}_j,\mathcal{G}_{\mathrm{frz}})\). Then
\begin{equation}
\label{eq:si-retained-exclusion}
\mathbb{E}
\!\left[
h(\mathcal{Z}_j,\mathcal{G}_{\mathrm{frz}})
\mid
\sigma(\tauL^{(j)}),
\mathcal{R}_j,
S^{(j)}=1,
\mathcal{G}_{\mathrm{frz}}
\right]
=
\mathbb{E}
\!\left[
h(\mathcal{Z}_j,\mathcal{G}_{\mathrm{frz}})
\mid
\mathcal{R}_j,
S^{(j)}=1,
\mathcal{G}_{\mathrm{frz}}
\right]
\quad\text{a.s.}
\end{equation}
In particular, the result applies to \(\Apre^{(j)}\) and to the frozen residual
endpoint
\[
\Apre^{\mathrm{resid},(j)}
=
\Apre^{(j)}
-
\widehat m_0^{(-k(j))}(\mathbf{X}_j),
\]
provided the comparator, folds, tuning rule, and residual array are frozen
without assigned-delay labels, delay-labelled residual diagnostics, or
assignment descendants that would invalidate the admissible reassignment
scheme. Equation~\eqref{eq:si-retained-exclusion} is a trial-local retained-sample
exclusion result. It does not by itself imply independence from the complete
retained assigned-delay vector or a zero expectation for a participant-level
ratio slope. The participant-level consequences are route-specific.

In the assignment-isolation route, the complete admissible reassignment law is
applied to the frozen residual array and retained design, as specified in
\Cref{sec:si-iso-route}. Under the joint retained-design and endpoint-array
invariance condition stated there, a within-stratum fixed-multiset assignment
law gives zero design expectation for the participant slope because the
stratum-centred assignment scores have expectation zero and the retained
leverage denominator is fixed by the multiset.

In the sequential route, the predictable-score argument of
\Cref{sec:si-sequential} makes each residual-weighted centred assignment
increment a martingale difference and supplies the route-specific e-value
calibration. That score-centering result does not by itself assert unbiasedness
of the reported ratio-form slope when its denominator is random.

For either route, positive assigned-delay leverage remains required, together
with prospectively fixed participant eligibility and non-estimability rules.
The slope functional must be locked, and aggregation must give equal weight to
participants. Population-level conclusions are additionally subject to the
participant-estimability qualification of
\Cref{sec:si-participant-estimability}.
\end{lemma}

\begin{proof}
Write
\[
q(\mathcal{Z}_j,\mathcal{R}_j,\mathcal{G}_{\mathrm{frz}})
=
\mathbb{P}
\!\left\{
S^{(j)}=1
\mid
\mathcal{Z}_j,\mathcal{R}_j,\mathcal{G}_{\mathrm{frz}}
\right\}.
\]
For bounded measurable \(\phi\) and integrable \(h\),
\[
\begin{aligned}
&
\mathbb{E}
\!\left[
h(\mathcal{Z}_j,\mathcal{G}_{\mathrm{frz}})
\phi(\tauL^{(j)})
\mathbf{1}_{\{S^{(j)}=1\}}
\mid
\mathcal{R}_j,\mathcal{G}_{\mathrm{frz}}
\right]
\\
&\quad =
\mathbb{E}
\!\left[
h(\mathcal{Z}_j,\mathcal{G}_{\mathrm{frz}})
\phi(\tauL^{(j)})
q(\mathcal{Z}_j,\mathcal{R}_j,\mathcal{G}_{\mathrm{frz}})
\mid
\mathcal{R}_j,\mathcal{G}_{\mathrm{frz}}
\right]
\\
&\quad =
\mathbb{E}
\!\left[
h(\mathcal{Z}_j,\mathcal{G}_{\mathrm{frz}})
q(\mathcal{Z}_j,\mathcal{R}_j,\mathcal{G}_{\mathrm{frz}})
\mid
\mathcal{R}_j,\mathcal{G}_{\mathrm{frz}}
\right]
\mathbb{E}
\!\left[
\phi(\tauL^{(j)})
\mid
\mathcal{R}_j,\mathcal{G}_{\mathrm{frz}}
\right].
\end{aligned}
\]
The first equality uses \(\mathrm{(R3^\star)}\). The second uses
\eqref{eq:si-frozen-independence}, because
\[
h(\mathcal{Z}_j,\mathcal{G}_{\mathrm{frz}})
q(\mathcal{Z}_j,\mathcal{R}_j,\mathcal{G}_{\mathrm{frz}})
\]
is measurable with respect to
\(\sigma(\mathcal{Z}_j,\mathcal{R}_j,\mathcal{G}_{\mathrm{frz}})\).

Taking \(h\equiv 1\) in the same display gives
\[
\mathbb{E}
\!\left[
\phi(\tauL^{(j)})
\mathbf{1}_{\{S^{(j)}=1\}}
\mid
\mathcal{R}_j,\mathcal{G}_{\mathrm{frz}}
\right]
=
\mathbb{P}
\!\left\{
S^{(j)}=1
\mid
\mathcal{R}_j,\mathcal{G}_{\mathrm{frz}}
\right\}
\mathbb{E}
\!\left[
\phi(\tauL^{(j)})
\mid
\mathcal{R}_j,\mathcal{G}_{\mathrm{frz}}
\right],
\]
so
\[
S^{(j)}
\ \perp\!\!\!\perp\
\tauL^{(j)}
\ \big|\ 
\big(
\mathcal{R}_j,\mathcal{G}_{\mathrm{frz}}
\big).
\]
This is the marginal delay-neutrality of retention implied by
\(\mathrm{(R3^\star)}\) and \eqref{eq:si-frozen-independence}.

Moreover,
\[
\mathbb{P}
\!\left\{
S^{(j)}=1
\mid
\mathcal{R}_j,\mathcal{G}_{\mathrm{frz}}
\right\}
=
\mathbb{E}
\!\left[
q(\mathcal{Z}_j,\mathcal{R}_j,\mathcal{G}_{\mathrm{frz}})
\mid
\mathcal{R}_j,\mathcal{G}_{\mathrm{frz}}
\right]
>0,
\]
Because the factorisation holds for every bounded measurable test function
\(\phi\) and every integrable prospectively fixed \(h\), division by this
positive conditional probability yields
\[
h(\mathcal{Z}_j,\mathcal{G}_{\mathrm{frz}})
\ \perp\!\!\!\perp\
\tauL^{(j)}
\ \big|\ 
\big(
\mathcal{R}_j,
S^{(j)}=1,
\mathcal{G}_{\mathrm{frz}}
\big).
\]
The retained-sample conditional-expectation equality
\eqref{eq:si-retained-exclusion} follows.

No conclusion about the participant slope is obtained by conditioning on the
realised full assigned-delay vector and then invoking the trial-local equality
\eqref{eq:si-retained-exclusion}. The two calibration routes instead supply
their own joint-design arguments. In the assignment-isolation route, the frozen
residual array is conditioned on and the complete participant- and
stratum-compatible assignment vector is regenerated under the admissible
reassignment law. The plus-one randomisation value is therefore calibrated by the
exchangeability or Monte Carlo rank argument for the observed statistic and its
admissible reassignment replicates. Where the design uses a fixed within-stratum
multiset, the centred score and its denominator are fixed by that design; more
general stochastic assignments are handled by the declared scheduler and
reassignment law rather than by a trialwise conditional-mean shortcut.

In the sequential route, the current frozen residual is measurable before the
current assignment draw and its product with the centred assignment increment
has conditional expectation zero by the martingale-difference argument in
\Cref{sec:si-sequential}. This establishes the centring required for the
sequential score and the corresponding e-value calibration. The denominator of
the reported sequential slope records the conditional assignment variance
supplied by the scheduler, but it may be random through the evolving
pre-assignment history. The martingale-difference result therefore does not
alone establish zero expectation of that ratio-form slope. The slope remains the
locked magnitude summary, whereas inferential validity is supplied by the
sequential e-value construction.

Equal-participant aggregation requires the separately declared
participant-estimability qualification. Thus the route-specific scheduler and reassignment law or predictable-score
argument, rather than
\eqref{eq:si-retained-exclusion} alone, supplies the relevant participant-level
calibration.

The marginal retention audit targets the necessary consequence
\(S^{(j)}\perp\!\!\!\perp\tauL^{(j)}
\mid(\mathcal{R}_j,\mathcal{G}_{\mathrm{frz}})\). The endpoint-by-delay collider
diagnostic targets a declared endpoint-by-delay interaction violation of
\(\mathrm{(R3^\star)}\). The selection-sensitivity gate bounds declared
violations within the sensitivity envelope. None of these procedures proves
\(\mathrm{(R3^\star)}\) globally.
\end{proof}

\subsection{Operational status of the frozen-comparator independence condition}
\label{sec:si-frozen-operational}

Because \eqref{eq:si-frozen-independence} conditions on the fitted object
\(\mathcal{G}_{\mathrm{frz}}\), it is not implied by the pre-selection
factorisation in \eqref{eq:si-exogeneity}. Conditioning on a fitted
object can reintroduce a dependence between \(\tauL^{(j)}\) and \(\mathcal{Z}_j\) through
either of two pathways. First, label contamination: if any assigned-delay label,
delay-labelled residual diagnostic, or delay-informed tuning choice enters the
construction of \(\mathcal{G}_{\mathrm{frz}}\), the frozen object becomes a
function of the labels and conditioning on it can break the independence.
Second, assignment descendants: if an endpoint used to fit the held-out residual
for trial \(j\) is a descendant of \(\tauL^{(j)}\) through carryover, then
\(\mathcal{G}_{\mathrm{frz}}\) can become a common descendant of
\(\tauL^{(j)}\) and \(\mathcal{Z}_j\), and conditioning on it can open the collider path
\[
\tauL^{(j)}
\longrightarrow
\mathcal{G}_{\mathrm{frz}}
\longleftarrow
\mathcal{Z}_j .
\]

The condition is therefore declared as a route-specific statistical
qualification. Its plausibility is assessed before unblinding by three locked
construction and diagnostic checks; those checks do not prove it globally. First, label invariance:
\(\mathcal{G}_{\mathrm{frz}}\), including fold assignment, nuisance fits, tuning
choices and residual construction, must be a fixed function of admissible
pre-assignment information. Injecting a placebo assigned-delay label into the
construction pipeline must leave the frozen comparator object and the residual
array unchanged. Second, assignment-descendant exclusion: the nuisance fit used
to construct the held-out residual for trial \(j\) must not use endpoint-derived
descendants of \(\tauL^{(j)}\). In the assignment-isolation route this is
operationally qualified by the reset, washout and lagged-delay diagnostics that
support endpoint-array invariance. Where carryover is possible, the analysis uses
prequential or participant-disjoint folds and the sequential martingale/e-value
route of \Cref{sec:si-sequential,sec:si-seq-calibration}. Third, route-specific
residual qualification is required: under assignment isolation, the frozen
residual array must be invariant under the declared admissible reassignment law;
under the sequential route, the current residual used at \((p,j)\) must be
predictable with respect to \(\F^{\mathrm{pre}}_{pj}\).

A dataset for which any of these checks fails is routed to diagnostic failure or
classified as operationally unqualified, never to support or to a clean
forward-only adequate null. These checks are sufficient operational
qualifications rather than necessary mathematical conditions. Like the other
retained-sample checks, they qualify
\eqref{eq:si-frozen-independence} within the declared operational envelope rather
than proving it globally.

\subsection{Scope of the statement}
\label{sec:si-scope}

Proposition~\ref{prop:si-exclusion} is a pre-selection boundary statement, not
a theorem that the comparator has captured every neural cause of anticipation.
It says that a temporally admissible past-adapted statistic cannot be ordered by
the post-endpoint assigned delay when the factorisation in
\eqref{eq:si-exogeneity} holds. The operational scheduler supplies the
post-commitment probe; the pre-selection exclusion follows from the past-adapted
factorisation, not from comparator completeness and not from temporal order
alone.

This distinction matters. A flexible comparator can absorb known
forward-accessible structure, such as foreperiod, hazard, CNV-like slow
potential, arousal, vigilance, motor preparation, and trial history. But the
pre-selection exclusion in \eqref{eq:si-fwd-null} would hold for any integrable
\(\F_{t_1}^{(j)}\)-measurable statistic, even without a perfect comparator.
The comparator improves precision and reduces conventional explanatory
loopholes; the operational randomisation boundary supplies the probe, while the
past-adapted factorisation supplies its pre-selection exclusion.

The retained-sample statement has narrower scope. It is not a claim that every
past-adapted statistic in \(\F_{t_1}^{(j)}\) remains delay-neutral after
conditioning on \(S^{(j)}=1\). The retained-sample theorem is deliberately
scoped to the committed endpoint and the frozen residual, because those are the
confirmatory Level~II-A objects. Mathematically, the factorisation closes only
when the target statistic is measurable with respect to the conditioning object
used in \(\mathrm{(R3^\star)}\), here \(\mathcal{Z}_j\), the stratum, and the frozen
comparator object, and when the fitted-object condition
\eqref{eq:si-frozen-independence} is operationally qualified as described in
\Cref{sec:si-frozen-operational}. A different past-adapted statistic
\(W^{(j)}\notin\sigma(\mathcal{Z}_j)\) could be
reweighted by a \(W^{(j)}\)-by-\(\tauL^{(j)}\) selection mechanism that cancels
on average after conditioning on \(\mathcal{Z}_j\), while still inducing retained-sample
delay dependence for \(W^{(j)}\). That stronger theorem is not needed for
Level~II-A and is not asserted.

Two clarifications make this role explicit. First, because \(Y^{(j)}\) is \(\F_{t_1}^{(j)}\)-measurable, the
past-adapted factorisation \eqref{eq:si-exogeneity} gives the stronger
conditional-independence statement
\[
Y^{(j)}
\ \perp\!\!\!\perp\
\tauL^{(j)}
\mid
\mathcal{R}_j .
\]
Thus the conditional law of \(Y^{(j)}\) given \(\tauL^{(j)}\) and
\(\mathcal{R}_j\) equals its conditional law given \(\mathcal{R}_j\).
Equation~\eqref{eq:si-fwd-null} records the mean-independence consequence
needed by the participant-slope estimand. Second,
Proposition~\ref{prop:si-exclusion} defines the pre-selection implication of
the forward-only null. In the assignment-isolation route, the finite-sample
exactness of the plus-one randomisation value in
\Cref{sec:si-randomisation-test} derives from the known scheduler law
\(P_{\mathrm{sch}}\), the design-compatible reassignment scheme and the
frozen-residual invariance condition under the sharp forward-only null, not from
the proposition itself.
In the sequential route, finite-sample calibration instead comes from the
martingale/e-value construction in \Cref{sec:si-seq-calibration}, using
predictable pre-assignment e-value eligibility, the declared current-trial
conditional scheduler law, residual predictability under the route's declared
conditioning set and the forward-only null.

A qualified material negative delay ordering of the committed endpoint or frozen
residual therefore localises to one of four classes:
\begin{enumerate}[leftmargin=2.0em,itemsep=2pt]
\item failure of the operational scheduler mechanism, through randomisation,
concealment, logging, scheduler or implementation failure;
\item failure of endpoint past-adaptedness or residual invariance, through
temporal leakage, post-\(t_1\) information entering the endpoint, comparator,
folds, nuisance fits, preprocessing path, or residual-freezing protocol;
\item failure of retained-sample delay-neutrality, through delivery error,
post-randomisation retention, missingness, exclusion, endpoint-by-delay
collider selection, analysis inclusion, or violation of
\(\mathrm{(R3^\star)}\);
\item failure of the past-adapted factorisation, with the operational scheduler
and the remaining qualifications intact, implying conditional insufficiency of
the declared class in the tested regime.
\end{enumerate}

The audit and sensitivity architecture is built to separate the first three
classes from the fourth. A dataset blocked by randomisation failure, temporal
leakage, delivery failure, implementation dependence, residual non-invariance,
or retained-sample selection is not evidence for a Level~II-A residual. It is
also not a clean forward-only adequate null. Only when the operational scheduler, endpoint lock, frozen-residual construction
and retained-sample delay-neutrality qualifications remain intact within the
declared sensitivity envelope can a surviving material negative slope be
interpreted as failure of the past-adapted factorisation and hence conditional
insufficiency of the declared class. A material
positive slope is screened against the same first three classes but remains a
prespecified opposite-direction diagnostic; even when those qualifications pass,
it carries no confirmatory class-insufficiency claim.


\section{Endpoint definition and preprocessing lock}
\label{sec:si-endpoint-preprocessing}

\subsection{Locked endpoint specification}
\label{sec:si-endpoint-spec}

The committed endpoint \(\Apre^{(j)}\) is a single-trial, signed,
baseline-corrected scalar feature computed on a pre-event window that closes at
\(t_1\), before the assigned-delay label is drawn or accessed. Because the
assigned label does not exist at \(t_1\), the imperative event is not yet
scheduled by that trial-specific delay when the endpoint window closes. The
endpoint is therefore a terminal pre-commitment slow-potential amplitude,
aligned to endpoint closure and not to the imperative event. The endpoint is
the object tested by Level~II-A, and it must be defined before any
delay-labelled residual plot, slope estimate, route-specific inferential
calibration, or audit outcome can influence the analysis.

The lock covers both the endpoint and the nuisance machinery used to construct
the frozen residual. Before delay labels are accessed, the protocol fixes the
recording specification, endpoint window, baseline interval, feature operation,
eligibility rules, comparator covariates, cross-fitting plan, loss function,
resolution floor, minimum participant count, assignment-isolation
randomisation replicate count, sequential e-value grid where that route is used,
and bootstrap settings. Once locked, these components are irrevocable for the
confirmatory analysis. Any change defines a new analysis regime and requires a
new prospectively declared protocol.

\Cref{tab:si-endpoint-lock} lists the required locked components. The table is
intended to prevent a common ambiguity in anticipatory analyses: the endpoint,
the preprocessing path, and the comparator are not allowed to evolve after
delay-ordered structure becomes visible. A supported residual is interpretable
only if it survives a feature definition and preprocessing chain that were
fixed while the assigned-delay labels were still unavailable.

\begin{table}[t]
\centering
\caption{Locked components of the committed pre-event endpoint and comparator.
All entries are fixed before assigned-delay labels are accessed. Any
post-label change defines a new analysis regime and cannot be part of the
confirmatory Level~II-A test.}
\label{tab:si-endpoint-lock}
\footnotesize
\begin{tabularx}{\textwidth}{@{}p{3.4cm}X@{}}
\toprule
\textbf{Component} & \textbf{Locked information} \\
\midrule
Recording &
Modality, sampling rate, channel or source set, montage, reference, acquisition
hardware, trigger pathway, and online filters, with online filters causal only. \\
Endpoint window &
Committed pre-event window \(I_{\mathrm{pre}}=[t_0,t_1]\) relative to the
warning cue or endpoint reference; baseline interval; endpoint closing time
\(t_1\); rule for handling boundary trials. \\
Feature &
Temporal operation, such as window average, mean slope, or signed component
score; aggregation rule across channels or sources; baseline correction; sign
convention; units of the scalar endpoint. \\
Eligibility &
Trial-inclusion and artefact criteria computable from
\(\F_{t_1}^{(j)}\); rejection thresholds; missing-data rules; participant
eligibility rules; minimum retained-delay leverage. \\
Comparator covariates &
Past-adapted covariate vector \(\mathbf{X}_j\) from \Cref{tab:si-comparator};
permitted transforms; permitted interactions; rules for standardisation,
imputation, and nuisance-feature construction; boundary conventions for
scheduler-derived covariates, including trials at or near the end of the
declared foreperiod support, with any truncation, winsorisation, exclusion,
or finite-limit substitution rule fixed in advance. \\
Folds and loss &
Cross-fitting fold structure, with assignment-relevant blocks kept intact;
comparator family; tuning rule; loss function; stopping rule; any
dimension-reduction step. \\
Decision constants &
Resolution floor \(\beta_{\min}\); minimum analysable participant count
\(N_{\min}\); assignment-isolation randomisation-replicate count \(R\);
sequential martingale/e-value grid where that route is used; bootstrap resample
count \(B\); one-sided level \(\alpha\); audit tolerances. \\
\bottomrule
\end{tabularx}
\end{table}

The endpoint lock does not require the endpoint to be simple. It may use a
scientifically motivated component, source-space summary, slow-potential
feature, or multichannel aggregation. The requirement is that the rule be fixed
before delay labels are available and that every operation used to compute the
endpoint be compatible with the pre-\(t_1\) information boundary.

\paragraph{Concrete instantiation.}
The abstract endpoint specification is instantiated concretely in the main-text
endpoint section; the description here uses the same instantiation and records
the label-blind calibration it presupposes. The representative choice is a
terminal CNV-like pre-assignment slow-potential endpoint: the signed,
baseline-corrected mean voltage over FCz and Cz during the final
\(250\,\mathrm{ms}\) before \(t_1\). The endpoint is baseline-corrected to a
prospectively fixed baseline interval, rereferenced under the locked montage,
and computed under strictly causal preprocessing so that no post-\(t_1\) sample
can enter the committed value. The sign is reversed so that larger \(\Apre\)
denotes greater late preparatory negativity. This anchor is used because the CNV
is a familiar anticipatory slow potential with known sensitivity to foreperiod,
hazard, temporal expectation, preparation and task history, not because it is
anomalous.

\paragraph{Label-blind calibration.}
The resolution floor \(\beta_{\min}\) in \eqref{eq:si-beta-min}, the retained
trial target, and the participant target cannot be declared from the endpoint
definition alone. They require quantities that must be estimated before any
confirmatory label is accessed, either in a dedicated label-blind calibration or
pilot dataset, or inherited from a locked calibration set. These quantities
include the within-participant residual scale
\(\sigma_{\mathrm{resid}}^{\mathrm{blind}}\) after the frozen forward-only
comparator, the retained-trial yield \(\bar n_{\mathrm{ret}}\) after causal
preprocessing and artefact rejection, participant-level slope heterogeneity,
assigned-to-measured timing error, delivery compliance, preprocessing losses and
exclusion losses. From these quantities the protocol fixes
\(\beta_{\min}\), the minimum analysable participant count \(N_{\min}\), and the
per-participant retained-trial target before the confirmatory dataset is opened.

The empirical assigned-delay support is set in the same label-blind
qualification stage to balance assigned-delay leverage against the foreperiod
and hazard structure the comparator must absorb. It is not the synthetic
benchmark support \(T_0=20\,\mathrm{ms}\), which is a short-support estimator
stress test rather than a proposed empirical anticipatory-EEG horizon. The
synthetic residual scale used in the benchmarks,
\(\sigma_{\mathrm{resid}}\approx1\,\mu\mathrm{V}\), is a simulation constant,
not an empirical EEG quantity, and does not substitute for empirical
calibration. The main text gives a worked feasibility calculation in which a
late-CNV endpoint with \(\sigma_{\mathrm{resid}}^{\mathrm{blind}}\approx
4\,\mu\mathrm{V}\), a \(600\,\mathrm{ms}\) empirical assigned-delay support, and
\(\bar n_{\mathrm{ret}}\approx300\) retained trials per participant yields a
single-participant resolution floor of about
\(2.7\,\mu\mathrm{V\,s^{-1}}\). That calculation is illustrative only; the
actual empirical floor, retained-trial target and \(N_{\min}\) must be fixed
from label-blind calibration in the intended acquisition regime.

\subsection{Strict causality and the preprocessing lock}
\label{sec:si-causality}

Every confirmatory operation that directly produces the endpoint or a comparator
covariate is strictly causal. An output at time \(t\le t_1\) may depend only on
samples at times \(u\le t\). The confirmatory endpoint may not depend on samples
after \(t_1\), even indirectly through filtering, detrending, artefact modelling,
baseline correction, interpolation, rejection, or feature normalisation.

The following operations are prohibited for confirmatory endpoint construction:
zero-phase filtering, bidirectional filtering, forward--reverse filtering,
symmetric smoothing, acausal time-shift correction, baseline or detrending steps
that use post-\(t_1\) samples, and any preprocessing, rejection, or inclusion
rule that depends on the assigned-delay label or a deterministic transform of
that label. These operations may be useful in exploratory EEG analysis, but they
are incompatible with a confirmatory test whose endpoint must be closed before
the delay label is accessed.

Label-blind artefact models are permitted only as nuisance procedures under the
same endpoint-lock constraint. Their training data, application window,
predictor set, thresholds, and rejection rules must be fixed before delay
access. They must also satisfy the temporal-leakage audit in
\Cref{sec:si-leakage-audit}: post-\(t_1\) activity must not be able to alter the
committed endpoint beyond the declared tolerance. If an artefact model uses
future samples, trial-level labels, or post-endpoint information in a way that
can change \(\Apre^{(j)}\), it is not admissible for the confirmatory endpoint.

The preprocessing lock also applies to comparator covariates. A covariate used
by the forward-only comparator may be rich, but it must be
\(\F_{t_1}^{(j)}\)-measurable. Covariates computed from later samples,
post-assignment delivery outcomes, retained-sample delay summaries, or
delay-labelled residual diagnostics are not admissible. The admissibility test
is informational: a quantity is allowed only if it would have been available
before the endpoint window closed and before the assigned-delay label was
accessed.

An acausal or label-dependent pipeline that fails these requirements may be
reported only as an exploratory robustness analysis, clearly separated from the
confirmatory result. It cannot rescue or replace the locked Level~II-A endpoint.
If the confirmatory endpoint fails the causal preprocessing lock, the dataset is
routed to diagnostic failure rather than to support or to a forward-only
adequate null.

\subsection{Temporal-leakage audit}
\label{sec:si-leakage-audit}

The temporal-leakage audit evaluates the complete locked pipeline for endpoint
and comparator construction before assigned-delay labels are accessed. The audit
uses synthetic records that are flat over \(I_{\mathrm{pre}}\) and contain
controlled post-\(t_1\) content. This content includes impulses, steps, ringing
patterns, and event-locked transients placed at the earliest and latest
admissible delivered latencies. These records pass through the same filtering,
baseline, artefact, endpoint, comparator-covariate and residual-construction
procedures as the confirmatory analysis.

The pipeline passes only if post-\(t_1\) inputs produce no committed pre-event
output beyond a declared tolerance \(\delta_{\mathrm{leak}}\). The tolerance is
fixed prospectively as a small fraction of the endpoint's label-blind noise
scale. The audit is evaluated before delay labels are accessed, and its result
is reported regardless of the final outcome. Failure invalidates the
confirmatory endpoint irrespective of the later fitted slope, route-specific
inferential value, or bootstrap bound.

This audit is deliberately mechanical. It does not ask whether leakage is likely
in the empirical dataset. It asks whether the locked processing chain is capable
of allowing post-endpoint information to enter the committed endpoint. If the
answer is yes beyond \(\delta_{\mathrm{leak}}\), the pre-event endpoint is no
longer protected by the randomisation boundary.

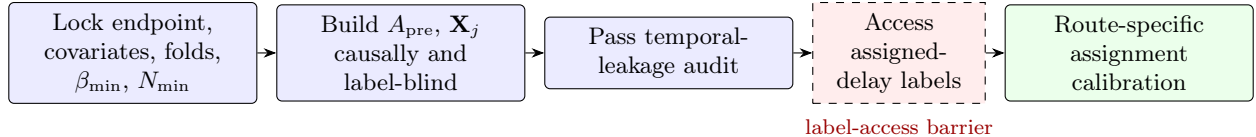
\begin{figure}[t]
\centering
\begin{tikzpicture}[>=Stealth,font=\footnotesize,node distance=2.5mm,
  blind/.style={draw,rounded corners=2pt,fill=blue!8,align=center,
                text width=3.0cm,minimum height=0.9cm,inner sep=4pt},
  conf/.style={draw,rounded corners=2pt,fill=green!8,align=center,
               text width=3.0cm,minimum height=0.9cm,inner sep=4pt},
  bar/.style={draw,dashed,fill=red!7,align=center,text width=2.0cm,
              minimum height=0.9cm,inner sep=4pt}]
\node[blind] (spec) {Lock endpoint, covariates, folds, \(\beta_{\min}\), \(N_{\min}\)};
\node[blind,right=of spec] (build) {Build \(\Apre\), \(\mathbf{X}_j\) causally and label-blind};
\node[blind,right=of build] (leak) {Pass temporal-leakage audit};
\node[bar,right=of leak] (barrier) {Access assigned-delay labels};
\node[conf,right=of barrier] (test) {Route-specific\\ assignment calibration};
\draw[->] (spec)--(build);
\draw[->] (build)--(leak);
\draw[->] (leak)--(barrier);
\draw[->] (barrier)--(test);
\node[below=2pt of barrier,font=\scriptsize,text=red!60!black]{label-access barrier};
\end{tikzpicture}
\caption{Endpoint and preprocessing lock. Endpoint construction, comparator
construction, fold assignment, decision constants, and temporal-leakage
validation occur label-blind. Assigned-delay labels are accessed only after the
locked pipeline passes the temporal-leakage audit.}
\label{fig:si-lock}
\end{figure}

\Cref{fig:si-lock} shows the ordering of the lock and the barrier it creates.
Endpoint construction, comparator construction, fold assignment, decision
constants, and leakage validation all occur before delay labels are accessed.
Only after the locked pipeline passes the audit does the analysis proceed to
delay-labelled route-specific assignment calibration.

\section{Forward-only comparator specification}
\label{sec:si-forward-comparator}

\subsection{Covariate class}
\label{sec:si-covariates}

The forward-only comparator is the declared past-adapted rival to a Level~II-A
residual. Its purpose is to absorb anticipatory structure that can be generated
from information already available before endpoint commitment: foreperiod,
hazard, temporal expectation, slow-potential build-up, arousal, motor
preparation, trial history, session structure, and preprocessing state. A
supported residual is therefore not a comparison against a weak null. It is a
comparison against a broad, prospectively locked past-adapted class.

Every comparator covariate must be \(\F_{t_1}^{(j)}\)-measurable for trial
\(j\). No assigned-delay label, realised event time, realised-delay stratum, or
deterministic transform of such a label may enter comparator fitting, tuning,
feature selection, preprocessing decisions, or fold construction. Scheduler
quantities are admissible only when they are known before \(t_1\), such as
block probabilities, elapsed foreperiod, conditional hazard, or the protocol
distribution available to the participant and experimenter. The specific
post-endpoint draw is not admissible.

This restriction is not a modelling preference. It is the temporal-admissibility
part of the past-adapted null; the statistical part is the factorisation in
\eqref{eq:si-exogeneity}. If a covariate could not have been known before the
committed endpoint was closed, including it in the comparator would leak the
object being tested. If a forward-accessible covariate group is omitted without prospective
justification, the residual could be dismissed as ordinary anticipation. \Cref{tab:si-comparator}
therefore lists the covariate groups that must be prospectively considered,
declared, and justified before delay-labelled endpoint analysis. A group that is
not recorded or is not applicable to the task may be omitted only when this is
stated before delay-label access; the resulting comparator is then reported as
less protective against that ordinary explanation.

\begin{table}[t]
\centering
\caption{Forward-only comparator covariate class. All entries must be
temporally admissible, meaning \(\F_{t_1}^{(j)}\)-measurable for trial \(j\),
and enter the past-adapted class under the factorisation in
\eqref{eq:si-exogeneity}. The realised
assigned-delay label and deterministic transforms of that label are excluded
from comparator fitting, tuning, and feature selection.}
\label{tab:si-comparator}
\footnotesize
\begin{tabularx}{\textwidth}{@{}p{3.6cm}X@{}}
\toprule
\textbf{Group} & \textbf{Covariates} \\
\midrule
Timing &
Elapsed foreperiod; conditional hazard; scheduler or block probability known
before \(t_1\); protocol timing variables available before endpoint commitment. \\
Slow potential &
CNV-like slow-potential amplitude or slope over a pre-\(t_1\) window disjoint
from the locked endpoint feature. \\
State &
Arousal or vigilance proxy, such as pupil diameter, alpha-band power, or a
sleep-stage or vigilance index; motor-preparation proxy declared before
analysis. \\
History &
Previous-trial timing and outcome; reaction-time history; previous assigned
delays when they are already realised from earlier trials; sequential indices;
time on task. \\
Structure &
Participant and session identifiers; block identifiers; acquisition covariates;
preprocessing covariates such as retained-channel counts, impedance summaries,
or artefact-burden summaries available before delay-labelled endpoint analysis. \\
Interactions &
Prospectively declared interactions and nonlinear transforms of the above,
including arousal-by-hazard, CNV-by-foreperiod, history-by-hazard, and
participant- or session-level smooth effects when allowed by the locked
comparator family. \\
\bottomrule
\end{tabularx}
\end{table}

The comparator may be intentionally rich, but it must remain past-adapted. A
high-capacity learner is allowed to model complex forward-accessible structure;
it is not allowed to learn the realised delay through timing leaks, fold
construction, post-randomisation exclusions, or features computed after the
assigned-delay label becomes available. The admissibility of a feature is
therefore decided by its information time, not by whether it improves
prediction.

\subsection{Family, cross-fitting and the frozen residual}
\label{sec:si-crossfit}

The comparator family may be high-capacity, including penalised regression,
gradient boosting, random forests, Gaussian-process regression, recurrent
predictors, or state-space predictors. The family, loss function, tuning rule,
feature dictionary, fold structure, stopping rule, and any dimension-reduction
step must be locked before assigned-delay access. Hyperparameter selection is
performed only inside the corresponding training partition. The held-out fold
is not used to tune the nuisance model that predicts it.

Outer cross-fitting folds are fixed prospectively. They keep randomisation
blocks and other assignment-relevant clusters intact. This prevents the
comparator from borrowing information across a boundary that the route-specific
inferential calibration later treats as fixed or predictable, following the
conditional cross-fitting logic of \citep{LuEtAl2025}.
For a trial \(j\) in fold \(k(j)\), the residual endpoint is
\begin{equation}
\label{eq:si-residual}
\Apre^{\mathrm{resid},(j)}
=
\Apre^{(j)}
-
\widehat{m}_0^{\,(-k(j))}(\mathbf{X}_j),
\qquad
\widehat{m}_0^{\,(-k)}
\approx
\mathbb{E}\!\left[\Apre \mid X,\ \mathrm{folds}\ne k\right].
\end{equation}
Here \(\mathbf{X}_j\) contains only the admissible past-adapted covariates described in
\Cref{tab:si-comparator}. The fitted value
\(\widehat{m}_0^{\,(-k(j))}(\mathbf{X}_j)\) is produced by a model trained without the
held-out fold containing trial \(j\).

Once held-out predictions have been generated for all trials, the residual
array is frozen. The route-specific inferential calibration, participant-slope
estimator, bootstrap bound, resolution-floor rule, audits, and selection diagnostics
all operate on this frozen residual object. Refitting, retuning, reselecting covariates,
changing fold assignments, changing preprocessing decisions, or changing the
endpoint after inspecting delay-ordered residuals is prohibited.

Freezing the residual has two purposes. First, it prevents nuisance-model
flexibility from becoming part of the inferential calibration. Second, in the
assignment-isolation route it makes the assignment test transparent: under the sharp forward-only null, the endpoint and frozen residual array are
treated as fixed, and only the assigned-delay vector is regenerated according to
the operational scheduler's admissible reassignment law. In the sequential route, the same frozen residual
object is instead required to be predictable with respect to the current
pre-assignment history. The comparator improves precision and guards against
conventional anticipatory explanations; it does not create the identifying
assignment contrast. Under the forward-only null, the precise statistical restriction required after
freezing is the frozen-comparator factorisation
\eqref{eq:si-frozen-independence},
\(\tauL^{(j)}\perp\!\!\!\perp \mathcal{Z}_j\mid(\mathcal{R}_j,\mathcal{G}_{\mathrm{frz}})\),
used in \Cref{lem:retained-exclusion}. It is not established by the scheduler
audits alone.

\subsection{Why residualisation does not identify the contrast}
\label{sec:si-design-based}

Residualisation is a precision and robustness device, not the source of
identification. The operational scheduler supplies the post-commitment probe;
the past-adapted factorisation, temporal integrity, frozen-object construction
and retained-sample conditions supply the no-ordering contrast. A flexible
comparator can remove forward-accessible structure, but it cannot by itself
establish either the scheduler mechanism or the statistical factorisation needed
for interpretation.

The fitted residual \(\Apre^{\mathrm{resid},(j)}\) is also not a simple
focal-trial measurement. A high-capacity nuisance model can borrow endpoint
information across training trials, and cross-fitting can make the held-out
prediction for trial \(j\) depend on the distribution of other endpoints in the
training folds. For that reason, the route-specific inferential calibration does
not treat each fitted residual as an independent draw from an i.i.d.\
superpopulation. In the assignment-isolation route, the entire frozen residual
array is held fixed and only the assigned-delay vector is regenerated under the
admissible reassignment law supplied by the operational scheduler and the sharp
forward-only null. In the sequential route, frozen-array reassignment is not
invoked; instead, each realised current-trial assignment increment is evaluated
against its declared pre-assignment conditional scheduler law under the
forward-only null.

For assignment isolation, covariate adjustment follows design-based
randomisation-inference logic
\citep{Freedman2008,Lin2013,ZhaoDing2021}.
Adjustment may improve precision and reduce sensitivity to forward-only
explanations, but the validity of that route comes from the design-compatible
randomisation distribution, not from a modelling claim that the residual is
conditionally independent under an i.i.d.\ superpopulation. The sequential route
uses the martingale/e-value calibration in \Cref{sec:si-seq-calibration}; it
requires predictability of the residual with respect to the current
pre-assignment history, correctness of the declared current-trial conditional
scheduler law, and the forward-only null for the residual-weighted score.

The sharp-null reassignment route is exact only when reassignment of
\(\tauL^{(j)}\) for trial \(j\) cannot change later committed endpoints. If
carryover from one assigned delay to later endpoints is possible, holding all
endpoints fixed while independently reassigning the current trial would no
longer represent the design. In that case the sequential martingale/e-value
route of \Cref{sec:si-estimand-inference,sec:si-seq-calibration} is used, so
that calibration follows the current-trial conditional scheduler law under the
forward-only null and does not invoke frozen-array reassignment.

The practical implication is simple. A supported Level~II-A residual must
survive the strongest admissible forward-only adjustment, but its
interpretation still depends on the design: proper randomisation, leakage
safety, delivery integrity, retained-sample delay-neutrality under the declared
\(\mathrm{(R3^\star)}\) envelope, and the declared reassignment law. A residual
that appears only after outcome-adaptive comparator choices, post-label
retuning, residual non-invariance, or invalid reassignment is not evidence
against the past-adapted account. It is an analysis failure.


\section{Confirmatory estimand and inference}
\label{sec:si-estimand-inference}

\subsection{Participant-level estimand}
\label{sec:si-estimand}

The confirmatory estimand is the equal-participant mean assigned-delay slope of
the frozen residual endpoint. The participant, not the trial, is the population
replication unit. This choice prevents participants with more retained trials
from dominating the estimand and aligns the inference with the intended
population claim: whether a typical eligible participant shows a negative
assigned-delay residual after the forward-only comparator has been frozen.

The corresponding slope on the unadjusted committed endpoint is a mandatory
adjustment-sensitivity report in every empirical application. Its point
estimate, sign and sign agreement with the frozen-residual slope are reported
alongside the confirmatory result. It is not part of the executable classifier
and does not alter the mutually exclusive outcome class. A supported
frozen-residual result accompanied by a non-negative unadjusted slope is marked
as adjustment-sensitive and is not described as robust to adjustment. Promoting
that diagnostic to a support condition or veto would require separate
operating-characteristic qualification of the resulting classifier.

Within participant \(p\), trial \(j\), and randomisation stratum
\(\mathcal{R}_{pj}\), the assigned delay is centred at the scheduler mean
\[
\bar\tau_{\mathcal{R}_{pj}}
=
\mathbb{E}_{P_{\mathrm{sch}}}
\!\left[
\tauL
\mid
\mathcal{R}_{pj}
\right].
\]
The participant-level working regression is
\begin{equation}
\label{eq:si-participant-slope}
\Apre^{\mathrm{resid},(pj)}
=
a_{\mathcal{R}_{pj}}
+
\beta_{\tau,p}
\bigl(
\tauL^{(pj)}-\bar\tau_{\mathcal{R}_{pj}}
\bigr)
+
\varepsilon_{pj},
\qquad
\beta_\tau
=
\mathbb{E}_p[\beta_{\tau,p}].
\end{equation}
Here \(\varepsilon_{pj}\) is the remainder from the
participant-specific, stratum-adjusted working projection; it is distinct from
the frozen comparator residual
\(\Apre^{\mathrm{resid},(pj)}\).
The stratum intercept \(a_{\mathcal{R}_{pj}}\) absorbs stratum-level endpoint
differences, including block, scheduler, or design features that are constant
within the randomisation stratum. The slope \(\beta_{\tau,p}\) is the
participant-specific residual association between the assigned delay and the
committed endpoint after label-blind forward-only adjustment. The target
\(\beta_\tau\) is the participant-population mean of these participant-specific
slopes.

The certified assignment-isolation benchmark implements the special case in
which participant is the reassignment block and no finer within-participant
randomisation stratum enters the slope. It centres the retained assigned-delay
vector at the retained participant mean before computing the slope and its
denominator leverage. With a participant intercept, this constant recentring
leaves the participant slope unchanged. Empirical implementations with finer
declared strata retain the corresponding stratum intercepts and require
separate operating-characteristic qualification.

The sign convention is fixed before analysis. Level~II-A directional support
requires a materially negative value of \(\beta_\tau\), meaning that shorter
assigned delays correspond to larger residual pre-event endpoints, or
equivalently that the residual endpoint weakens as assigned delay increases.
A positive material slope is not support with the sign reversed; it is routed to
the opposite-direction outcome.

Each eligible and estimable participant contributes one slope with equal weight.
Participants whose slope is non-estimable, for example because too few
assigned-delay levels remain after exclusions or because the denominator has
insufficient leverage, are handled by a prospectively fixed rule that is applied
identically in every Monte Carlo replicate and every empirical dataset. Such
participants are not removed or retained based on the sign or magnitude of their
estimated slope. Equal weighting preserves the population interpretation but
need not be variance-optimal: participants who barely clear the locked leverage
threshold can contribute more variable ratio slopes. The minimum-leverage rule,
participant-estimability sensitivity and whole-participant resampling are the
prespecified protections against that instability.

Estimability is nevertheless a post-assignment selection property. Whether a
participant retains enough assigned-delay leverage can depend on delivery,
artefact rejection, preprocessing success, exclusion and inclusion, and these
steps can depend on the committed endpoint or on participant-level properties
associated with the endpoint. The equal-participant mean over the estimable set
is therefore a population-participant estimand only under an additional
participant-level qualification: estimability must not select participants in a
way that is associated with the participant-level assigned-delay slope after
conditioning on the locked eligibility and calibration variables. This
participant-level qualification is the extra condition needed to aggregate the
trial-level retained-sample exclusion of \Cref{lem:retained-exclusion} into the
equal-participant mean null. It is not implied by the trial-level
\(\mathrm{(R3^\star)}\), which concerns trial inclusion given
\((\mathcal{Z}_j,\mathcal{R}_j,\mathcal{G}_{\mathrm{frz}})\).

The protocol therefore reports the non-estimable fraction, the reasons for
non-estimability, the retained-trial yield and assigned-delay leverage by
participant. It also reports label-blind endpoint and residual-scale summaries
for estimable and non-estimable participants. The participant-level estimability
sensitivity analysis in \Cref{sec:si-participant-estimability} is applied before
any population-level magnitude conclusion is drawn. If the estimable fraction is too low, if estimable and
non-estimable participants differ materially on locked label-blind summaries, or
if the sensitivity analysis cannot rule out a conclusion-changing
participant-level selection effect, the dataset is routed to
selection-limited or inconclusive classification rather than to support or a
clean forward-only adequate null. If the number of eligible and estimable
participants falls below \(N_{\min}\), the assignment-calibrated slope may be
reported descriptively, but no confirmatory population-level magnitude conclusion is
drawn.

\subsection{Intention-to-treat indexing}
\label{sec:si-itt}

The primary analysis is indexed by the assigned delay label \(\tauL^{(j)}\), not
by the measured realised latency
\[
\widetilde{\tau}_L^{(j)}
=
t_{\mathrm{event}}^{(j)}-t_1 .
\]
Here \(t_{\mathrm{event}}^{(j)}\) is the logged onset time of the delivered
imperative event on trial \(j\).
This is an intention-to-treat convention for the randomised timing label. It
preserves the randomisation guarantee because \(\tauL^{(j)}\) is the quantity
drawn by the scheduler and protected by concealment. By contrast, the measured
latency can be affected by operating-system jitter, display latency, audio
latency, dropped frames, trigger delay, device buffering, or other delivery
imperfections occurring after assignment.

The measured latency is still essential, but it has a different role. It is
retained for the delivery audit, compliance summaries, latency-error bounds,
and descriptive reporting of assigned-to-delivered timing accuracy
(\Cref{sec:si-leakage-controls}). It is not substituted for the assigned label
in the primary estimand. Substitution would change the estimand from an
assignment-protected contrast to a delivered-latency contrast, and would make
the analysis vulnerable to post-assignment delivery mechanisms.

This separation also clarifies how timing failures are handled. Small,
prospectively bounded delivery error is reported as compliance information
around the assigned-label analysis. Systematic or stratum-dependent delivery
error that exceeds the declared tolerance is not corrected by switching to
\(\widetilde{\tau}_L^{(j)}\) as the primary label. It is routed through the
delivery audit and, when support-blocking, prevents a Level~II-A support claim.

\subsection{Sequential martingale/e-value route under carryover}
\label{sec:si-sequential}

When carryover makes isolated label reshuffling invalid, the route below
preserves the trial chronology by comparing each realised delay with the
assignment distribution available immediately before that trial's draw.

The assignment-isolation randomisation route treats the committed endpoint array
as invariant under reassignment of the current trial's assigned delay. That
route is valid only when the assigned delay on trial \(j\) cannot affect later
committed endpoints in a way that would be changed by reassignment. If such
carryover is possible, the sharp-null endpoint-array invariance used by the
simple reassignment test is no longer exact.

The default in that case is the sequential martingale/e-value route. The
current endpoint and frozen residual are treated as part of the pre-assignment
history for the current draw, and the assignment increment is evaluated
conditionally on that history. Here \(\F_{pj}^{\mathrm{pre}}\) denotes the information available
immediately before the assigned-delay draw for participant \(p\), trial \(j\),
including the declared randomisation stratum \(\mathcal{R}_{pj}\) and any
fixed-multiset, without-replacement, adaptive-blocking, or scheduler-state
constraints that affect the conditional law of the current draw. Define
\[
\mu_{pj}
=
\mathbb{E}_{P_{\mathrm{sch}}}
\!\left(
\tauL
\mid
\F_{pj}^{\mathrm{pre}}
\right),
\qquad
\Delta\tau_{pj}
=
\tauL^{(pj)}-\mu_{pj},
\qquad
v_{pj}
=
\operatorname{Var}_{P_{\mathrm{sch}}}
\!\left(
\tauL
\mid
\F_{pj}^{\mathrm{pre}}
\right).
\]
In the special case of independent draws within a fixed stratum, these reduce
to the stratum-conditional mean and variance.

Let \(\mathcal{J}_p\) denote the retained, eligible trials for participant \(p\)
under the locked eligibility and non-estimability rules. The participant
sequential-score slope is
\begin{equation}
\label{eq:si-sequential-slope}
\widehat{\beta}_{\tau,p}^{\,\mathrm{seq}}
=
\frac{
\sum_{j\in\mathcal{J}_p}
\Apre^{\mathrm{resid},(pj)}\,\Delta\tau_{pj}
}{
\sum_{j\in\mathcal{J}_p}
v_{pj}
}.
\end{equation}
The numerator is a sum of residual endpoints multiplied by centred assignment
increments. By construction of \(\mu_{pj}\) from the operational current-trial
scheduler law,
\[
\mathbb{E}_{P_{\mathrm{sch}}}
\!\left[
\Delta\tau_{pj}
\mid
\F_{pj}^{\mathrm{pre}}
\right]
=
0.
\]
This mean-zero centring is a property of the scheduler law; it is not itself the
scientific null. Under the forward-only null, if the frozen endpoint residual is
\(\F_{pj}^{\mathrm{pre}}\)-measurable, then
\[
\mathbb{E}_0
\!\left[
\Apre^{\mathrm{resid},(pj)}\,\Delta\tau_{pj}
\mid
\F_{pj}^{\mathrm{pre}}
\right]
=
\Apre^{\mathrm{resid},(pj)}
\mathbb{E}_{P_{\mathrm{sch}}}
\!\left[
\Delta\tau_{pj}
\mid
\F_{pj}^{\mathrm{pre}}
\right]
=
0.
\]
Thus the martingale-difference property of the residual-weighted score is the
joint consequence of the operational scheduler law, residual predictability and
the forward-only null. For a prequential construction, predictability can be
established directly with respect to \(\F_{pj}^{\mathrm{pre}}\): the nuisance
fit, including fold assignment, standardisation, imputation, tuning and
dimension reduction, must not use any endpoint or derived quantity that is a
descendant of \(\tauL^{(pj)}\) and is not already contained in
\(\F_{pj}^{\mathrm{pre}}\). A prequential fold trained only on information
available before the current draw satisfies this condition.

Participant-disjoint outer folds use a different conditioning argument. The
residual for participant \(p\) is constructed from a prospectively fixed
external-fold nuisance object trained without data from participant \(p\)'s
fold. The sequential mean-one argument is then conditional on that frozen
external-fold object in addition to \(\F_{pj}^{\mathrm{pre}}\). Experimental
independence must ensure that conditioning on the external-fold object does not
alter the declared current-trial scheduler law for participant \(p\); conditional
on that object, the current residual is fixed before the current assignment is
evaluated.

If the locked fold structure does not establish the required predictability and
conditional-scheduler-law conditions, the sequential route is marked invalid
and the executable outcome is inconclusive rather than support or a
forward-only adequate null. A separately detected operational audit failure
retains its own diagnostic-failure classification. The
denominator in \eqref{eq:si-sequential-slope} records the conditional assignment
variance actually supplied by the scheduler within the eligible trials of
participant \(p\). Participants with insufficient assignment variance are
handled by the same prospectively fixed non-estimability rule used for the
participant-level estimand.

The sequential route preserves the design logic without imposing an
incorrect sharp-null reassignment. It uses the conditional scheduler law at
each trial, given the information available before that assignment. The current
implementation obtains frozen residuals in prospectively fixed,
participant-disjoint outer folds. Each participant belongs to one fold, and the
comparator producing that participant's residuals is trained without any data
from the participant's fold. Within each fold, the one-step assignment factors
are multiplied in the declared order and mixed over the fixed
\(\lambda\)-grid. The resulting fold e-values are combined by a second fixed
convex mixture. They are never multiplied across folds. At the declared
reporting look, the final fold mixture is an e-value. This route does not form a
frozen-array conditional-randomisation distribution and is not reported as an
exact plus-one randomisation test.

The assignment-isolation route is permitted only when predeclared reset or
washout diagnostics justify endpoint-array invariance. These diagnostics include
lagged endpoint-on-previous-delay checks, declared pass thresholds, minimum
retained-trial yield, and sufficient assigned-delay leverage. If these
diagnostics fail, the analysis reverts to the sequential martingale/e-value
route. If the sequential route cannot be implemented because adequate
participant count, assignment variance, frozen-residual predictability or
another route-validity requirement cannot be established, the resulting
classification is inconclusive rather than support or a forward-only adequate
null. A failed operational audit is classified separately under the
non-compensatory diagnostic-failure rule.


\section{Route-specific assignment calibration}
\label{sec:si-randomisation-test}

\subsection{Assignment-isolation route}
\label{sec:si-iso-route}

In the assignment-isolation route, the confirmatory randomisation value is
design-based. It does not treat trials as independent draws from an i.i.d.\
superpopulation, and it does not rely on the forward-only comparator being a
correctly specified outcome model. The route conditions on the frozen residual
array and asks whether the observed ordering of those residuals by assigned
delay is unusually negative under the sharp forward-only null, calibrated by
the operational scheduler law actually used.

The test statistic is the equal-participant average slope
\[
\widehat{\beta}_{\tau}
=
\frac{1}{N}
\sum_{p=1}^{N}
\widehat{\beta}_{\tau,p},
\]
where \(N\) is the number of eligible and estimable participants entering the
confirmatory estimator. The statistic is computed from the frozen residual
endpoints \(\Apre^{\mathrm{resid},(pj)}\). Each eligible participant contributes
one slope, so participants with more retained trials do not dominate the
statistic. The negative direction is fixed prospectively: smaller values of
\(\widehat{\beta}_{\tau}\) give stronger evidence for the declared Level~II-A
directional residual.

The assignment-isolation route separates the operational replicate mechanism
from the sharp statistical null. The scheduler supplies the admissible
reassignment distribution. The sharp forward-only null additionally requires
that each committed residual endpoint remain invariant when the assigned-delay
vector is regenerated under that distribution. Every replicate therefore keeps
the residual array fixed and regenerates only the assigned-delay vector. If the
design used fixed multisets, the multiset is permuted only within its declared
block or stratum. If the design used stochastic draws, the draw is repeated from
the declared scheduler distribution
\(P_{\mathrm{sch}}(\mathrm{d}\tauL\mid\mathcal{R}_{pj})\). No replicate
moves an assignment across participant, session, block, device condition,
scheduler stratum or any other declared assignment stratum.

The comparator is not refitted inside an assignment-isolation randomisation
replicate. Residuals, participant eligibility, fold assignments, endpoint
definitions, and preprocessing decisions remain fixed. This restriction prevents
the nuisance model from becoming part of the reassignment calibration. The only random object in a replicate is the assignment label regenerated under
the design-compatible scheduler law; validity also requires the sharp-null
invariance of the frozen objects held fixed in that replicate.

The validity condition for this route is joint rather than trial-local.
Conditional on the locked retained trial set \(\mathcal{J}_p\), the retained
stratum pattern \((\mathcal{R}_{pj})_{j\in\mathcal{J}_p}\), the frozen comparator
object \(\mathcal{G}_{\mathrm{frz}}\), the complete frozen residual array
\((\Apre^{\mathrm{resid},(pj)})_{j\in\mathcal{J}_p}\), and the locked participant
eligibility and estimability decisions, the retained assigned-delay vector must
still follow the declared admissible reassignment law. Equivalently, conditioning
on those frozen objects must not alter the within-stratum fixed-multiset
permutation law, or the scheduler law \(P_{\mathrm{sch}}\) used to generate the
replicates. This retained-design and endpoint-array invariance condition is the
vector-level analogue of the retained-sample delay-neutrality condition
\(\mathrm{(R3^\star)}\) of \Cref{def:r3star}: it stands to the trial-local
equality \eqref{eq:si-retained-exclusion} as the vector exclusion stands to the
marginal one, and it is strictly stronger than applying
\eqref{eq:si-retained-exclusion} separately to each retained trial. It is a route-specific statistical condition qualified by the locked
construction and diagnostics, not an operational scheduler audit and not a
corollary of randomisation alone.

Under this condition, and for a within-stratum fixed-multiset reassignment law,
the route delivers the participant-slope design expectation directly. The law is
exchangeable within each retained stratum, so the design-expected label at each
retained position equals that stratum's retained multiset mean, and the
stratum-centred assignment score at each retained position has design expectation
zero. The frozen residual array is held fixed, and the assigned-delay leverage
and slope denominator are fixed by the retained multiset. The participant-slope
numerator is therefore a fixed linear combination of frozen residuals whose
coefficients have design expectation zero within each retained stratum, so
\[
\mathbb{E}_{P_{\mathrm{sch}}}
\!\left[
\widehat{\beta}_{\tau,p}
\ \middle|\
\big(\Apre^{\mathrm{resid},(pj)}\big)_{j\in\mathcal{J}_p},
\mathcal{J}_p,
\big(\mathcal{R}_{pj}\big)_{j\in\mathcal{J}_p},
\mathcal{G}_{\mathrm{frz}}
\right]
=
0
\]
for every eligible and estimable participant whose retained design satisfies the
declared fixed-multiset reassignment conditions. The conditioning set is exactly
the frozen retained design of the joint condition above; the equal-participant
estimator then has design expectation zero by linearity, subject to the separate
participant-level qualification in \Cref{sec:si-participant-estimability}.

For a stochastic-draw scheduler law the leverage denominator is itself random,
so the slope is a ratio of random quantities and a zero design expectation does
not follow from route-specific calibration alone. Exact calibration of the
one-sided value still holds by regenerating assignments under
\(P_{\mathrm{sch}}\), but a zero design expectation for the ratio-form slope is
asserted only once the corresponding conditional-unbiasedness result for the
locked slope functional has been established; the predictable-score construction
of \Cref{sec:si-seq-calibration} is the route for that case.

The assignment-isolation route is used only when endpoint-array invariance is
defensible. If the assigned delay on one trial may affect later committed
endpoints, the sharp-null reassignment of isolated labels no longer represents
the protocol. In that case the sequential martingale/e-value route of
\Cref{sec:si-seq-calibration} is used instead.

\subsection{One-sided assignment-isolation randomisation value}
\label{sec:si-pvalue}

For the assignment-isolation route only, let
\(\widehat{\beta}_{\tau}^{(r)}\), \(r=1,\dots,R\), be the replicate statistics
generated under the admissible reassignment law. The one-sided plus-one
randomisation value for the negative-direction test is
\begin{equation}
\label{eq:si-pvalue}
p_{\mathrm{rand}}
=
\frac{
1+\#\{r:\widehat{\beta}_{\tau}^{(r)}
\le
\widehat{\beta}_{\tau}\}
}{
1+R
}.
\end{equation}
The inequality is in the negative-support direction. A replicate statistic
counts as at least as extreme as the observed statistic when it is no larger
than the observed slope.

The number of replicates \(R\) is declared prospectively for the
assignment-isolation route. Where the admissible reassignment set is small
enough, exhaustive enumeration replaces Monte Carlo sampling and gives the exact
randomisation distribution. Otherwise, Monte Carlo replicates are generated from the declared admissible
scheduler law. The assignment-isolation criterion passes when
\(p_{\mathrm{rand}}\le\alpha\) for the declared one-sided level \(\alpha\)
\citep{Harris2023}. The symbol \(p_{\mathrm{rand}}\) is reserved for this
assignment-isolation calibration. Carryover-sensitive sequential analyses use
the e-value-derived quantity \(p_{\mathrm{seq}}\) defined in
\Cref{sec:si-seq-calibration}; both routes feed the common route-neutral
decision field \(p_{\mathrm{infer}}\).

The route-specific inferential value is one decision object, not the whole
decision rule. In the assignment-isolation route it is \(p_{\mathrm{rand}}\);
in the sequential route it is \(p_{\mathrm{seq}}\). In both cases, it asks
whether the observed signed delay score is unusually negative under the
route-specific forward-only null, using the operational scheduler law for
calibration. Directional support additionally requires the
participant-bootstrap upper bound to clear \(-\beta_{\min}\), the analysable
participant count to meet \(N_{\min}\), and the audit, collider-diagnostic, and
selection-sensitivity gates to pass. A small route-specific inferential value
does not rescue a failed magnitude bound or a failed audit.

\subsection{Sequential martingale/e-value calibration}
\label{sec:si-seq-calibration}

The equations below turn the trial-by-trial assignment comparison into
calibrated evidence. The operational scheduler supplies the current-trial
conditional support and assignment probabilities. The mean-one property of the
resulting one-step factor is a statistical statement under the forward-only
null and the route-specific predictability conditions, not an audit of the
scheduler mechanism. Repeated directional alignment between the frozen residual
and the component of the assigned delay that was unpredictable before the
current draw can then accumulate evidence against that null.

When a current assignment may affect later committed endpoints, the
assignment-isolation sharp-null argument is not exact because regenerating an
earlier label while holding later endpoints fixed no longer represents the
declared data-generating sequence. The sequential route therefore does not
reshuffle earlier assigned labels while treating later endpoints as invariant.
Instead, it evaluates each realised assignment against the conditional scheduler
law available immediately before that draw.

Let \(f(p)\in\{1,\ldots,F\}\) denote the prospectively fixed outer fold for
participant \(p\), with every participant belonging to exactly one e-value fold.
For participant-disjoint fitting, let
\(\mathcal{G}^{\mathrm{ext}}_{f}\) denote the frozen nuisance object trained
without data from fold \(f\). For each trial admitted by the predictable
pre-assignment e-value eligibility rule, let
\[
r_{pj}
=
\Apre^{\mathrm{resid},(pj)}
\]
be the frozen residual. Under a prequential construction, \(r_{pj}\) is formed
from information available before the current draw. Under the
participant-disjoint construction used by the benchmark, \(r_{pj}\) is a fixed
function of the current participant's admissible pre-assignment information and
\(\mathcal{G}^{\mathrm{ext}}_{f(p)}\).

E-value eligibility itself must be predictable. Whether trial \((p,j)\)
contributes a one-step factor is fixed without using the realised current-trial
assignment or any of its descendants. The eligibility rule includes a valid
conditional scheduler law, a finite frozen residual, a finite conditional
assignment mean and positive conditional assignment variance. Post-assignment
retention and the later participant-estimability filter do not determine
membership in the e-value product. The latter is used only for the reported
participant slope and population-magnitude calculations. Equivalently, a trial
that is not predictably eligible contributes factor one rather than being
selected after its current assignment is observed.

Let
\[
\mathcal{S}_{pj},\qquad
\pi_{pj}(s)
=
P_{\mathrm{sch}}\!\left(
\tauL=s
\mid
\F_{pj}^{\mathrm{pre}}
\right),
\qquad
\mu_{pj}
=
\sum_{s\in\mathcal{S}_{pj}}
s\,\pi_{pj}(s)
\]
be the declared conditional support, mass function and mean for the current
assigned-delay draw. The history \(\F_{pj}^{\mathrm{pre}}\) contains the
randomisation stratum, scheduler state, remaining fixed-multiset counts and any
other prospectively declared design information that determines the conditional
law of the current draw. It excludes the realised current-trial assigned delay.

The displayed construction assumes the discrete conditional scheduler used in
the benchmark. For a continuous scheduler law, the normalising sum below is
replaced by the corresponding conditional integral; such an implementation
requires separate route-specific qualification.

For the negative and positive directions, respectively, define
\[
X_{pj}^{(-)}
=
-r_{pj}\{\tauL^{(pj)}-\mu_{pj}\},
\qquad
X_{pj}^{(+)}
=
r_{pj}\{\tauL^{(pj)}-\mu_{pj}\}.
\]
For \(d\in\{-,+\}\), let \(Z_{pj}^{(d)}(s)\) denote the corresponding support
value obtained by replacing the realised current assignment in
\(X_{pj}^{(d)}\) by \(s\in\mathcal{S}_{pj}\), while keeping the frozen residual
and all pre-assignment objects fixed. This support substitution is used only to
evaluate the one-step normalising factor; it does not regenerate, replace or
counterfactually reshuffle the committed endpoint or frozen residual.

For every prospectively declared \(\lambda>0\), define the one-step factor
\[
E_{pj}^{(d)}(\lambda)
=
\frac{
\exp\{\lambda X_{pj}^{(d)}\}
}{
\displaystyle
\sum_{s\in\mathcal{S}_{pj}}
\pi_{pj}(s)
\exp\{\lambda Z_{pj}^{(d)}(s)\}
}.
\]

For a prequential construction, if the frozen residual and conditional
scheduler law are \(\F_{pj}^{\mathrm{pre}}\)-measurable, then under the
forward-only null
\[
\mathbb{E}_0\!\left[
E_{pj}^{(d)}(\lambda)
\mid
\F_{pj}^{\mathrm{pre}}
\right]
=
1.
\]
Indeed, conditional on \(\F_{pj}^{\mathrm{pre}}\),
\[
\mathbb{E}_0\!\left[
E_{pj}^{(d)}(\lambda)
\mid
\F_{pj}^{\mathrm{pre}}
\right]
=
\frac{
\displaystyle
\sum_{s\in\mathcal{S}_{pj}}
\pi_{pj}(s)
\exp\{\lambda Z_{pj}^{(d)}(s)\}
}{
\displaystyle
\sum_{s\in\mathcal{S}_{pj}}
\pi_{pj}(s)
\exp\{\lambda Z_{pj}^{(d)}(s)\}
}
=
1.
\]

Participant-disjoint outer folds require a different conditioning argument.
For participant \(p\), the frozen residual is constructed using
\(\mathcal{G}^{\mathrm{ext}}_{f(p)}\), which contains training information from
outside that participant's fold and therefore need not itself belong to
\(\F_{pj}^{\mathrm{pre}}\). The required route-specific scheduler-separation
condition is
\[
\mathbb{P}_0\!\left(
\tauL^{(pj)}=s
\mid
\F_{pj}^{\mathrm{pre}},
\mathcal{G}^{\mathrm{ext}}_{f(p)}
\right)
=
P_{\mathrm{sch}}\!\left(
\tauL=s
\mid
\F_{pj}^{\mathrm{pre}}
\right)
=
\pi_{pj}(s),
\qquad
s\in\mathcal{S}_{pj}.
\]
Thus conditioning on the frozen external-fold nuisance object must not change
the declared current-participant scheduler law. Experimental independence
between participants, together with scheduler separation across participants,
is the sufficient condition used by the participant-disjoint benchmark
construction.

Conditional on
\(\F_{pj}^{\mathrm{pre}}\) and
\(\mathcal{G}^{\mathrm{ext}}_{f(p)}\), the residual, support and scheduler
probabilities are fixed before the current assignment is evaluated. Hence,
under the forward-only null,
\[
\mathbb{E}_0\!\left[
E_{pj}^{(d)}(\lambda)
\mid
\F_{pj}^{\mathrm{pre}},
\mathcal{G}^{\mathrm{ext}}_{f(p)}
\right]
=
1,
\]
with the same numerator-over-normaliser calculation as above. The prequential
and participant-disjoint constructions therefore use different conditioning
sets but the same mean-one one-step factor.

Let \(\mathcal{I}^{\mathrm{ev}}_f\) be the prospectively and predictably ordered
eligible trials whose participants belong to fold \(f\). In that fixed order,
define
\[
E_f^{(d)}(\lambda)
=
\prod_{(p,j)\in\mathcal{I}^{\mathrm{ev}}_f}
E_{pj}^{(d)}(\lambda).
\]
Because eligibility is predictable and every included factor has conditional
mean one under the appropriate route-specific conditioning law, the finite
within-fold product is a non-negative test martingale with unit initial value,
and hence an e-value at the declared reporting look.

The executable benchmark uses the fixed grid
\[
\Lambda
=
\{1,2,5,10,20,50,100,200\}.
\]
Under the executable unit convention, residuals are expressed in
\(\mu\mathrm{V}\) and assigned delays in seconds. Hence
\(X_{pj}^{(d)}\) has units \(\mu\mathrm{V\,s}\), and every grid value
\(\lambda\) has units \((\mu\mathrm{V\,s})^{-1}\), so the exponent is
dimensionless. The numerical grid is tied to this convention. Any change of
amplitude or time units requires reciprocal rescaling of the grid; any empirical
retuning must be fixed using label-blind calibration and separately qualified.

The benchmark uses equal \(\lambda\)-weights. It first forms the fixed convex
mixture within each fold,
\[
E_f^{(d)}
=
\sum_{\lambda\in\Lambda}
w_\lambda E_f^{(d)}(\lambda),
\qquad
w_\lambda=|\Lambda|^{-1},
\]
and then combines folds through a second prospectively fixed convex mixture,
\[
E_{\mathrm{mix}}^{(d)}
=
\sum_{f=1}^{F}
\omega_f E_f^{(d)},
\qquad
\omega_f\geq0,
\qquad
\sum_{f=1}^{F}\omega_f=1.
\]
The locked benchmark uses equal fold weights. Fold e-values are not multiplied.
For each fold,
\(\mathbb{E}_0[E_f^{(d)}]\leq1\); therefore linearity of expectation gives
\[
\mathbb{E}_0\!\left[
E_{\mathrm{mix}}^{(d)}
\right]
=
\sum_{f=1}^{F}
\omega_f
\mathbb{E}_0\!\left[
E_f^{(d)}
\right]
\leq
1,
\]
without requiring independence among fold contributions. The fixed convex
mixture is therefore itself an e-value.

At the declared fixed reporting look, the route-specific inferential quantity is
\[
p_{\mathrm{seq}}^{(d)}
=
\min\!\left\{
1,
\frac{1}{E_{\mathrm{mix}}^{(d)}}
\right\}.
\]
By Markov's inequality,
\[
\Pr_0\!\left(
p_{\mathrm{seq}}^{(d)}
\leq\alpha
\right)
=
\Pr_0\!\left(
E_{\mathrm{mix}}^{(d)}
\geq\alpha^{-1}
\right)
\leq
\alpha.
\]
Thus \(p_{\mathrm{seq}}^{(d)}\) is an e-value-derived, fixed-look
route-specific inferential quantity rather than the plus-one randomisation value
used by assignment isolation
\citep{VovkWang2021,Shafer2021}. It enters the common decision field
\(p_{\mathrm{infer}}\) when the selected inference route is sequential. The
construction is not claimed to supply one global anytime-valid e-process across
adaptively selected reporting looks.

The sequential participant-level slope reported for estimation and magnitude
assessment remains the conditional-score slope
\[
\widehat{\beta}_{\tau,p}^{\,\mathrm{seq}}
=
\frac{
\displaystyle
\sum_{j\in\mathcal{J}_p}
\Apre^{\mathrm{resid},(pj)}
\{\tauL^{(pj)}-\mu_{pj}\}
}{
\displaystyle
\sum_{j\in\mathcal{J}_p}
\operatorname{Var}_{P_{\mathrm{sch}}}
\!\left(
\tauL
\mid
\F_{pj}^{\mathrm{pre}}
\right)
}.
\]
The slope set \(\mathcal{J}_p\) and the predictable e-value set
\(\mathcal{I}^{\mathrm{ev}}_f\) have different roles and need not coincide.
Later participant-estimability decisions affect the reported slope and
population-magnitude analysis but do not retroactively alter the e-value
product. Trials with zero current conditional assignment variance do not
contribute to the sequential score or slope, and participants with no remaining
positive conditional assignment variance are non-estimable for this route.

The martingale argument calibrates the signed assignment score; it does not by
itself establish zero expectation of the reported ratio-form participant slope
when its denominator is random through the evolving pre-assignment history.
The slope remains the locked magnitude summary, while inferential validity is
supplied by the e-value construction above.

The critical route conditions are therefore predictable eligibility, residual
predictability under the declared conditioning set and correctness of the
current-trial conditional scheduler law. Under carryover, the current endpoint
or residual may depend on earlier assigned delays because those earlier
assignments are already realised history. What is prohibited is dependence of
the current residual, its nuisance construction or its e-value eligibility
decision on the realised current-trial assignment or on assignment descendants
not contained in the admissible conditioning set.

If predictable eligibility, residual predictability or correctness of the
current-trial conditional scheduler law cannot be established under the locked
construction, the sequential route is marked invalid and the executable outcome
is inconclusive. A separately detected operational audit failure retains its
declared diagnostic-failure classification. An invalid sequential route cannot
yield either directional support or a clean forward-only adequate conclusion.

\begin{figure}[t]
\centering
\begin{tikzpicture}[
  >=Stealth,
  font=\footnotesize,
  node distance=5mm,
  s/.style={
    draw,
    rounded corners=2pt,
    fill=gray!8,
    align=center,
    text width=3.35cm,
    minimum height=0.95cm,
    inner sep=4pt
  }
]

\node[s] (route) {Declared inference route};

\node[s,below left=9mm and 8mm of route]
  (iso) {Assignment isolation};

\node[s,below right=9mm and 8mm of route]
  (seq) {Sequential martingale/e-value};

\node[s,below=8mm of iso]
  (isostat) {Frozen residual array;\\
             admissible reassignment law};

\node[s,below=8mm of seq]
  (seqstat) {Predictable eligibility and residual;\\
             current-trial conditional law};

\node[s,below=8mm of isostat]
  (prand) {Plus-one \(p_{\mathrm{rand}}\)};

\node[s,below=8mm of seqstat]
  (pseq) {Fixed-look e-value-derived\\
          \(p_{\mathrm{seq}}
          =\min\{1,1/E_{\mathrm{mix}}\}\)};

\path (prand) -- (pseq)
  node[midway,below=11mm,s,text width=3.8cm]
  (pinfer) {Route-neutral \(p_{\mathrm{infer}}\)\\
            enters decision rule};

\draw[->] (route)--(iso);
\draw[->] (route)--(seq);

\draw[->] (iso)--(isostat);
\draw[->] (seq)--(seqstat);

\draw[->] (isostat)--(prand);
\draw[->] (seqstat)--(pseq);

\draw[->] (prand.south)
  to[bend right=15] (pinfer.west);

\draw[->] (pseq.south)
  to[bend left=15] (pinfer.east);

\end{tikzpicture}

\caption{Route-specific inferential calibration. Directional superscripts are
suppressed for readability; the construction is applied separately to the
declared negative and positive directions. Assignment isolation holds the
frozen endpoint or residual array fixed and obtains a plus-one randomisation
value under the admissible reassignment law. The sequential route instead uses
predictably eligible one-step factors, a residual fixed under the route-specific
conditioning set, and the declared current-trial conditional scheduler law
under the forward-only null. It does not assert frozen-array invariance or
reshuffle committed endpoints under carryover. For participant-disjoint nuisance
fitting, residual predictability is conditional on the frozen external-fold
object described in \Cref{sec:si-seq-calibration}. Both routes supply the common
route-neutral inferential field \(p_{\mathrm{infer}}\) used by the decision
rule.}
\label{fig:si-randtest}
\end{figure}
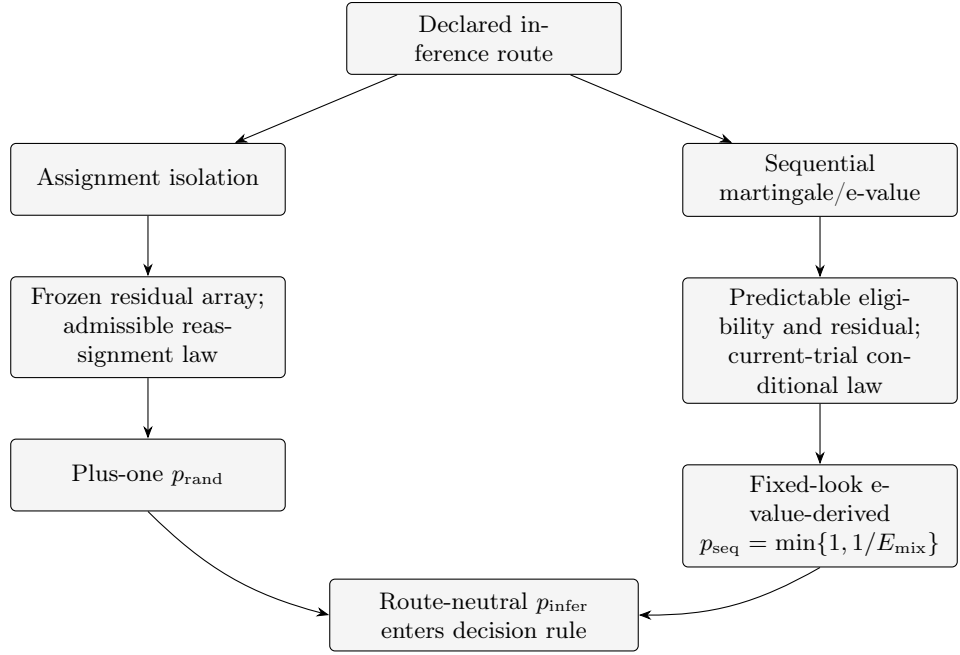

\Cref{fig:si-randtest} summarises the route-specific workflow. Assignment
isolation uses a plus-one frozen-array randomisation value. The sequential route
uses the predictable one-step construction above and the declared current-trial
conditional scheduler law under the forward-only null. Both routes supply the
common inferential decision field \(p_{\mathrm{infer}}\), but they do not share
the same calibration argument.

\subsection{Finite-sample calibration and choice of replicate count}
\label{sec:si-pvalue-finite}

The confirmatory pipeline uses route-specific finite-sample calibration. In the
assignment-isolation route, the frozen endpoint or residual array is held fixed
and only the assigned-delay labels are regenerated under the declared sharp-null
reassignment law. The plus-one randomisation value in
\eqref{eq:si-pvalue} is therefore calibrated by exchangeability of the observed
statistic and the replicate statistics under that reassignment law. In
carryover-sensitive settings, this frozen-array reassignment argument is not
invoked. Those analyses use the sequential martingale/e-value route established
in \Cref{sec:si-seq-calibration}.

For the sequential route, \Cref{sec:si-seq-calibration} establishes the
route-specific finite-sample calibration. Predictable e-value eligibility,
residual predictability under the declared conditioning set, and correctness of
the current-trial conditional scheduler law give conditional mean one for each
included one-step factor under the forward-only null. The resulting within-fold
products are e-values, and the prospectively fixed convex mixture across folds
satisfies
\[
\mathbb{E}_0\!\left[
E_{\mathrm{mix}}^{(d)}
\right]
\leq 1,
\qquad d\in\{-,+\}.
\]
Consequently, at the declared reporting look,
\[
p_{\mathrm{seq}}^{(d)}
=
\min\!\left\{
1,\frac{1}{E_{\mathrm{mix}}^{(d)}}
\right\}
\]
is a valid fixed-look e-value-derived inferential quantity
\citep{Ramdas2023,GrunwalddeHeideKoolen2024,VovkWang2021,Shafer2021}.
The construction is not claimed to define one global anytime-valid e-process
across adaptively selected reporting looks.

Fold e-values are combined by the declared convex mixture rather than
multiplied across folds. Cross-fitting alone does not establish the joint
conditional structure needed to validate a cross-fold product, and no such
product is assumed here. A global product could be justified under additional
structure, for example by using a nuisance object trained and frozen on a
calibration sample independent of the entire confirmatory sample, or by using a
fully prequential construction whose prediction at each trial depends only on
admissible earlier information.

The implementation is additionally guarded by an exhaustive clean-null
calibration test. The test enumerates all \(2^{10}\) assignments in a finite
design, verifies that the final fold combination equals the arithmetic mean of
the fold e-values, and confirms that its exact assignment expectation is at
most one. This is a finite enumeration check rather than a Monte Carlo
approximation. It verifies the implementation on the declared finite test
design; it does not replace the analytical conditional-mean argument in
\Cref{sec:si-seq-calibration}. The convex fold mixture may sacrifice power
relative to a global product that could itself be validly justified, but such a
product requires the additional joint structure just described. The sequential
route is not an exact frozen-array randomisation test and does not use the
plus-one Monte Carlo grid.

For assignment isolation, the plus-one construction gives finite-sample
super-uniformity on the Monte Carlo randomisation grid. For attainable grid
points \(k/(R+1)\),
\begin{equation}
\label{eq:si-pvalue-superuniform}
\mathbb{P}_0
\!\left(
p_{\mathrm{rand}}
\leq
\frac{k}{R+1}
\right)
\leq
\frac{k}{R+1},
\qquad
k=1,\ldots,R+1.
\end{equation}
For an arbitrary \(\alpha\in(0,1)\), if
\(\alpha<1/(R+1)\), then
\(\mathbb{P}_0(p_{\mathrm{rand}}\leq\alpha)=0\).
Otherwise, let
\[
k_{\alpha}
=
\left\lfloor
\alpha(R+1)
\right\rfloor .
\]
Because \(p_{\mathrm{rand}}\) takes values on the plus-one grid,
\[
\begin{aligned}
\mathbb{P}_0
\!\left(
p_{\mathrm{rand}}\leq\alpha
\right)
&=
\mathbb{P}_0
\!\left(
p_{\mathrm{rand}}
\leq
\frac{k_{\alpha}}{R+1}
\right)
\\
&\leq
\frac{k_{\alpha}}{R+1}
\leq
\alpha .
\end{aligned}
\]
If exhaustive enumeration is used, calibration is with respect to the exact
assignment distribution rather than to a Monte Carlo sample from it
\citep{PhipsonSmyth2010}.

The \(+1\) in both numerator and denominator preserves the finite-sample
Monte Carlo rank calibration. Omitting it permits
\(p_{\mathrm{rand}}=0\) when none of the simulated reassignment statistics is
as extreme as the observed statistic, even though the observed assignment is
itself one draw from the same null assignment law and must be represented in
the rank calculation. The resulting finite-sample validity follows from the
declared scheduler law together with sharp-null exchangeability of the observed
and replicate statistics, not from a large-sample normal approximation.

The randomisation replicate count \(R\) is declared before analysis for the
assignment-isolation route. The executable benchmark uses
\[
R=999,
\]
so the smallest attainable plus-one randomisation value is
\[
p_{\min}
=
\frac{1}{R+1}
=
10^{-3}.
\]
This grid is finer than the declared \(\alpha=0.05\) level, although it is not
an exhaustive-enumeration grid.

A second finite-\(R\) consideration is the Monte Carlo variation around the
ideal assignment-tail probability \(p^{\star}\). Conditional on the observed
statistic and locked reassignment law, if the number \(K\) of replicate
statistics at least as extreme as the observed statistic is approximated as
\[
K\sim\mathrm{Binomial}(R,p^{\star}),
\]
then
\[
p_{\mathrm{rand}}
=
\frac{1+K}{R+1}
\]
has approximate Monte Carlo standard deviation
\begin{equation}
\label{eq:si-pvalue-mc-se}
\widehat{\mathrm{se}}(p_{\mathrm{rand}})
\approx
\frac{
\sqrt{R\,p^{\star}(1-p^{\star})}
}{
R+1
}.
\end{equation}
Under the same approximation, the plus-one construction has the finite-\(R\)
mean shift
\[
\mathbb{E}\!\left[p_{\mathrm{rand}}\right]-p^{\star}
=
\frac{1-p^{\star}}{R+1},
\]
which is distinct from the Monte Carlo standard deviation in
\eqref{eq:si-pvalue-mc-se}.

Near \(p^{\star}\approx\alpha=0.05\), \(R=999\) gives
\[
\widehat{\mathrm{se}}
\approx
\frac{\sqrt{999\cdot0.05\cdot0.95}}{1000}
\approx
6.9\times10^{-3},
\]
about \(14\%\) of \(\alpha\). The corresponding finite-\(R\) mean shift is
approximately
\[
\frac{1-0.05}{1000}
=
9.5\times10^{-4}.
\]
These Monte Carlo-resolution quantities apply only to assignment-isolation
randomisation calibration. Carryover-sensitive analyses use the sequential
martingale/e-value route and are not counted as frozen-array randomisation
calibrations. Where the admissible reassignment set is small enough, exhaustive
enumeration removes the Monte Carlo component in
\eqref{eq:si-pvalue-mc-se} and is preferred.


\section{Participant-level bootstrap and resolution floor}
\label{sec:si-bootstrap-materiality}

\subsection{Studentised participant bootstrap}
\label{sec:si-bootstrap}

The confirmatory magnitude bound is computed at the participant level because
Level~II-A estimates an equal-participant mean of participant-specific assigned-delay
slopes. The independent inferential unit is therefore the participant, not the
trial. Each bootstrap resample draws participants with replacement and carries all
of the selected participant's trials, sessions, strata, assigned-delay labels,
retention indicators, and frozen residuals intact. The comparator is not refitted
inside the bootstrap. Resampling operates only on the locked residual objects that
enter the confirmatory estimator.

For each bootstrap resample, recompute the participant-level slope estimate
\(\widehat{\beta}^{*}_{\tau}\) and its corresponding standard error
\(\widehat{\mathrm{se}}^{*}\). The studentised bootstrap statistic is
\begin{equation}
\label{eq:si-bootstrap-tstar}
t^{*}
=
\frac{\widehat{\beta}^{*}_{\tau}-\widehat{\beta}_{\tau}}
     {\widehat{\mathrm{se}}^{*}} .
\end{equation}
Let \(q^{*}_{0.05}\) denote the \(5\%\) quantile of the bootstrap distribution
of \(t^{*}\). The one-sided upper confidence bound used by the negative-support
rule is
\begin{equation}
\label{eq:si-ucb}
\UCB(\widehat{\beta}_{\tau})
=
\widehat{\beta}_{\tau}
-
q^{*}_{0.05}\,\widehat{\mathrm{se}},
\end{equation}
where \(\widehat{\mathrm{se}}\) is the participant-level standard error in the
observed dataset. Because \(q^{*}_{0.05}\) is typically negative, the bound in
\eqref{eq:si-ucb} lies above the observed negative slope. A negative residual is
therefore resolved only if this upper bound remains below the declared floor,
\(\UCB(\widehat{\beta}_{\tau})< -\beta_{\min}\).

This separation is deliberate. The route-specific inferential calibration asks
whether the observed signed delay score is unusually negative under the
route-specific forward-only null, calibrated by the declared scheduler law. In
the assignment-isolation route this is a plus-one
randomisation value with the frozen residual array held fixed; in the sequential
route it is the e-value-derived quantity from
\Cref{sec:si-seq-calibration}. The participant bootstrap asks whether the
population participant-level slope is resolved far enough below zero to clear
the declared floor. A negative point estimate alone is not support, and a small
route-specific inferential value alone is not support. The decision requires
both assignment-calibrated evidence and a participant-level upper bound that
clears \(-\beta_{\min}\).

Trial-level resampling is not used. It would treat within-participant trials as
independent replications, understate participant-level uncertainty, and inflate
the apparent precision of \(\widehat{\beta}_{\tau}\). This is especially important
for EEG designs, where trial noise, session structure, preprocessing losses,
within-participant dependence, and participant heterogeneity can all be
substantial. Whole-participant resampling preserves the level at which the final
population claim is made.

A bias-corrected and accelerated (BCa) participant bootstrap and a conventional
participant \(t\)-interval are reported as sensitivity analyses. They are not
additional confirmatory decision rules. Agreement among the studentised
bootstrap-\(t\), BCa, and participant \(t\)-intervals supports the stability of
the bound. Material disagreement among them is treated as evidence that the
participant sample is too small, too skewed, or too heavy-tailed for a confident
population-level magnitude conclusion.

\subsection{Resolution floor}
\label{sec:si-beta-min}

The resolution criterion is
\(\UCB(\widehat{\beta}_{\tau})< -\beta_{\min}\) for a prospectively declared
\(\beta_{\min}\ge 0\). The floor is needed because a small negative slope can be
statistically unusual under the route-specific assignment calibration without
being large enough to resolve against the residual noise scale of the planned
design. Conversely, a large negative point estimate is not enough if
participant-level uncertainty leaves its upper bound above the floor.

Because no validated biological effect-size scale exists for an assigned-delay
residual in this setting, \(\beta_{\min}\) is defined as a common resolution
floor anchored to label-blind single-participant resolvability,
\begin{equation}
\label{eq:si-beta-min}
\beta_{\min}
=
\kappa\,
\frac{\sigma_{\mathrm{resid}}^{\mathrm{blind}}}
     {\sigma_\tau\sqrt{\bar n_{\mathrm{ret}}}},
\end{equation}
where \(\sigma_{\mathrm{resid}}^{\mathrm{blind}}>0\) is the
prospectively specified residual-scale summary,
\(\sigma_\tau>0\) is the route-specific effective retained assignment scale,
\(\bar n_{\mathrm{ret}}>0\) is the arithmetic mean retained usable-trial yield
per estimable participant, and \(\kappa>0\) is a dimensionless multiplier fixed
before analysis. Each empirical implementation must declare these estimators,
their participant-aggregation rules and the resulting construction of
\(\sigma_\tau\sqrt{\bar n_{\mathrm{ret}}}\).

In the certified benchmark,
\(\sigma_{\mathrm{resid}}^{\mathrm{blind}}\) is the pooled sample standard
deviation of the participant-demeaned frozen residual values on the locked
analysis set. Under assignment isolation, \(\sigma_\tau\) is the
root-mean-square of the participant-centred retained assigned-delay values
selected by that analysis set. Under the sequential route, it is the square
root of the pooled mean current-trial conditional assignment variance. It
follows that
\(\sigma_\tau\sqrt{\bar n_{\mathrm{ret}}}\) is the square root of the mean
participant slope-denominator leverage under the selected route. In a balanced
assignment-isolation design, this reduces to the common within-participant
assigned-delay standard deviation multiplied by the square root of the retained
trial count per participant. Participant-specific minimum-yield, delay-support
and denominator-leverage requirements remain separate locked estimability
conditions.

At the planning stage, \(\sigma_\tau\) and
\(\bar n_{\mathrm{ret}}\) are supplied by the declared scheduler and a
conservative retained-yield target. In the executable analysis, the realised
floor uses the route-specific effective assignment scale and retained yield of
the qualified participant set. If the analysis-set residual scale is nonfinite
or nonpositive, the code reapplies the same participant-demeaned pooled-scale
calculation to all retained frozen residuals. If the route-specific effective
assignment scale is nonfinite or nonpositive, the population standard deviation
of the declared design grid is retained as a computational contingency. These
branches do not override route-validity or participant-estimability failures.
Participants who fail the locked minimum-yield, delay-support or
denominator-leverage requirements remain non-estimable, and an unqualified
population conclusion is routed away from support and forward-only adequacy.

The form of \eqref{eq:si-beta-min} follows the scale of a
within-participant slope. For fixed residual noise, the slope is easier to
resolve when the retained assigned-delay leverage is larger and when more usable
trials contribute to that leverage. It is harder to resolve when the residual
scale is larger, the effective retained delay support is narrow, retention is
poor, or assigned-delay bins collapse within participants. The use of the
within-participant residual scale is deliberate: the participant-slope estimand
centres assigned delay within participant and is invariant to participant
offsets, so \(\sigma_{\mathrm{resid}}^{\mathrm{blind}}\) is the dispersion the
slope must resolve against after label-blind forward-only adjustment.

The frozen residual values and their demeaning within participants are computed
without assigned-delay labels. The locked analysis set may depend on
prospectively specified requirements for delay support and denominator leverage.
It may not depend on the sign or magnitude of a participant slope, or on a
delay-ordered pattern in the endpoint or residual. Retained yield and
route-specific assignment leverage are assignment and design quantities
assessed under fixed rules; they are not outcome-adaptive choices. None of the
inputs to \eqref{eq:si-beta-min} is selected from the observed
\(\widehat{\beta}_{\tau}\), an unblinded pilot or the injected synthetic
departure used in the benchmark. The estimators, analysis mask, aggregation
rules and contingency conventions must be fixed before confirmatory label
access. Changing them after observing the residual slope would turn the
resolution floor into an outcome-adaptive filter and invalidate the declared
decision architecture.

\(\beta_{\min}\) is anchored to label-blind single-participant resolvability,
not to a fixed population \(\alpha\)-level threshold and not to a claimed
biological importance threshold. Applied to the equal-participant population
slope, however, the criterion
\(\UCB(\widehat{\beta}_{\tau})<-\beta_{\min}\) also sets a minimum resolvable
and reportable magnitude. As the number of participants increases,
the participant-level standard error can decrease, so the upper-bound criterion
approaches a requirement that the population slope itself lie below
\(-\beta_{\min}\). Thus the floor is set by the retained-trial yield,
assigned-delay leverage and residual scale that determine single-participant
slope resolution; it is not made more stringent merely by adding participants.

This dual role is intentional. The floor prevents the analysis from declaring
support for slopes that are statistically detectable at the population level but
below the predeclared single-participant resolution scale. Whether a resolved
slope is large enough, reproducible enough, or structurally informative enough
to justify downstream modelling remains a separate scientific question.

The support decision is a non-compensatory intersection. The route-specific
inferential calibration must pass, the participant-bootstrap upper bound must
lie below \(-\beta_{\min}\), the analysable participant count must meet
\(N_{\min}\), and the audit, collider-diagnostic, and selection-sensitivity
gates must not block the result. No single component can compensate for failure of
another. A very small \(p\)-value does not compensate for a bound that fails to
clear the floor; a large negative point estimate does not compensate for a failed
audit; and a clean audit record does not compensate for inadequate
participant-level resolution.

\paragraph{Normal-approximation orientation diagnostic.}
For orientation only, suppose that the upper-bound component behaved like a
normal one-sided participant-level bound with standard-error scale
\(\mathrm{se}_{\mathrm{pop}}\). This component alone would then have
approximate null probability
\begin{equation}
\label{eq:si-stringency}
\Pr\!\left\{\UCB(\widehat{\beta}_{\tau})< -\beta_{\min}\right\}
\approx
\Phi\!\left[
-\left\{
\frac{\beta_{\min}}{\mathrm{se}_{\mathrm{pop}}}
+1.645
\right\}
\right],
\end{equation}
where \(\Phi\) is the standard normal distribution function. This expression is
not the declared test size, and it does not replace the route-specific
assignment calibration. It is a diagnostic approximation showing why the
upper-bound component becomes more or less stringent as participant
heterogeneity, retained-trial yield, assigned-delay support, and residual scale
change. The confirmatory calibration remains design-based: the route-specific
assignment calibration, the participant-bootstrap bound, \(N_{\min}\), the
non-compensatory audit hierarchy, and the realised synthetic operating
characteristics, not \eqref{eq:si-stringency}, determine whether the locked
pipeline is qualified at the declared design point.

\subsection{Minimum participant count and operating characteristics}
\label{sec:si-nmin}

The minimum analysable participant count \(N_{\min}\) is fixed before analysis. In
the executable benchmark package version~1.2.0 used to generate the
certified run, the value is \(N_{\min}=10\) eligible and estimable participants.
The canonical operating characteristics use \(P=24\), so
this value is a hard information floor rather than a claim that every design with
ten participants has qualified bootstrap or population-bound calibration. It is
not a convenience threshold applied after exclusions. Its role is to mark the
point below which the participant-level slope distribution, bootstrap quantiles,
and magnitude conclusion are too unstable to support a population claim. If
fewer than \(N_{\min}\) eligible participants remain, the assignment-calibrated
slope and descriptive intervals may still be reported, but no confirmatory
population-magnitude conclusion is drawn.

The value of \(N_{\min}\) is chosen by simulation under the planned design:
participant heterogeneity, endpoint variance, scheduler, assigned-delay support,
retained-trial yield, retention mechanism, skewness, tail behaviour, and
comparator performance. The selected value must control the false-support
decision rate under the sharp forward-only null and the adversarial forward-only
null, while recovering the resolution-boundary injected departure with the
planned sensitivity. In this sense \(N_{\min}\) belongs to the same locked
operating-characteristic qualification as the route-specific inferential
calibration, bootstrap bound, resolution floor, and audit hierarchy.

A failure to meet \(N_{\min}\) is not evidence for the null. It is an
insufficient-information outcome. This distinction matters because post hoc
attrition, preprocessing exclusions, delivery failures, or retention imbalance
can simultaneously reduce participant count and alter the retained endpoint
distribution. The classifier therefore separates underpowered or
selection-limited analyses from forward-only adequate nulls.

\subsection{Finite-sample rationale and diagnostics for the bootstrap-\texorpdfstring{$t$}{t} bound}
\label{sec:si-bootstrap-finite}

The confirmatory object is a one-sided bound on the participant-level estimand
in a finite-participant regime. The analysis therefore uses the studentised
participant bootstrap-\(t\). Pivoting on the statistic
\begin{equation}
\label{eq:si-studentised-pivot}
t
=
\frac{\widehat{\beta}_{\tau}-\beta_{\tau}}
     {\widehat{\mathrm{se}}}
\end{equation}
lets the bootstrap distribution adapt to skewness, heavy tails, and unequal
participant influence in the empirical participant-slope distribution. Under
regularity conditions, the studentised bootstrap has one-sided coverage error of
order \(O(N^{-1})\), compared with \(O(N^{-1/2})\) for the normal approximation
and for basic or percentile bootstrap intervals
\citep{Hall1992,DiCiccioEfron1996}. With modest participant counts, this
difference is practically relevant because the decision depends on whether the
upper bound crosses a fixed resolution floor. No standalone exact finite-sample
coverage claim is made for this bound; its use in Level~II-A is qualified through
the complete classifier operating characteristics at the declared design point.

The bootstrap-\(t\) quantile \(q^{*}_{0.05}\) is not forced to equal the normal
value \(-1.645\). It moves with the empirical participant-slope distribution.
When the participant slopes are approximately symmetric and light-tailed, the
studentised quantile will often be close to the normal value. When the slopes are
skewed, heavy-tailed, or dominated by a small number of influential participants,
the quantile can move substantially. This is a feature of the procedure:
substituting \(-1.645\) by default would suppress precisely the finite-sample
information the bootstrap is intended to capture.

Three finite-sample requirements follow. First, the per-resample standard error
\(\widehat{\mathrm{se}}^{*}\) must be stable, because it appears in the denominator
of every bootstrap statistic \(t^{*}\). Let \(B\) denote the prospectively
declared number of bootstrap resamples. The executable benchmark uses
\[
B=999
\]
participant bootstrap resamples. This count fixes the Monte Carlo resolution of
the estimated one-sided bootstrap quantile, but it cannot rescue a participant
sample whose slope distribution is too sparse or unstable. Second, the cluster
structure must be preserved: participants are resampled with all of their trials,
sessions, strata, and frozen residuals intact. Third, the hard floor \(N_{\min}\)
is required because below that count the bootstrap-\(t\) quantile is not
sufficiently stable for a population-level magnitude decision.

The BCa participant bootstrap, with acceleration estimated from the participant
jackknife, and the conventional participant \(t\)-interval are reported alongside
the studentised bound as sensitivity analyses \citep{EfronTibshirani1993}. These
summaries help diagnose whether the one-sided bound is being driven by skewness,
tail behaviour, participant influence, or small-participant-count instability.
Material disagreement among the studentised bootstrap, BCa, and participant
\(t\)-interval is treated as a warning that the bound is not stable enough for a
confident confirmatory claim. The leave-one-participant-out range of
\(\widehat{\beta}_\tau\) is also reported as a descriptive influence
diagnostic; it does not enter the classifier. The worked example in
\Cref{sec:si-worked-example} illustrates these objects for representative
simulated datasets from the locked benchmark run.

\section{Leakage audits and negative controls}
\label{sec:si-leakage-controls}

\subsection{Audit battery}
\label{sec:si-audit-battery}

The Level~II-A decision rule is interpretable only when the operational
scheduler in (R1), temporal integrity in (R2), and the retained-sample and
frozen-object qualifications are adequate for the declared contrast. The audit
battery qualifies those operational and assignment-to-analysis conditions; it
does not verify the past-adapted factorisation, which is the scientific null
interrogated by the confirmatory statistic. The battery is therefore part of the
confirmatory qualification procedure, not an optional quality-control appendix.
Its purpose is to separate three cases that would otherwise be conflated: a
qualified forward-only null, a qualified negative residual, and an
implementation or retained-sample failure that makes the contrast unqualified.

Five audit classes are reported regardless of the final outcome:
randomisation, temporal leakage, delivery, retention and implementation swap.
They do not all play the same role. Randomisation and implementation-swap checks assess the operational scheduler
mechanism and its software or hardware realisation; they do not test the
full-filtration factorisation in \eqref{eq:si-exogeneity}. Temporal-leakage checks assess the closed pre-endpoint measurement
window. Delivery and retention checks assess the realised
assignment-to-analysis chain, including whether the retained sample remains
qualified under \(\mathrm{(R3^\star)}\) within the declared sensitivity
envelope. \Cref{tab:si-audits} lists each audit, the condition it primarily
assesses, its pass criterion and the corresponding failure route.

\begin{table}[t]
\centering
\caption{Audit and failure-mode checklist. The listed condition is the primary
condition assessed by the audit; in practice the audits jointly qualify the
operational scheduler, temporal boundary and retained-sample delay-neutrality
condition \(\mathrm{(R3^\star)}\) within the declared
operational envelope. Passing an audit qualifies interpretation within that
envelope; it does not prove the corresponding independence condition globally.}
\label{tab:si-audits}
\footnotesize
\setlength{\tabcolsep}{4pt}
\renewcommand{\arraystretch}{1.08}
\begin{tabularx}{\textwidth}{@{}
    >{\raggedright\arraybackslash}p{2.7cm}
    >{\centering\arraybackslash}p{1.7cm}
    >{\raggedright\arraybackslash}X
    >{\raggedright\arraybackslash}p{3.1cm}
@{}}
\toprule
\textbf{Audit} &
\textbf{Assesses} &
\textbf{Pass criterion} &
\textbf{Failure classification} \\
\midrule

Randomisation &
(R1) &
Realised assignments lie in the legal support of the declared scheduler law;
fixed-multiset, blocking or restricted-randomisation counts are satisfied where
those constraints were prospectively imposed; and seed, scheduler, log,
timestamp and concealment integrity are verified before endpoint-labelled
analysis. Under unrestricted stochastic assignment, chance imbalance is reported
descriptively and is not by itself an audit failure. &
Invalid randomisation contrast; diagnostic failure. \\

Temporal leakage &
(R2) &
Injected post-\(t_1\) impulses or steps produce no committed pre-event output
beyond \(\delta_{\mathrm{leak}}\), and delay-correlated probes do not predict
the committed endpoint before the allowed access point. &
Invalid endpoint; diagnostic failure. \\

Delivery &
\(\mathrm{(R3^\star)}\) qualification &
Assigned-to-measured latency error is within the declared tolerance, and
non-compliance or delivery failure is bounded by stratum. &
Compliance-limited result; diagnostic failure if the declared bound is
exceeded. \\

Retention &
\(\mathrm{(R3^\star)}\) qualification &
Stratum-specific delivery, rejection, missingness and analysis inclusion are
recorded; retained-vs-excluded endpoint distributions are logged before
unblinding; marginal retention imbalance and endpoint-dependent retention are
evaluated. &
Selection-limited result when the retained sample is not qualified under the
declared \(\mathrm{(R3^\star)}\) envelope. \\

Implementation swap &
(R1) &
The association is not tied to a particular generator, command pathway,
logging pathway, timing implementation or analysis implementation across
declared swaps. &
Implementation-confounded result; diagnostic failure. \\

\bottomrule
\end{tabularx}
\end{table}

The audits are non-compensatory. A strong negative slope cannot compensate for
a failed temporal-leakage audit, failed delivery audit, failed implementation
swap or support-blocking retention diagnostic. Conversely, a dataset with a
failed retention or collider check is not converted into a forward-only
adequate null simply because the final support criterion does not pass. Such a
dataset is routed to a selection-limited or diagnostic-failure outcome,
depending on which part of the assignment-to-analysis chain has failed.

\subsection{Negative controls as two-sided diagnostics}
\label{sec:si-negative-controls}

The assigned delay is the primary negative-control exposure for the
past-adapted factorisation. Before retained-sample selection, the operational
scheduler in (R1), temporal integrity in (R2), and the past-adapted
factorisation imply no systematic association of either sign between a
committed pre-selection statistic and \(\tauL\). In the retained analysis
sample, the corresponding statement for the committed endpoint and frozen
residual additionally requires retained-sample delay-neutrality
\(\mathrm{(R3^\star)}\), retained-support positivity and the
frozen-comparator independence condition in
\Cref{lem:retained-exclusion}. Participant-level slope diagnostics additionally
inherit the selected route's joint-design or predictable-score conditions and
the participant-estimability qualification. This diagnostic statement is
two-sided \citep{Lipsitch2010,Shi2020}. A material negative slope, material
positive slope, or association on a control endpoint can indicate failure of
the operational scheduler, temporal or frozen-object integrity, retained-sample
qualification, route-specific calibration conditions, participant-level
qualification, or the past-adapted factorisation itself.

The support claim, however, is one-sided. Level~II-A support is reserved for a
materially negative assigned-delay slope that survives the declared
route-specific inferential calibration, participant-bootstrap upper-bound
criterion, resolution floor, audit battery, collider diagnostic and
selection-sensitivity gate. A material positive slope is therefore not support
with the sign reversed. It is a negative-control departure in the unpredicted
direction and is routed separately. This separation is deliberate: the design
must be sign-blind when detecting departures from the past-adapted
factorisation, but
sign-specific when assigning directional Level~II-A support.

Auxiliary negative controls should be declared before analysis and evaluated
with the same sign-blind discipline. Recommended controls include:
\begin{enumerate}[leftmargin=2.0em,itemsep=2pt]
\item a pre-cue baseline endpoint computed on a window preceding any
anticipatory build-up, which should be unrelated to \(\tauL\);
\item a future-only placebo assignment generated by the scheduler but never
delivered, which should not order any committed endpoint;
\item channels, components or sources outside the locked endpoint set and with
no declared anticipatory role;
\item implementation-specific probes, such as timing logs, command queues or
trigger paths, that should not carry endpoint-relevant delay information before
the allowed access point.
\end{enumerate}

A material association on any auxiliary negative control of either sign sends
the analysis back to the audit layer before any substantive claim is made. The
result is not rescued by matching the expected sign, and it is not dismissed
because the sign is unexpected. The question is first whether the
randomisation, leakage, delivery, implementation and retained-sample
delay-neutrality conditions still qualify the contrast for interpretation.

\subsection{Opposite-direction departures}
\label{sec:si-opposite}

A material positive slope is reported separately as an opposite-direction
departure. It does not support the proposed Level~II-A residual, because the
directional support rule is pre-specified for negative slopes. It is also not
eligible for a clean forward-only-adequate classification. Before retained-sample
selection, under (R1)--(R2) and the past-adapted factorisation, the
endpoint-level no-ordering implication yields a zero coefficient for the
registered linear projection. In the retained analysis sample, the analogous
statement for the committed endpoint and frozen residual additionally requires
\(\mathrm{(R3^\star)}\), retained-support positivity and frozen-comparator
independence. Participant-level slope interpretation further requires the
selected route's joint-design or predictable-score conditions and the
participant-estimability qualification. A resolved material association of the
opposite sign is therefore unexpected under the declared no-ordering conditions
and requires diagnostic investigation, but the positive tail is not a second
confirmatory level-\(\alpha\) class-rejection route.

The analysis applies the same qualification discipline used for a negative
departure. It reviews randomisation integrity, temporal leakage, delivery
accuracy, implementation dependence, retention imbalance, endpoint-dependent
selection, auxiliary negative controls, the endpoint-by-delay collider
diagnostic and the selection-sensitivity bound. If any support-blocking condition
fails, the result is routed to diagnostic failure or selection-limited
classification. If the positive departure survives those checks, it is reported
as a prespecified opposite-direction diagnostic that blocks a forward-only-
adequate conclusion and requires independent replication and investigation. It
carries neither directional support nor a confirmatory class-insufficiency claim.

This distinction protects both sides of the decision. A positive slope cannot be
counted as evidence for the proposed residual, but it also cannot be dismissed as
artefact merely because its sign is inconvenient. Sign is relevant to the
substantive Level~II-A support label and to the confirmatory error architecture;
it is not by itself an audit. The same implementation and selection standards
therefore adjudicate negative and positive associations, while the final
classifier preserves their different inferential status.

\section{Selection and omitted-pathway sensitivity analysis}
\label{sec:si-sensitivity-analysis}

\subsection{The post-randomisation selection problem}
\label{sec:si-selection-problem}

Condition \(\mathrm{(R3^\star)}\) cannot be proved by inspection. Even under
perfect post-endpoint randomisation, the retained analysis sample can fail to
inherit the pre-selection boundary if inclusion depends on the assigned delay
after endpoint commitment. One route is an unmeasured or incompletely
conditioned pre-event state \(U^{(j)}\) that affects both the committed endpoint
and a delay-dependent selection event \(S^{(j)}\), such as delivery compliance,
artefact rejection, missingness, preprocessing success, exclusion, or inclusion.
Another route is a direct endpoint-by-delay selection rule, in which inclusion
depends on the joint configuration of \(\Apre^{(j)}\) and \(\tauL^{(j)}\). In
either case, conditioning the analysis on \(S^{(j)}=1\) can induce an
association between \(\Apre^{(j)}\) and \(\tauL^{(j)}\) that is absent before
selection and that can mimic a residual. The sensitivity analysis does not
prove \(\mathrm{(R3^\star)}\). It bounds how large a declared selection pathway
would have to be to manufacture the observed retained-sample slope, and compares
that requirement with the imbalance still compatible with the audit records.

The construction is deliberately reported on the native endpoint-slope scale.
It is conceptually aligned with modern sensitivity-analysis summaries that ask
how strong an unmeasured pathway would have to be to explain an observed
association, including the E-value for unmeasured confounding and
selection-bias bounds~\citep{VanderWeeleDing2017,SmithVanderWeele2019}. Unlike
a risk-ratio E-value, however, the Level~II-A gate is calibrated to the slope
that a delay-dependent retained-sample pathway could manufacture under the
declared selection-model class.

\subsection{Required versus audited imbalance}
\label{sec:si-required-audited}

Let \(\Delta_{\mathrm{sel}}\) denote the probability-scale differential
retention or inclusion contrast used by the declared selection-model class. The
model maps this probability-scale contrast into the endpoint-slope scale, so
all comparisons in the gate are made after the declared mapping has been
applied. Two quantities are computed:
\begin{itemize}[leftmargin=1.4em,itemsep=2pt]
\item the \textit{required} imbalance
\(\Delta_{\mathrm{sel}}^{\mathrm{req}}\): the smallest probability-scale
differential retention or inclusion contrast that, under the declared
selection-model class, could reproduce the observed
\(\widehat{\beta}_\tau\) under an otherwise past-adapted account;
\item the \textit{audited} imbalance
\(\Delta_{\mathrm{sel}}^{\mathrm{aud}}\): the largest probability-scale
selection contrast consistent with the delivery, retention, preprocessing,
inclusion and exclusion records within each stratum. This includes endpoint
distributions for retained and excluded trials available before unblinding.
\end{itemize}
The mapping from imbalance to induced slope uses monotone sharp bounds under a
monotone-selection assumption \citep{Lee2009} and worst-case bounds without
monotonicity \citep{Manski1990}, in the sensitivity tradition for observational
threats \citep{Rosenbaum2002}.

A minimal implementable audit compares retained and excluded trials within each
declared stratum. The protocol first computes, without using the outcome
decision, how different the committed endpoint is between retained and excluded
trials. It then computes how unevenly retention varies across assigned-delay
labels within the same stratum. The declared selection model maps these
quantities into a conservative endpoint-slope bound for delay-dependent
retention. This audited selection bound is compared with the slope required to
explain the observed assigned-delay ordering. More conservative monotone or
worst-case bounds may be used instead, but the mapping from probability-scale
retention imbalance to possible endpoint shift must be declared before
unblinding and reported together with the confirmatory result.

\subsection{Confidence-limit gate}
\label{sec:si-gate}

The \(\LCB\) and \(\UCB\) in this gate are the prospectively locked
selection-model bounds for the two imbalance quantities. They are distinct from
the whole-participant bootstrap bounds used for
\(\widehat{\beta}_\tau\). The selection gate compares confidence limits, not point estimates,
so sampling uncertainty in both quantities enters the decision:
\begin{equation}
\label{eq:si-selection-gate}
\LCB\!\left(\Delta_{\mathrm{sel}}^{\mathrm{req}}\right)
>\UCB\!\left(\Delta_{\mathrm{sel}}^{\mathrm{aud}}\right).
\end{equation}
If \eqref{eq:si-selection-gate} holds, the observed slope is larger than the
declared marginal-imbalance selection pathway allowed by the audit records, and
the scalar selection gate passes. This does not prove
\(\mathrm{(R3^\star)}\) globally. If the slope is material but
\eqref{eq:si-selection-gate} fails, the result is reported as
selection-limited, not as support. The selection-model class, the imbalance
summary, and the bound estimators are locked prospectively.

\subsection{Scope boundary: endpoint-by-delay collider selection}
\label{sec:si-collider}

The gate \eqref{eq:si-selection-gate} is built on the marginal retention
imbalance across delay bins and on a declared selection-model class. It therefore
qualifies only the selection pathways represented by that audited imbalance
summary and that class. Marginal retention balance is not sufficient to
establish \(\mathrm{(R3^\star)}\). The relevant causal structure is the standard
collider-stratification or structural selection-bias problem: conditioning on
inclusion can open an association between variables that were independent
before selection, because the analysis sample is defined by a common effect or
interaction-defined selection variable \citep{Hernan2004,Greenland2003}.
Consider a pure endpoint-by-delay collider in which the inclusion indicator
\(S_{pj}\) depends on the joint configuration of the committed endpoint
\(\Apre^{(pj)}\) and the assigned delay \(\tauL^{(pj)}\) through their
interaction, with no main effect of delay on retention:
\begin{equation}
\label{eq:si-collider}
\operatorname{logit}\Pr(S_{pj}=1)
=
c_0
+
\gamma\,
z_{A,pj}\,
z_{\tau,pj}.
\end{equation}
Here \(c_0\) is the intercept, \(\gamma\) is the endpoint-by-delay interaction
coefficient, \(z_{A,pj}\) is the centred and standardised committed endpoint,
and \(z_{\tau,pj}\) is the centred and standardised assigned delay. Because
\(z_{A,pj}\) is centred and the model has no delay main effect, the first-order
contribution to the marginal retention rate averages to zero under a symmetric
endpoint distribution; the marginal retention rate by delay bin shifts only at
second order in \(\gamma z_{\tau,pj}\). The retained endpoint distribution,
however, shifts linearly across delay bins, manufacturing a retained-sample
slope. The full pre-selection sample still satisfies the post-endpoint
randomisation boundary of \S1: \(\Apre^{(pj)}\) is independent of
\(\tauL^{(pj)}\) before selection. The induced association arises only after
conditioning on the retained analysis sample, so this is a scope test of the
selection gate, not a failure of the randomisation boundary.

Because the committed endpoint is computed for every trial before the delay is
assigned, the declared endpoint-by-delay collider mechanism can be targeted by
diagnostics distinct from the marginal imbalance summary. The first diagnostic
fits inclusion with participant fixed effects, a cubic basis for the committed
endpoint, categorical assigned-delay main effects and endpoint-by-delay
interaction terms. The interaction block is tested with a participant-clustered
CR1 sandwich covariance. The null model therefore allows participant-specific
inclusion levels, nonlinear endpoint dependence and arbitrary categorical delay
main effects; only the endpoint-by-delay interaction is diagnostic.

The second diagnostic compares retained and excluded committed endpoints within
each delay bin. It first forms a retained-minus-excluded contrast within each
participant and bin, then applies a two-sided one-sample \(t\)-test across
participants with Bonferroni control across bins. The trial-level standardised
mean difference is reported descriptively and does not determine the fire. A
numerically invalid interaction or retained-versus-excluded fit fires
conservatively. The retained-sample rank association is also descriptive because
a genuine endpoint-level departure can produce the same retained-endpoint
ordering.

A genuine endpoint-level residual with random retention should not activate
these inclusion diagnostics. The declared collider generator should activate
them because inclusion depends on the endpoint within delay strata. The
interaction threshold is guarded by a clustered nonlinear endpoint-only null
test whose realised firing rate must match the declared two-sided normal size
within the prespecified Monte Carlo tolerance. A diagnostic fire is
support-blocking and is interpreted as a selection-path qualification failure.

The decision rule adds a hard support-blocking condition. If the endpoint-by-delay
interaction diagnostic fires, or if the retained-sample slope is compatible with an
uncovered collider of the observed magnitude under approximately balanced marginal
retention, the dataset cannot receive supported-residual classification; it is
classified selection-limited. The
manufactured collider slope is bounded by the retention rate, so at a high resolution
floor and high retention it may remain sub-material and be blocked incidentally by
the resolution criterion; at the realistic artefact-rejection rates and resolution floors where it
becomes material, the scalar gate \eqref{eq:si-selection-gate} passes and only this
diagnostic prevents a supported classification (\S9).

\subsection{Worked selection-sensitivity calculation}
\label{sec:si-selection-worked}

We make the gate \eqref{eq:si-selection-gate} concrete using the current
representative supported draw. Its certified run is \LevelIIARunHash. The
locked analysis returns
\[
\widehat{\beta}_\tau
=-60.1246\,\mu\mathrm{V\,s^{-1}},
\qquad
\sigma_{\mathrm{resid}}^{\mathrm{blind}}
=1.0999\,\mu\mathrm{V}.
\]
Across the \(T_0=0.020\,\mathrm{s}\) support, the observed slope corresponds to
an extreme-bin shift of
\[
|\widehat{\beta}_\tau|T_0
=1.2025\,\mu\mathrm{V}.
\]
The representative draw retains 2289 of 2880 assigned trials, giving overall
retention \(0.7948\). The configured reference retention used by the gate is
\(p_{\mathrm{high}}=0.80\), and the endpoint is Winsorised at the label-blind
\(\pm3\sigma_{\mathrm{resid}}^{\mathrm{blind}}\) limits for the bounded-support
calculation.

\paragraph{Manski worst-case bound.}
Without a monotonicity restriction, the missing mass can be placed
adversarially at the Winsorising limits. At the declared retention reference,
each bin mean can differ from its retained mean by
\[
(1-p_{\mathrm{high}})
3\sigma_{\mathrm{resid}}^{\mathrm{blind}}
=0.6599\,\mu\mathrm{V}.
\]
The resulting between-bin bound is \(1.3199\,\mu\mathrm{V}\), equivalent to
\(66.0\,\mu\mathrm{V\,s^{-1}}\) across the registered support
\citep{Manski1990}. This broad non-monotone bound exceeds the observed slope and
therefore does not exclude every adversarial selection mechanism from marginal
retention information alone.

\paragraph{Audited differential retention.}
The retained counts across the five delay bins are
\[
(458,443,464,458,466).
\]
The corresponding retention rates are
\[
(0.795,0.769,0.806,0.795,0.809).
\]
The executable audit uses the maximum absolute binwise deviation from overall
retention,
\[
\Delta_{\mathrm{sel}}^{\mathrm{aud}}
=
\max_k|\widehat p_k-\widehat{\bar p}|
=0.025694.
\]
This is the scalar supplied to the selection gate. It is not the
shortest-minus-longest retention contrast.

\paragraph{Lee monotone bound at the audited imbalance.}
Under the declared one-sided monotone trimming model, the audited imbalance
corresponds to trim fraction
\[
t_{\mathrm{aud}}
=
\frac{\Delta_{\mathrm{sel}}^{\mathrm{aud}}}{p_{\mathrm{high}}}
=0.03212.
\]
For a Gaussian residual, \(c=\Phi^{-1}(t_{\mathrm{aud}})=-1.851\), and the
one-sided induced mean shift is
\[
\sigma_{\mathrm{resid}}^{\mathrm{blind}}
\frac{\phi(c)}{1-\Phi(c)}
=0.0818\,\mu\mathrm{V},
\]
which corresponds to \(4.09\,\mu\mathrm{V\,s^{-1}}\) across the delay support
\citep{Lee2009}. The one-sided upper confidence limit for the audited imbalance
is \(0.05645\); applying the same map at that limit gives an induced slope of
approximately \(7.99\,\mu\mathrm{V\,s^{-1}}\).

\paragraph{Required imbalance and the gate.}
To manufacture the full \(1.2025\,\mu\mathrm{V}\) shift, the inverse-Mills
relation requires
\[
\frac{\phi(c)}{1-\Phi(c)}
=
\frac{1.2025}{1.0999}
=1.0933.
\]
The solution is \(c=0.434\), with kept fraction \(0.3321\) and trim fraction
\(0.6679\). Multiplication by \(p_{\mathrm{high}}=0.80\) gives
\[
\Delta_{\mathrm{sel}}^{\mathrm{req}}
=0.53433.
\]
The executable confidence calculation uses the pooled retained count per bin,
\(n_{\mathrm{bin}}=2289/5=457.8\), and yields
\[
\LCB\!\left(\Delta_{\mathrm{sel}}^{\mathrm{req}}\right)
=0.50358,
\qquad
\UCB\!\left(\Delta_{\mathrm{sel}}^{\mathrm{aud}}\right)
=0.05645.
\]
The registered gate therefore passes by a wide margin.

This calculation qualifies the declared monotone marginal-imbalance class. The
Manski bound retains the broader caveat that marginal retention rates alone do
not exclude every non-monotone selection mechanism. The targeted
endpoint-by-delay collider diagnostic addresses the registered interaction
class that can preserve marginal balance while inducing a retained-sample
slope.

\subsection{Participant-level estimability sensitivity}
\label{sec:si-participant-estimability}

The trial-level retained-sample condition \(\mathrm{(R3^\star)}\) qualifies
trial inclusion for the committed endpoint and frozen residual. It does not by
itself qualify the participant-level set over which the equal-participant slope
is averaged, and therefore does not by itself complete the aggregation step from
\Cref{lem:retained-exclusion} to the population-level magnitude claim. A participant
can become non-estimable if too few retained trials, too few assigned-delay
levels, or too little assigned-delay variance remain after delivery, artefact
rejection, preprocessing, exclusion and inclusion. This is a participant-level
selection step.

The primary report therefore includes the number and fraction of eligible
participants who are non-estimable, the reason for non-estimability, retained
trials per assigned-delay bin, participant-level assigned-delay variance,
label-blind endpoint scale, label-blind residual scale, preprocessing burden,
delivery failures and retention summaries. These quantities are computed without
using the sign or magnitude of the fitted assigned-delay slope.

The participant-level sensitivity analysis asks whether the confirmatory
conclusion would survive plausible values for the non-estimable participants.
For support, the sensitivity analysis asks whether adding the non-estimable
participants back into the equal-participant mean under the declared
participant-level selection model could move the population slope above the
resolution boundary. For a forward-only adequate null, it asks whether the
non-estimable participants could hide a material negative or opposite-direction
departure. The calculation is reported on the same native slope scale as the
confirmatory estimand.

The default bounded analysis uses the label-blind participant-level residual
scale and assigned-delay support to define a prospectively declared plausible
slope range for non-estimable participants. A worst-case bound assigns
non-estimable participants the conclusion-opposing values within that range. A
monotone bound may also be reported when the protocol declares a monotone
relationship between an observed label-blind participant summary, such as
retained-trial yield or residual scale, and the participant slope. The
monotone and worst-case summaries play the same role as the trial-level
selection-sensitivity gate: they do not prove that participant-level selection
is absent, but they bound how large such selection would have to be to change
the decision.

A dataset cannot receive supported-residual classification if the
participant-level estimability bound can move the equal-participant slope above
\(-\beta_{\min}\), or if the bound makes the participant-level upper confidence
limit fail the declared resolution criterion. A dataset cannot receive
forward-only adequate classification if the non-estimable participants could
hide a material negative or opposite-direction departure under the declared
bound. In either case the result is routed to selection-limited or inconclusive
classification, depending on whether the limiting feature is participant-level
selection or insufficient participant-level information.

\section{Synthetic benchmark design}
\label{sec:si-synthetic-benchmarks}

\subsection{Role and the seven required behaviours}
\label{sec:si-benchmark-role}

The synthetic benchmarks qualify the Level~II-A decision pipeline before any
human EEG claim is made. They are not evidence for an anticipatory neural
effect, and they are not a substitute for preregistered empirical replication.
Their role is narrower and more stringent: they ask whether the locked
estimator, route-specific inferential calibration, participant-bootstrap bound,
resolution floor, audits, selection checks, collider diagnostic and
non-compensatory classifier behave as intended when the data-generating truth is
known.

The benchmark suite is designed around seven required behaviours. Each
behaviour corresponds to a distinct reviewer-relevant failure mode. A pipeline
that fails one of these behaviours is operationally unqualified at the declared
synthetic operating point; the failure is not interpreted as evidence for or
against the Level~II-A sufficiency claim in human data.

\begin{enumerate}[leftmargin=2.0em,itemsep=3pt]
\item \textbf{Forward-only null.} Under a past-adapted data-generating process
with no assigned-delay dependence, the route-specific inferential calibration
should pass at approximately its nominal one-sided operating rate, but the
complete support rule should not return a supported residual. This checks that
assignment-calibrated evidence alone is not enough: the one-sided bound must
also clear the resolution floor, and the implementation audits must pass.

\item \textbf{Injected residual.} Under a negative endpoint-level injected
residual, the pipeline should recover the declared sign and approximate
magnitude with the planned endpoint, comparator, residual-freezing rule,
participant-level estimand, route-specific inferential calibration and bootstrap
bound. This is the positive operating check: the rule should not be so conservative
that a declared material residual at the synthetic anchor is routinely missed.

\item \textbf{Leakage detection.} Under an injected post-\(t_1\) leak, assigned
delay becomes available to a path that should be inaccessible before endpoint
commitment. The temporal-leakage audit should flag the violation and support
should be blocked. This checks that an apparent residual is not accepted after
the randomisation boundary has been compromised.

\item \textbf{Standard selection.} Under a delay-dependent retention artefact
with marginal imbalance, the retention audit should fire and the
selection-sensitivity gate in \eqref{eq:si-selection-gate} should move as the
imbalance is varied. This checks the ordinary assignment-to-analysis failure in
which some assigned-delay levels are preferentially retained or excluded.

\item \textbf{Endpoint-by-delay collider.} Under a pure endpoint-by-delay
collider with approximately balanced marginal retention, the retained sample
can show a negative slope even when the marginal retention audit and scalar
selection gate do not block it. The endpoint-by-delay diagnostic should fire,
and the dataset should be classified as selection-limited rather than
supported. This checks the scope boundary of marginal retention diagnostics
(\Cref{sec:si-collider}).

\item \textbf{Adversarial forward-only null.} Under a forward-only generator
with nonlinear hazard structure, heavy-tailed participant heterogeneity,
heteroskedastic and autocorrelated residuals, carryover, and comparator
misspecification, the pipeline should still avoid supported-residual
classification. This is not a proof of universal robustness, but it is a
deliberately rough null designed to stress the comparator and inference stack.

\item \textbf{Opposite-direction injection.} Under a positive injected slope,
the result should be classified as an opposite-direction departure, not as
directional support. This checks that the implementation respects the
one-sided Level~II-A support claim while still treating a material association
of the wrong sign as a distinct warning signal.
\end{enumerate}

The seven scenarios are operating-characteristic scenarios, not an exhaustive
enumeration of every software branch in the classifier. Branch-level unit tests
also exercise the non-compensatory decision logic directly, including cases in
which a material slope is present but a support-blocking gate fails. The
scenario suite and the unit tests therefore serve complementary purposes: the
scenario suite tests end-to-end behaviour under interpretable generators, while
the branch tests ensure that the executable classifier implements the declared
hierarchy.

\Cref{sec:si-operating} reports the realised operating characteristics from
the public benchmark package. \Cref{sec:si-worked-example} follows a single
representative supported draw and a single representative clean-null draw
through the decision objects. Together, these sections show both the aggregate
behaviour of the locked pipeline and the mechanics of an individual
classification.

\subsection{Data-generating processes}
\label{sec:si-dgp}

All synthetic generators share the same basic architecture. The endpoint is
generated from participant structure, trial-level covariates, scheduler
quantities and residual noise, then passed through the same label-blind
comparator, cross-fitted residualisation step, participant-level slope
estimator, route-specific inferential calibration, bootstrap bound and
classifier used in the benchmark. The generators differ only in the mechanism
they are intended to stress.

The clean forward-only generator builds the endpoint from comparator-accessible
covariates and past-adapted noise, with no term in \(\tauL\). Under this
generator, assigned delay is conditionally irrelevant to the committed
endpoint after the declared forward-accessible structure has been accounted
for. Any supported residual in this scenario would therefore indicate a
false-support failure of the pipeline at the synthetic anchor.

The injected-residual generator adds an endpoint-level term
\[
\beta^{\mathrm{inj}}
\bigl(
\tauL-\bar\tau_{\mathcal{R}}
\bigr),
\qquad
\bar\tau_{\mathcal{R}}
=
\mathbb{E}_{P_{\mathrm{sch}}}
\!\left[
\tauL
\mid
\mathcal{R}
\right],
\]
where \(\beta^{\mathrm{inj}}\) is the injected endpoint-level slope and
\(\bar\tau_{\mathcal{R}}\) is the scheduler mean within the corresponding
synthetic randomisation stratum. This placement is deliberate:
the residual is not inserted after analysis as a convenient slope. It is
inserted into the endpoint-generating process, then the same comparator and
residual-freezing rule are applied. The benchmark therefore tests whether the
full locked pipeline can recover a material negative residual after the
forward-only adjustment has had the same opportunity to absorb
forward-accessible structure as it would in the null scenarios.

The leakage generator adds a post-\(t_1\), delay-correlated component to the
committed endpoint and to a pre-endpoint probe used by the temporal-leakage
audit. This scenario is not meant to mimic a subtle empirical confound. It is a
positive-control violation of the randomisation boundary. The expected result
is diagnostic failure rather than support: once delay information leaks into a
pre-endpoint path, the Level~II-A interpretation is invalid even if the fitted
slope has the predicted sign.

The standard selection generator imposes a delay-dependent inclusion rule with
marginal imbalance across assigned-delay levels. This represents the most
direct assignment-to-analysis threat: the retained sample is no longer
qualified as delay-neutral under the declared \(\mathrm{(R3^\star)}\) envelope.
The retention audit and scalar selection gate are expected to block support
when the imbalance is large enough to explain or qualify the apparent slope.

The endpoint-by-delay collider generator imposes a selection rule in which
retention depends jointly on the endpoint and the assigned-delay label while
preserving approximately balanced marginal retention. This generator is
included because marginal balance is not sufficient to guarantee a valid
retained analysis sample. A collider can manufacture a slope inside the
retained subset even when each assigned-delay level has a similar overall
retention rate. The endpoint-by-delay diagnostic is therefore a separate
support-blocking object, not a redundant restatement of the retention audit.

The adversarial forward-only generator keeps the assigned delay causally
irrelevant to the endpoint but makes the forward-only problem deliberately
unfriendly. It adds nonlinear hazard structure, heavy-tailed participant
heterogeneity, heteroskedastic and autocorrelated residuals, carryover from the
previous assigned delay, and a comparator structure that the linear reference
model cannot fully absorb. Its purpose is to test whether ordinary modelling
stress can be converted into directional support by the decision rule. Passing
this scenario is not universal robustness; it is evidence that the locked
pipeline is not fragile to this declared mixture of forward-only stresses.

The opposite-direction generator injects a material positive slope at the
endpoint-generating level. This scenario checks that the classifier does not
confuse any material assigned-delay association with Level~II-A support. Under the
Level~II-A rule, support and confirmatory class rejection are directional. A
positive material slope is reported as an opposite-direction diagnostic that
blocks a forward-only-adequate outcome but carries no confirmatory
class-insufficiency claim.

\subsection{Synthetic anchor and planning grid}
\label{sec:si-anchor}

The reference benchmark uses a short-horizon, origin-coded synthetic support
\(T_0=20\,\mathrm{ms}\) with five assigned-delay analysis bins
\[
\tauL \in \{0,5,10,15,20\}\,\mathrm{ms}.
\]
The \(0\) ms bin is a synthetic analysis origin, not a claim that an empirical
apparatus can randomise, command, deliver and log an imperative event with zero
physical latency after endpoint closure. In an empirical implementation, the
assigned-delay support must start at a strictly positive, logged and auditable
minimum interval after the assigned label has been generated and concealed up to
\(t_1\). If an empirical grid is displayed after subtracting this minimum
interval, the axis must be labelled as an offset from the empirical minimum rather
than as an absolute post-endpoint latency. The same origin-coded grid is used
across the anchor scenarios so that operating characteristics can be compared
without changing the assigned-delay leverage. Where a residual is injected, it is
added as a signed endpoint-level slope on the native residual-amplitude scale
before comparator fitting and residualisation.

The anchor is a pipeline-qualification point, not an empirical EEG design
recommendation. It fixes an interpretable combination of assigned-delay
support, participant count, retained-trial yield and residual scale so that the
decision rule can be exercised under known truth. Empirical protocols should
not copy the anchor mechanically. They must estimate the label-blind residual
variance, expected retention yield, participant heterogeneity, delivery error
and preprocessing losses in their own acquisition regime before declaring a
resolution floor or retained-trial target.

For a future confirmatory human-EEG implementation, including any later
Registered Report implementation, these feasibility quantities
should be estimated with the same locked endpoint definition and causal
preprocessing planned for confirmation. For example, a terminal CNV-like implementation
would first declare the single-trial endpoint, such as the mean signed
slow-potential amplitude over a fronto-central cluster in a terminal
pre-assignment window closing at \(t_1\)
\citep{Walter1964,Brunia2012,NobreVanEde2018}),
the fronto-central electrode cluster, baseline interval,
pre-endpoint averaging window, rereferencing, filter family, artefact rules,
minimum usable-trial rule, comparator class and fold structure without access to
confirmatory assigned-delay labels. A label-blind calibration dataset would then
estimate the single-trial residual scale, retained-trial yield and
participant-level slope variability after the frozen forward-only comparator.
Those estimates determine whether the declared resolution floor is empirically
resolvable and whether the proposed participant and trial counts can distinguish
the six decision outcomes. Calibration may tune feasibility quantities; it may
not optimise the endpoint, comparator or exclusion rules against delay-ordered
confirmatory residuals.

\Cref{tab:si-power} reports planning power for the injected
resolution-boundary effect as a function of retained trials per bin and the
relative residual-noise setting \(\nu_\varepsilon\), the synthetic residual-noise
multiplier used in the planning grid. Here \(n_{\mathrm{rep}}\) denotes retained
trials per assigned-delay bin and \(N_{\mathrm{ret}}\) denotes the corresponding
total retained trials across the five-bin synthetic support. With moderate
residual noise, the benchmark reaches approximately \(80\%\) power by about five
retained trials per bin, corresponding to
\(N_{\mathrm{ret}}\approx25\). Under the more conservative
\(\nu_\varepsilon=0.50\) setting, the same operating point requires about twenty
retained trials per bin, corresponding to \(N_{\mathrm{ret}}\approx100\). These
are synthetic retained-sample requirements at the benchmark anchor. They are not
participant-level EEG guarantees and do not replace empirical pilot estimates.

\begin{table}[t]
\centering
\caption{Planning power for the injected resolution-boundary effect at the
synthetic anchor (\(T_0=20\,\mathrm{ms}\), five bins). Values are illustrative
planning targets from the locked Monte Carlo benchmark, not empirical results.}
\label{tab:si-power}
\footnotesize
\begin{tabular}{lccc}
\toprule
Retained trials/bin \(n_{\mathrm{rep}}\) & \(N_{\mathrm{ret}}\) &
Power (\(\nu_\varepsilon=0.30\)) & Power (\(\nu_\varepsilon=0.50\)) \\
\midrule
5  & 25  & \(\approx 0.80\) & \(\approx 0.45\) \\
10 & 50  & \(\approx 0.95\) & \(\approx 0.63\) \\
20 & 100 & \(>0.99\)        & \(\approx 0.80\) \\
\bottomrule
\end{tabular}
\end{table}

The power table should be read together with the resolution-floor formula. For
fixed assigned-delay support, the slope resolution improves with lower
label-blind residual scale and larger retained-trial yield, and worsens when
participant-level residual variability is high or retention is poor. Increasing
the assigned-delay support can improve slope leverage, but it also increases
the burden on the forward-only comparator because ordinary foreperiod, hazard
and temporal-expectation structure have more room to operate. The synthetic
anchor therefore demonstrates pipeline behaviour at one declared operating
point; it does not remove the need for a prospective empirical design
calculation.

\subsection{Support as a design variable}
\label{sec:si-support}

The assigned-delay support is itself a design variable. A wider support
increases \(\sigma_\tau^2\), giving the slope estimator more leverage, but it
also lengthens the interval over which ordinary forward-accessible temporal
structure can influence the endpoint. That larger interval increases the
burden on the comparator and on the implementation audits. A narrower support
keeps the contrast closer to the endpoint and reduces ordinary foreperiod and
hazard variation, but it also drives the assigned-delay variance toward zero
and makes any slope harder to resolve.

The support should therefore be fixed prospectively in the label-blind
qualification stage. It should not be widened, narrowed, rebinned or shifted
after inspecting delay-ordered residuals. The benchmark anchor uses an origin-coded synthetic support
\([0,20]\,\mathrm{ms}\) because it gives a short-horizon coordinate on which the
randomisation boundary, resolution floor, collider diagnostic and audit structure
can be tested transparently. This synthetic origin does not define the empirical
minimum post-assignment latency, which must be strictly positive, logged and fixed
before delay-label access. Other empirical regimes may justify a
different support, but the same principle applies: the support must be chosen
before delay-labelled endpoint residuals are analysed, and the comparator must
be capable of absorbing the ordinary forward-accessible structure implied by
that support.

\subsection{Reporting per benchmark}
\label{sec:si-benchmark-reporting}

For each generator and grid point, the benchmark report should include the
declared \(T_0\), the assigned-delay bin grid, the number of Monte Carlo
datasets, the participant count, the retained-trial yield, the residual scale,
the resolution floor, and the classifier version or run hash. The primary
operating quantities are the mutually exclusive final outcome rates:
supported residual, forward-only adequate null, diagnostic failure,
selection-limited result, opposite-direction departure and inconclusive result.
These outcome rates should be generated by the executable classifier rather than
reconstructed by hand.

Diagnostic columns should be reported separately from final outcomes. For
leakage and delivery scenarios, report temporal-leakage, delivery and
assignment-balance audit firing rates. For selection scenarios, report retention
audit firing rates, selection-gate pass or failure rates, and the
endpoint-by-delay collider diagnostic. For the adversarial forward-only null,
report false-support rate, route-specific calibration pass rate,
resolution-pass rate and the distribution of participant-level slopes and
bounds. When the adversarial generator includes carryover, this calibration is
the sequential martingale/e-value route rather than frozen-array
randomisation. For material endpoint-level departure generators, report the route-, direction-
and magnitude-indexed false-adequacy rate
\(\widehat{\mathrm{FA}}_{r,d}(\delta)\), defined as the fraction of datasets
classified as forward-only adequate despite a genuine endpoint-level
assigned-delay slope of magnitude \(\delta\). Pointwise Wilson intervals are
descriptive. Certification uses a one-sided, Bonferroni-adjusted
Clopper--Pearson upper bound over the complete declared family of route,
direction and magnitude cells. For each route and direction, the candidate
magnitude is evaluated by the maximum simultaneous upper bound at that
magnitude and all larger evaluated magnitudes. A cell passes only when both its
false-adequacy point estimate and that upper-tail familywise envelope are at or
below \(p_{\mathrm{FA,max}}=0.05\). The report also records whether the certified
boundary equals the smallest evaluated magnitude, because that would indicate
lower-grid censoring. For the opposite-direction generator, report the
opposite-direction classification rate, the corresponding false-adequacy curve,
and verify that positive departures are not counted as directional support.

Participant-level estimability should be reported as its own diagnostic layer.
For every scenario, report the eligible-participant count, the estimable
participant count, the non-estimable participant fraction, the reasons for
non-estimability, the retained assigned-delay leverage among estimable
participants, and whether the participant-level estimability sensitivity bound
passes. For injected-residual and opposite-direction scenarios, report the
fraction of datasets in which participant-level estimability selection is
conclusion-changing under the declared sensitivity bound. For null scenarios,
report the false-support rate before and after the participant-level
estimability gate. These columns should be diagnostic columns, not replacements
for the mutually exclusive final outcomes.

One-sided \(95\%\) bootstrap upper bounds are the confirmatory decision objects
for negative support. Two-sided intervals may be reported as descriptive
summaries, but they do not replace the declared one-sided decision rule.
Similarly, route-specific inferential values should be reported as
assignment-calibrated diagnostics for the slope, not as sufficient evidence for
support. A supported classification requires the full non-compensatory
intersection: route-specific inferential evidence, magnitude resolution,
sufficient eligible and estimable participant count, implementation integrity,
frozen-residual construction and frozen-comparator independence qualification,
retained-sample qualification under \(\mathrm{(R3^\star)}\),
collider-diagnostic clearance, trial-level selection-gate clearance and
participant-level estimability-gate clearance.

All reported operating characteristics should be interpreted as realised Monte
Carlo behaviour at the declared synthetic operating point. A zero count, such as
\(0/1200\) support under a null scenario, means that no support occurred in that
run; it does not prove that the underlying support probability is exactly zero.
Conversely, a high injected-residual support rate demonstrates sensitivity to
the declared synthetic injection; it does not predict empirical EEG power
without a separate label-blind design calculation.

\subsection{Realised operating characteristics}
\label{sec:si-operating}

This subsection reports the realised operating characteristics of the locked
Level~II-A pipeline under known synthetic generators. The active parent run is
\LevelIIARunHash, with \(M=\LevelIIAM\) datasets per scenario, \(P=24\)
participants, 24 planned trials per assigned-delay bin, \(R=999\)
assignment-isolation draws where that route is used and \(B=999\) participant
bootstrap resamples. The assigned-delay support is \([0,20]~\mathrm{ms}\). The
adversarial carryover scenario uses the sequential e-value route. The other
canonical scenarios use assignment isolation. The collider stress test uses
\(\kappa=1\); the remaining canonical scenarios use \(\kappa=2\).

The machine-readable parent summary is
\path{outputs/0cd4cac11153c546/summary/operating_characteristics.csv}.
The two generated tables separate design information from decision behaviour.
\Cref{tab:si-oc-design} reports the generator, planned trials, realised
resolution floor, injected slope and slope summaries. Panel~A of
\Cref{tab:si-oc-outcomes} reports diagnostic and qualification rates; these
columns can overlap and are not intended to sum to one. Panel~B reports the six
mutually exclusive final outcome classes, whose counts sum to \(M\) within each
scenario.

In Panel~A, ``rand. pass'' is the route-specific negative-direction inferential
pass indicator: a plus-one randomisation value for assignment isolation and the
e-value-derived threshold for the sequential route. Under a null generator, a
low value in this column is expected because negative-direction evidence should
rarely cross its route-specific threshold. ``Resol. pass'' records whether the
slope clears the dataset-specific resolution floor, ``comp. dis.'' records
route-component disagreement that is sent to the inconclusive class, and
``est. block'' records the rate at which participant-level estimability
qualification is conclusion-changing. The audit, selection-gate and
estimability columns are therefore diagnostic layers; Panel~B gives the final
classifier output after the non-compensatory hierarchy has been applied.

\begin{table}[t]
\centering
\caption{Benchmark design and slope-resolution summaries for the locked Level II-A pipeline on simulated data. Rates and decision outcomes are reported separately in Table~\ref{tab:si-oc-outcomes}. Run hash \texttt{0cd4cac11153c546}; $M=1200$ datasets per scenario; $P=24$ participants; assigned-delay support $[0,20]$ ms; $\sigma_{\mathrm{resid}}=1\,\mu$V is the nominal generator residual-noise scale; the displayed $\beta_{\min}$ is recomputed for each scenario from the label-blind realised residual scale, retained-trial yield and effective retained assigned-delay leverage through the locked materiality formula, so its variation across rows is expected; not human EEG.}
\label{tab:si-oc-design}
\footnotesize
\setlength{\tabcolsep}{3pt}
\resizebox{\linewidth}{!}{%
\begin{tabular}{@{}llrrrrrr@{}}
\toprule
Scenario & Generator & $n$/bin & $\beta_{\min}$ & $\beta^{\mathrm{inj}}$ & Mean $\widehat{\beta}_\tau$ & Med. UCB & Med. $N_{\mathrm{est}}$ \\
\midrule
clean\_null & Forward-only null & 24 & 28.8 & 0 & 0.000 & 5.0 & 24.0 \\
injected\_residual & Negative endpoint-level residual & 24 & 31.4 & -60.0 & -60.1 & -53.5 & 24.0 \\
leakage & Forward-only null with temporal leak & 24 & 33.6 & 0 & -84.9 & -79.9 & 24.0 \\
selection\_standard & Forward-only null with monotone selection & 24 & 33.4 & 0 & -37.6 & -31.5 & 21.0 \\
collider\_selection & Forward-only null with endpoint-by-delay collider & 24 & 17.5 & 0 & -35.0 & -23.1 & 8.0 \\
adversarial\_null & Adversarial forward-only null & 24 & 33.8 & 0 & 0.078 & 6.2 & 24.0 \\
opposite\_direction & Positive endpoint-level residual & 24 & 31.4 & 60.0 & 60.0 & 66.5 & 24.0 \\
\bottomrule
\end{tabular}%
}
\end{table}

\begin{table}[t]
\centering
\caption{Diagnostic rates and mutually exclusive decision outcomes for the locked Level II-A pipeline on simulated data. Design constants and slope summaries are reported in Table~\ref{tab:si-oc-design}. Run hash \texttt{0cd4cac11153c546}; $M=1200$ Monte Carlo datasets per scenario. Panel~A reports diagnostic and qualification rates. Panel~B reports exact counts with rates in parentheses for the six mutually exclusive outcomes. The negative tail is the sole confirmatory level-$\alpha$ hypothesis; the positive tail is a prespecified diagnostic. Component disagreement is routed to the inconclusive class. $^{a}$The scalar selection gate is evaluated only for a resolved material departure; the displayed rate is conditional on applicability. Not human EEG.}
\label{tab:si-oc-outcomes}
\scriptsize
\setlength{\tabcolsep}{3.5pt}
\textbf{A. Diagnostic and qualification rates}\par\smallskip
\resizebox{\linewidth}{!}{%
\begin{tabular}{@{}lrrrrrrrr@{}}
\toprule
Scenario & rand. pass & resol. pass & comp. dis. & reten. fire & leak fire & gate pass$^{a}$ & collider fire & est. block \\
\midrule
clean\_null & 0.043 & 0.000 & 0.098 & 0.000 & 0.000 & n/a & 0.008 & 0.000 \\
injected\_residual & 1.000 & 1.000 & 0.000 & 0.002 & 0.000 & 1.000 & 0.008 & 0.000 \\
leakage & 1.000 & 1.000 & 0.000 & 0.002 & 1.000 & 1.000 & 0.008 & 0.000 \\
selection\_standard & 1.000 & 0.295 & 0.705 & 1.000 & 0.003 & 1.000 & 1.000 & 0.941 \\
collider\_selection & 0.998 & 0.730 & 0.268 & 0.009 & 0.001 & 0.993 & 1.000 & 0.998 \\
adversarial\_null & 0.001 & 0.000 & 0.002 & 0.002 & 0.001 & n/a & 0.008 & 0.000 \\
opposite\_direction & 0.000 & 0.000 & 0.000 & 0.001 & 0.001 & 1.000 & 0.006 & 0.000 \\
\bottomrule
\end{tabular}%
}

\medskip
\textbf{B. Mutually exclusive decision outcomes}\par\smallskip
\resizebox{\linewidth}{!}{%
\begin{tabular}{@{}lrrrrrr@{}}
\toprule
Scenario & support & sel.-lim. & diag. fail & null & opp. diag. & inconcl. \\
\midrule
clean\_null & 0/1200 (0.000) & 9/1200 (0.008) & 0/1200 (0.000) & 1073/1200 (0.894) & 0/1200 (0.000) & 118/1200 (0.098) \\
injected\_residual & 1189/1200 (0.991) & 11/1200 (0.009) & 0/1200 (0.000) & 0/1200 (0.000) & 0/1200 (0.000) & 0/1200 (0.000) \\
leakage & 0/1200 (0.000) & 0/1200 (0.000) & 1200/1200 (1.000) & 0/1200 (0.000) & 0/1200 (0.000) & 0/1200 (0.000) \\
selection\_standard & 0/1200 (0.000) & 1196/1200 (0.997) & 4/1200 (0.003) & 0/1200 (0.000) & 0/1200 (0.000) & 0/1200 (0.000) \\
collider\_selection & 0/1200 (0.000) & 1199/1200 (0.999) & 1/1200 (0.001) & 0/1200 (0.000) & 0/1200 (0.000) & 0/1200 (0.000) \\
adversarial\_null & 0/1200 (0.000) & 12/1200 (0.010) & 1/1200 (0.001) & 1185/1200 (0.988) & 0/1200 (0.000) & 2/1200 (0.002) \\
opposite\_direction & 0/1200 (0.000) & 8/1200 (0.007) & 1/1200 (0.001) & 0/1200 (0.000) & 1191/1200 (0.993) & 0/1200 (0.000) \\
\bottomrule
\end{tabular}%
}
\end{table}

The clean anchor produced no supported outcomes
(\(\LevelIIAAnchorSupportCountRate\)). It was classified as forward-only
adequate in \(\LevelIIAAnchorAdequateCountRate\) datasets, inconclusive in
\(\LevelIIAAnchorInconclusiveCountRate\), and selection-limited in
\(\LevelIIAAnchorSelectionLimitedCountRate\); no diagnostic failure occurred.
These inconclusive and selection-limited outcomes are a conservative
decision-yield cost, not a false-positive rate. The
adversarial forward-only null also produced no support
(\(\LevelIIAAdversarialSupportCountRate\)). Under the sequential route it
yielded \(\LevelIIAAdversarialAdequateCountRate\) forward-only-adequate,
\(\LevelIIAAdversarialInconclusiveCountRate\) inconclusive,
\(\LevelIIAAdversarialSelectionLimitedCountRate\) selection-limited and
\(\LevelIIAAdversarialDiagnosticFailureCountRate\) diagnostic-failure outcomes.
These rare unqualified outcomes are distinct from false support.

The injected \(-60\,\mu\mathrm{V\,s^{-1}}\) residual was supported in
\(\LevelIIAInjectedSupportCountRate\) datasets. The remaining
\(\LevelIIAInjectedSelectionLimitedCountRate\) datasets were selection-limited,
and none was classified as forward-only adequate. The positive injection was
classified as an opposite-direction departure in
\(\LevelIIAOppositeOppositeDirectionCountRate\), with
\(\LevelIIAOppositeSelectionLimitedCountRate\) selection-limited and
\(\LevelIIAOppositeDiagnosticFailureCountRate\) diagnostic-failure outcomes.
It was never classified as directional support or as forward-only adequate.
Thus the two material-injection scenarios distinguish sensitivity to the
registered negative signature from detection of a material association in the
wrong direction.

The temporal-leakage generator fired the leakage audit in every dataset.
All datasets were therefore diagnostic failures:
\(\LevelIIALeakageDiagnosticFailureCountRate\).
The standard-selection generator fired the retention audit in every dataset
and produced \(\LevelIIASelectionSelectionLimitedCountRate\)
selection-limited and \(\LevelIIASelectionDiagnosticFailureCountRate\)
diagnostic-failure outcomes. In that scenario, participant-level estimability
qualification was conclusion-changing in \(0.941\) of datasets, consistent with
the deliberate loss of usable assigned-delay leverage.

The endpoint-by-delay collider generated a mean retained-sample slope of
\(-35.0\,\mu\mathrm{V\,s^{-1}}\), median upper bound
\(-23.1\,\mu\mathrm{V\,s^{-1}}\), and resolution-pass rate \(0.730\), while the
marginal retention audit fired in only \(0.009\) of datasets. Conditional on
applicability, the scalar gate passed in \(0.993\). The clustered interaction
diagnostic fired in every dataset, and participant-level estimability
qualification was conclusion-changing in \(0.998\). The final classifier
therefore returned \(\LevelIIAColliderSelectionLimitedCountRate\)
selection-limited outcomes, one diagnostic failure and no support. This is the
intended separation between marginal retention checks and the dedicated
endpoint-by-delay collider guard.

The canonical clean and adversarial rows cannot by themselves isolate a
generator effect because they use different valid inference routes. The
auxiliary route-matched experiment makes that comparison explicit. Its two
clean rows use identical generated datasets replicate by replicate. Clean
sequential adequacy exceeded clean assignment-isolation adequacy by \(+0.101\),
with \(95\%\) interval \([+0.083,+0.119]\). Under the common sequential route,
adversarial adequacy exceeded clean adequacy by \(+0.007\), with interval
\([-0.001,+0.015]\). The larger canonical clean-versus-adversarial difference
is therefore primarily attributable to inference route rather than to the
adversarial generator. The adversarial assignment-isolation cell is excluded by
design because carryover violates endpoint-array invariance. Support and
opposite-direction counts were zero in all three auxiliary cells and are
omitted from the table body. This auxiliary experiment has its own identifier
and does not replace the canonical seven-scenario counts.

\begin{table}[t]
\centering
\caption{Validity-matched null-generator and inference-route comparison. The two clean rows use identical generated datasets replicate by replicate. The full adversarial carryover generator is evaluated only under the sequential e-value route because assignment isolation requires endpoint-array invariance. Outcome cells report counts and rates from \(M=1200\) datasets; support and opposite-direction counts were zero in all three cells. Contrast intervals use the locked paired or independent procedure appropriate to each comparison.}
\label{tab:si-route-matched-null-comparison}
\scriptsize
\setlength{\tabcolsep}{2.5pt}
\resizebox{\linewidth}{!}{%
\begin{tabular}{@{}llrrrrr@{}}
\toprule
Generator & Route & Adequate & Inconclusive & Selection-limited & Diagnostic failure & Component disagreement \\
\midrule
Clean & assignment isolation & 1064/1200 (0.887) & 126/1200 (0.105) & 10/1200 (0.008) & 0/1200 (0.000) & 128/1200 (0.107) \\
Clean & sequential e-value & 1185/1200 (0.988) & 5/1200 (0.004) & 10/1200 (0.008) & 0/1200 (0.000) & 5/1200 (0.004) \\
Adversarial & sequential e-value & 1193/1200 (0.994) & 2/1200 (0.002) & 4/1200 (0.003) & 1/1200 (0.001) & 2/1200 (0.002) \\
\bottomrule
\end{tabular}%
}

\medskip
\begin{tabular}{@{}lrrp{0.42\textwidth}@{}}
\toprule
Contrast & Estimate & 95\% interval & Interpretation \\
\midrule
Clean sequential $-$ clean assignment & +0.101 & [+0.083, +0.119] & Paired route contrast on identical clean datasets \\
Adversarial sequential $-$ clean sequential & +0.007 & [-0.001, +0.015] & Generator contrast under the common sequential route \\
\bottomrule
\end{tabular}
\end{table}

Operating rates are realised Monte Carlo rates and retain binomial uncertainty.
A zero count means that no event occurred in the realised run; it does not set
the underlying probability to zero. Pointwise Wilson intervals for selected
canonical rates, including the displayed \(\pm60\,\mu\mathrm{V\,s^{-1}}\)
false-adequacy points, are generated directly from the active parent run.

\begin{table}[t]
\centering
\caption{Selected realised operating rates with pointwise Wilson 95\% intervals.}
\label{tab:si-oc-intervals}
\footnotesize
\begin{tabular}{lrrr}
\toprule
Rate & Count & Estimate & Wilson 95\% interval \\
\midrule
Anchor support & 0/1200 & 0.000 & [0.000,0.003] \\
Adversarial-null support & 0/1200 & 0.000 & [0.000,0.003] \\
Injected-residual support & 1189/1200 & 0.991 & [0.984,0.995] \\
Injected-residual false adequacy & 0/1200 & 0.000 & [0.000,0.003] \\
Leakage audit fire & 1200/1200 & 1.000 & [0.997,1.000] \\
Standard-selection retention audit fire & 1200/1200 & 1.000 & [0.997,1.000] \\
Standard-selection selection-limited classification & 1196/1200 & 0.997 & [0.991,0.999] \\
Collider diagnostic fire & 1200/1200 & 1.000 & [0.997,1.000] \\
Collider selection-limited classification & 1199/1200 & 0.999 & [0.995,1.000] \\
Opposite-direction classification & 1191/1200 & 0.993 & [0.986,0.996] \\
Opposite-direction false adequacy & 0/1200 & 0.000 & [0.000,0.003] \\
\bottomrule
\end{tabular}
\end{table}

The displayed false-adequacy rows are stored in
\path{outputs/0cd4cac11153c546/summary/false_adequacy_rates.csv}. The
route-general magnitude-indexed certificate is stored separately in
\path{outputs/0cd4cac11153c546/summary/adequacy_operating_characteristic.csv}.
Pointwise intervals in \Cref{tab:si-oc-intervals} are descriptive; the
familywise certification rule is defined next.

\subsection{Adequacy operating characteristic for affirmative-null qualification}
\label{sec:si-adequacy-oc}

This subsection asks a different question from recovery power. Recovery power
asks how often the pipeline supports a declared negative injection.
False-adequacy qualification asks how often the same pipeline would instead call
a dataset forward-only adequate when a genuine additive endpoint-level linear
assigned-delay injection is present. The certificate is conditional on that
evaluated injection family, inference route, nuisance grid and retained-sample
envelope; it does not qualify arbitrary nonlinear, local, time-varying or mixture
alternatives. A small false-adequacy rate is what makes an affirmative null
informative; failure to qualify at a smaller magnitude does not imply support,
but means that a forward-only-adequate classification is not yet reliably
discriminating at that magnitude.

False adequacy is indexed by inference route, direction and injected
magnitude. Let \(r\in\{\mathrm{AI},\mathrm{SEQ}\}\) denote assignment isolation
or sequential e-value inference, and let \(d\in\{-,+\}\). For the declared
magnitude grid \(\Delta=\{\delta_1,\ldots,\delta_K\}\), define
\[
\widehat{\mathrm{FA}}_{r,d}(\delta_k)
=
\frac{1}{M}\sum_{m=1}^{M}
\mathbf{1}\{C_{m,r,d}(\delta_k)=\text{forward-only adequate}\}.
\]
Inconclusive, selection-limited, diagnostic-failure and opposite-direction
outcomes do not count as forward-only adequacy.

The certification family contains all
\(\LevelIIAAdequacyFamilySize\) declared cells. Pointwise Wilson intervals are
descriptive. Let
\(U^{\mathrm{CP,B}}_{r,d}(\delta_k)\) denote the one-sided
Bonferroni-adjusted Clopper--Pearson upper bound for a cell, and define the
route-direction envelope
\[
U^{\mathrm{env}}_{r,d}(\delta_k)
=
\max_{j\geq k}
U^{\mathrm{CP,B}}_{r,d}(\delta_j).
\]
No monotonicity of \(\widehat{\mathrm{FA}}_{r,d}(\delta)\) is assumed.
The maximum over \(j\geq k\) protects against empirical reversals, while the
Bonferroni adjustment provides simultaneous coverage over the declared
\(\LevelIIAAdequacyFamilySize\)-cell family. A shape-constrained envelope
would define a different certificate and require separate qualification.

The route-specific certified boundary is
\[
\delta^\star_{r,d}
=
\min\left\{
\delta_k:
\widehat{\mathrm{FA}}_{r,d}(\delta_k)\leq p_{\mathrm{FA,max}},
\quad
U^{\mathrm{env}}_{r,d}(\delta_k)\leq p_{\mathrm{FA,max}}
\right\},
\]
with \(p_{\mathrm{FA,max}}=\LevelIIAAdequacyPFaMax\). The envelope prevents a
local pass at one magnitude from qualifying the route when a larger evaluated
magnitude fails. The lower-grid-censor flag records whether the first pass
occurs at the smallest magnitude tested.

\begin{table}[t]
\centering
\caption{Route-specific adequacy operating characteristic for the assignment-isolation route. Point intervals are descriptive Wilson intervals; certification uses the displayed one-sided, Bonferroni-adjusted Clopper--Pearson familywise upper-bound envelope.}
\label{tab:si-adequacy-assignment-isolation}
\footnotesize
\setlength{\tabcolsep}{3pt}
\begin{tabular}{llrrrrr}
\toprule
Direction & $\delta$ & False adequacy & Wilson 95\% & simultaneous UCB & envelope UCB & verdict \\
\midrule
negative & 5.0 & 567/1200 (0.472) & [0.4444,0.5008] & 0.5165 & 0.5165 & fail \\
negative & 10.0 & 105/1200 (0.087) & [0.0728,0.1048] & 0.1148 & 0.1148 & fail \\
negative & 15.0 & 3/1200 (0.003) & [0.0009,0.0073] & 0.0106 & 0.0106 & pass \\
negative & 20.0 & 0/1200 (0.000) & [0.0000,0.0032] & 0.0056 & 0.0056 & pass \\
negative & 30.0 & 0/1200 (0.000) & [0.0000,0.0032] & 0.0056 & 0.0056 & pass \\
negative & 40.0 & 0/1200 (0.000) & [0.0000,0.0032] & 0.0056 & 0.0056 & pass \\
negative & 50.0 & 0/1200 (0.000) & [0.0000,0.0032] & 0.0056 & 0.0056 & pass \\
negative & 60.0 & 0/1200 (0.000) & [0.0000,0.0032] & 0.0056 & 0.0056 & pass \\
negative & 75.0 & 0/1200 (0.000) & [0.0000,0.0032] & 0.0056 & 0.0056 & pass \\
negative & 90.0 & 0/1200 (0.000) & [0.0000,0.0032] & 0.0056 & 0.0056 & pass \\
positive & 5.0 & 602/1200 (0.502) & [0.4734,0.5299] & 0.5456 & 0.5456 & fail \\
positive & 10.0 & 109/1200 (0.091) & [0.0759,0.1084] & 0.1186 & 0.1186 & fail \\
positive & 15.0 & 4/1200 (0.003) & [0.0013,0.0085] & 0.0120 & 0.0120 & pass \\
positive & 20.0 & 0/1200 (0.000) & [0.0000,0.0032] & 0.0056 & 0.0056 & pass \\
positive & 30.0 & 0/1200 (0.000) & [0.0000,0.0032] & 0.0056 & 0.0056 & pass \\
positive & 40.0 & 0/1200 (0.000) & [0.0000,0.0032] & 0.0056 & 0.0056 & pass \\
positive & 50.0 & 0/1200 (0.000) & [0.0000,0.0032] & 0.0056 & 0.0056 & pass \\
positive & 60.0 & 0/1200 (0.000) & [0.0000,0.0032] & 0.0056 & 0.0056 & pass \\
positive & 75.0 & 0/1200 (0.000) & [0.0000,0.0032] & 0.0056 & 0.0056 & pass \\
positive & 90.0 & 0/1200 (0.000) & [0.0000,0.0032] & 0.0056 & 0.0056 & pass \\
\bottomrule
\end{tabular}
\end{table}

\begin{table}[t]
\centering
\caption{Route-specific adequacy operating characteristic for the sequential e-value route. Point intervals are descriptive Wilson intervals; certification uses the displayed one-sided, Bonferroni-adjusted Clopper--Pearson familywise upper-bound envelope.}
\label{tab:si-adequacy-sequential-evalue}
\footnotesize
\setlength{\tabcolsep}{3pt}
\begin{tabular}{llrrrrr}
\toprule
Direction & $\delta$ & False adequacy & Wilson 95\% & simultaneous UCB & envelope UCB & verdict \\
\midrule
negative & 5.0 & 1133/1200 (0.944) & [0.9297,0.9558] & 0.9622 & 0.9622 & fail \\
negative & 10.0 & 1035/1200 (0.863) & [0.8419,0.8808] & 0.8910 & 0.8910 & fail \\
negative & 15.0 & 783/1200 (0.652) & [0.6251,0.6789] & 0.6936 & 0.6936 & fail \\
negative & 20.0 & 381/1200 (0.318) & [0.2918,0.3444] & 0.3595 & 0.3595 & fail \\
negative & 30.0 & 14/1200 (0.012) & [0.0070,0.0195] & 0.0244 & 0.0244 & pass \\
negative & 40.0 & 0/1200 (0.000) & [0.0000,0.0032] & 0.0056 & 0.0056 & pass \\
negative & 50.0 & 0/1200 (0.000) & [0.0000,0.0032] & 0.0056 & 0.0056 & pass \\
negative & 60.0 & 0/1200 (0.000) & [0.0000,0.0032] & 0.0056 & 0.0056 & pass \\
negative & 75.0 & 0/1200 (0.000) & [0.0000,0.0032] & 0.0056 & 0.0056 & pass \\
negative & 90.0 & 0/1200 (0.000) & [0.0000,0.0032] & 0.0056 & 0.0056 & pass \\
positive & 5.0 & 1146/1200 (0.955) & [0.9417,0.9653] & 0.9711 & 0.9711 & fail \\
positive & 10.0 & 1029/1200 (0.858) & [0.8366,0.8761] & 0.8864 & 0.8864 & fail \\
positive & 15.0 & 770/1200 (0.642) & [0.6141,0.6683] & 0.6831 & 0.6831 & fail \\
positive & 20.0 & 354/1200 (0.295) & [0.2699,0.3214] & 0.3363 & 0.3363 & fail \\
positive & 30.0 & 12/1200 (0.010) & [0.0057,0.0174] & 0.0221 & 0.0221 & pass \\
positive & 40.0 & 0/1200 (0.000) & [0.0000,0.0032] & 0.0056 & 0.0056 & pass \\
positive & 50.0 & 0/1200 (0.000) & [0.0000,0.0032] & 0.0056 & 0.0056 & pass \\
positive & 60.0 & 0/1200 (0.000) & [0.0000,0.0032] & 0.0056 & 0.0056 & pass \\
positive & 75.0 & 0/1200 (0.000) & [0.0000,0.0032] & 0.0056 & 0.0056 & pass \\
positive & 90.0 & 0/1200 (0.000) & [0.0000,0.0032] & 0.0056 & 0.0056 & pass \\
\bottomrule
\end{tabular}
\end{table}

The assignment-isolation route first passes in both directions at
\[
\delta^\star_{\mathrm{AI},-}
=
\delta^\star_{\mathrm{AI},+}
=
\LevelIIAAssignmentIsolationNegativeCertifiedDelta
\,\mu\mathrm{V\,s^{-1}}.
\]
At \(15\,\mu\mathrm{V\,s^{-1}}\), the negative and positive envelope upper
bounds are \(0.0106\) and \(0.0120\), respectively. The
\(10\,\mu\mathrm{V\,s^{-1}}\) cells fail in both directions. The sequential
e-value route first passes in both directions at
\[
\delta^\star_{\mathrm{SEQ},-}
=
\delta^\star_{\mathrm{SEQ},+}
=
\LevelIIASequentialNegativeCertifiedDelta
\,\mu\mathrm{V\,s^{-1}}.
\]
At \(20\,\mu\mathrm{V\,s^{-1}}\), false adequacy is \(0.318\) in the negative
direction and \(0.295\) in the positive direction, with envelope upper bounds
\(0.3595\) and \(0.3363\). At \(30\,\mu\mathrm{V\,s^{-1}}\), the corresponding
rates are \(14/1200=0.012\) and \(12/1200=0.010\), with envelope upper bounds
\(0.0244\) and \(0.0221\). Neither route is lower-grid censored.

The certificate is therefore route-dependent. A single
\(\delta^\star_{\pm}\) would discard information required to interpret the
classifier. Assignment isolation qualifies false adequacy from
\(15\,\mu\mathrm{V\,s^{-1}}\), while the deployed sequential fold-mixture route
requires \(30\,\mu\mathrm{V\,s^{-1}}\) under the evaluated generators. This is a
false-adequacy resolution boundary for the complete classifier within the
evaluated additive endpoint-level linear-injection family, route, nuisance grid
and retained-sample envelope. It is not a
biological materiality threshold, a minimum detectable effect, or a claim that
smaller departures are absent. The dataset-specific \(\beta_{\min}\) remains the
magnitude floor used in the support decision; the route-specific
\(\delta^\star_{r,d}\) values qualify the affirmative-null classifier through
simulation. At \(60\,\mu\mathrm{V\,s^{-1}}\), all four route-direction cells
have false adequacy \(0/1200\), with simultaneous upper bound \(0.0056\).

\Cref{tab:si-collider-sweep} gives a finer view of the collider diagnostic.
Across \(\gamma=-0.5\) to \(-3.2\), the median manufactured slope grows from
\(-15.9\) to \(-38.0\,\mu\mathrm{V\,s^{-1}}\), while marginal retention
imbalance remains between \(0.025\) and \(0.036\). The retention-audit firing
rate ranges from \(0\) to \(0.020\), and the resolution-pass rate rises from
\(0.055\) to \(0.875\). The endpoint-by-delay interaction diagnostic fires in
every dataset at every evaluated strength. The final selection-limited rate is
\(1.000\) through \(\gamma=-2.0\) and \(0.990\) at the two strongest settings.

The sweep shows why the interaction diagnostic is not redundant with marginal
retention balance. At the principal collider operating point, the scalar gate
passes in \(0.993\) of applicable datasets, while the interaction diagnostic
still blocks the manufactured association.

\begin{table}[t]
\centering
\caption{Collider-selection scope sweep (simulated data). The manufactured retained-sample slope grows with collider strength while the marginal retention imbalance stays small; the endpoint-by-delay interaction diagnostic fires throughout.}
\label{tab:si-collider-sweep}
\footnotesize
\begin{tabular}{lrrrrrr}
\toprule
$\gamma$ & med.\ $\widehat{\beta}_\tau$ & marg.\ imbal. & reten.\ fire & resol.\ pass & interaction fire & sel.-lim. \\
\midrule
-0.500 & -15.9 & 0.025 & 0 & 0.055 & 1.0 & 1.0 \\
-0.900 & -24.9 & 0.028 & 0.005 & 0.225 & 1.0 & 1.0 \\
-1.4 & -31.1 & 0.030 & 0.010 & 0.515 & 1.0 & 1.0 \\
-2.0 & -34.6 & 0.033 & 0.020 & 0.730 & 1.0 & 1.0 \\
-2.6 & -36.7 & 0.034 & 0.020 & 0.780 & 1.0 & 0.990 \\
-3.2 & -38.0 & 0.036 & 0.020 & 0.875 & 1.0 & 0.990 \\
\bottomrule
\end{tabular}
\end{table}

\FloatBarrier

\subsection{Worked numeric example: one dataset through every decision object}
\label{sec:si-worked-example}

This subsection walks two representative simulated datasets through the
non-compensatory decision rule. Its purpose is explanatory rather than
evidential: the examples show how the frozen endpoint residuals, the
participant-level slope, the route-specific inferential calibration, the
participant-bootstrap upper bound, the resolution floor, the audit outcomes,
the selection gate and the final classifier combine in one concrete trace. The
examples illustrate the per-dataset classifier; the magnitude-indexed adequacy
operating characteristic above is the object used for affirmative-null
qualification. The numbers are outputs of the locked synthetic benchmark
pipeline; they are not empirical EEG results.

\providecommand{\LevelIIAWorkedRepSupported}{657}
\providecommand{\LevelIIAWorkedSeedSupported}{102}
\providecommand{\LevelIIAWorkedRepClean}{438}
\providecommand{\LevelIIAWorkedSeedClean}{101}
\providecommand{\LevelIIAWorkedP}{24}
\providecommand{\LevelIIAWorkedSigmaTau}{0.00708}
\providecommand{\LevelIIAWorkedSigmaBlind}{1.10}
\providecommand{\LevelIIAWorkedNbarRet}{95.4}
\providecommand{\LevelIIAWorkedBetaMin}{31.8}
\providecommand{\LevelIIAWorkedSlopeMean}{-60.1}
\providecommand{\LevelIIAWorkedSlopeSE}{3.7}

\paragraph{Design and locked constants.}
The worked examples are generated from the same frozen certified benchmark
output directory as \Cref{tab:si-oc-design,tab:si-oc-outcomes}, under run hash
\LevelIIARunHash, by \path{scripts/make_worked_example.py}. The representative
indices are read from
\path{outputs/0cd4cac11153c546/summary/representative_index.json}: the
supported injected-residual draw is replicate
\(\LevelIIAWorkedRepSupported\), with base seed
\(\LevelIIAWorkedSeedSupported\), and the clean-null draw is replicate
\(\LevelIIAWorkedRepClean\), with base seed \(\LevelIIAWorkedSeedClean\).
Headline decision objects are read from the frozen raw per-replicate rows as
per-draw values, not scenario averages. The bin means and participant slopes
are recomputed deterministically from the same representative replicate. The
trace therefore remains tied to the run, classifier hierarchy and outcome
columns used throughout \Cref{sec:si-operating}.

Both examples use \(P=\LevelIIAWorkedP\) participants and the assigned-delay
grid \(\tau_L\in\{0,5,10,15,20\}\,\mathrm{ms}\), with
\(\sigma_\tau=\LevelIIAWorkedSigmaTau\,\mathrm{s}\) on the second scale used by
the slope estimator. In the supported draw, \(\kappa=2\), the label-blind
residual scale is
\(\sigma^{\mathrm{blind}}_{\mathrm{resid}}=
\LevelIIAWorkedSigmaBlind\,\mu\mathrm{V}\), the retained-trial yield is
\(\bar n_{\mathrm{ret}}=\LevelIIAWorkedNbarRet\), and the registered
resolution floor is
\[
\beta_{\min}
=
\kappa\,
\frac{\sigma^{\mathrm{blind}}_{\mathrm{resid}}}
{\sigma_\tau\sqrt{\bar n_{\mathrm{ret}}}}
=
\LevelIIAWorkedBetaMin\,\mu\mathrm{V\,s^{-1}} .
\]
This floor is a resolvability threshold for the participant-level slope, not a
biological importance threshold and not a population \(p\)-value.

\paragraph{Supported injected-residual draw.}
The injected-residual generator adds a negative endpoint-level slope before the
comparator is fitted. The forward-only comparator is label-blind, the
cross-fitted residuals are frozen, and the full-benchmark decision rule is then
applied. The retained residual bin means in \Cref{tab:si-worked-bins} show the
delay-ordered residual pattern for this draw. The participant slopes in
\Cref{tab:si-worked-slopes} are the independent units entering the
equal-participant estimand, with mean
\(\widehat\beta_\tau=
\LevelIIAWorkedSlopeMean\,\mu\mathrm{V\,s^{-1}}\) and participant-level
standard error
\(\LevelIIAWorkedSlopeSE\,\mu\mathrm{V\,s^{-1}}\). As a descriptive
influence check, the leave-one-participant-out mean slope ranges from
\(-61.4\) to \(-58.6\,\mu\mathrm{V\,s^{-1}}\); no single participant
changes the sign or material conclusion in this representative draw. The five displayed bin means
are a transparent descriptive trace of the residual ordering; the registered
estimand and uncertainty calculation are based on the participant slopes, not
on treating the five aggregate bin means as independent observations.

\begin{table}[H]
\centering
\caption{Retained residual-endpoint means by assigned-delay bin for the representative supported injected-residual draw. Means and standard errors are on the frozen residual endpoint scale in \(\mu\mathrm{V}\).}
\label{tab:si-worked-bins}
\footnotesize
\setlength{\tabcolsep}{4pt}
\begin{tabular}{lccccc}
\toprule
Assigned delay \(\tau_L\) (ms) & 0 & 5 & 10 & 15 & 20 \\
Centred \(\Delta\tau\) (ms) & -10 & -5 & 0 & 5 & 10 \\
\midrule
Retained trials & 458 & 443 & 464 & 458 & 466 \\
Mean \(A^{\mathrm{resid}}_{\mathrm{pre}}\) & 0.63 & 0.21 & 0.06 & -0.29 & -0.61 \\
SE & 0.05 & 0.05 & 0.05 & 0.05 & 0.05 \\
\bottomrule
\end{tabular}
\end{table}

\begin{table}[H]
\centering
\caption{Participant slopes \(\widehat\beta_{\tau,p}\) for the representative supported injected-residual draw. Units are \(\mu\mathrm{V\,s^{-1}}\). The equal-participant mean is \(\widehat\beta_\tau=-60.1\).}
\label{tab:si-worked-slopes}
\footnotesize
\begin{tabular}{rrrrrrrr}
\toprule
-95.4 & -84.0 & -34.9 & -30.6 & -53.4 & -60.6 & -46.7 & -67.0 \\
-52.8 & -88.1 & -77.2 & -49.3 & -65.2 & -68.5 & -48.8 & -84.2 \\
-73.4 & -61.6 & -48.5 & -32.7 & -41.2 & -46.3 & -52.9 & -79.7 \\
\bottomrule
\end{tabular}
\end{table}

For this draw, the negative-direction randomisation value is \(0.001\), the
bootstrap-\(t\) upper bound is
\(-54.0\,\mu\mathrm{V\,s^{-1}}\), and
\(-\beta_{\min}=-31.8\,\mu\mathrm{V\,s^{-1}}\). The inferential criterion and
the magnitude-resolution criterion therefore both pass. All
\(\LevelIIAWorkedP\) participants are estimable, component disagreement and the
estimability conclusion-change indicator are zero, the support-blocking audits
do not fire, and the selection gate passes. The executable classifier returns
\texttt{supported}.

\paragraph{Clean-null draw and contrast.}
The clean-null draw uses the same endpoint construction, comparator,
residual-freezing rule, inference, audits and classifier, but its generator
contains no assigned-delay residual. Its finite-sample slope is
\(-0.2\,\mu\mathrm{V\,s^{-1}}\), with negative-direction randomisation value
\(0.474\), bootstrap-\(t\) upper bound
\(4.2\,\mu\mathrm{V\,s^{-1}}\), and dataset-specific resolution floor
\(28.2\,\mu\mathrm{V\,s^{-1}}\). It therefore fails the directional support
rule while retaining a valid, fully estimable analysis sample. The side-by-side
trace in \Cref{tab:si-worked-decision} shows the same locked pipeline returning
\texttt{supported} for the injected-residual draw and
\texttt{forward\_\allowbreak only\_\allowbreak adequate} for the clean-null
draw. The contrast is illustrative of classifier mechanics; it does not replace
the \(M=\LevelIIAM\) operating-characteristic results above.

\begin{table}[H]
\centering
\caption{Every decision object available in the frozen per-replicate benchmark rows for the representative supported injected-residual draw and the representative clean-null draw. Slope, bound and floor units are \(\mu\mathrm{V\,s^{-1}}\).}
\label{tab:si-worked-decision}
\footnotesize
\setlength{\tabcolsep}{3pt}
\begin{tabularx}{\textwidth}{@{}
    >{\raggedright\arraybackslash}p{4.0cm}
    >{\raggedright\arraybackslash}X
    >{\raggedright\arraybackslash}p{3.0cm}
    >{\raggedright\arraybackslash}p{3.0cm}
@{}}
\toprule
Decision object & Criterion & Supported draw & Clean-null draw \\
\midrule
Scenario & recorded generator & injected residual & clean null \\
Replicate & representative index & \(657\) & \(438\) \\
Decision & executable classifier & \texttt{supported} &
\texttt{forward\_\allowbreak only\_\allowbreak adequate} \\
Estimate \(\widehat\beta_\tau\) & reported & \(-60.1\) & \(-0.2\) \\
Bootstrap-\(t\) \(\mathrm{UCB}_{0.95}\) & \(<-\beta_{\min}\) & \(-54.0\) & \(4.2\) \\
Bootstrap-\(t\) \(\mathrm{LCB}_{0.95}\) & reported & \(-67.4\) & \(-4.8\) \\
Resolution floor \(\beta_{\min}\) & registered formula & \(31.8\) & \(28.2\) \\
Randomisation \(p\), negative direction & \(\le 0.05\) & \(0.001\) & \(0.474\) \\
Randomisation \(p\), positive direction & prespecified diagnostic & \(1.000\) & \(0.527\) \\
Estimable participants & \(N_{\mathrm{est}}\) & \(24\) & \(24\) \\
Non-estimable fraction & reported & \(0.000\) & \(0.000\) \\
Component disagreement & routes to inconclusive & \(0\) & \(0\) \\
Estimability conclusion-changing & selection qualification & \(0\) & \(0\) \\
Support indicator & final outcome class & \(1\) & \(0\) \\
Selection-limited indicator & final outcome class & \(0\) & \(0\) \\
Diagnostic-failure indicator & final outcome class & \(0\) & \(0\) \\
Opposite-direction indicator & final outcome class & \(0\) & \(0\) \\
Null indicator & final outcome class & \(0\) & \(1\) \\
Inconclusive indicator & final outcome class & \(0\) & \(0\) \\
\bottomrule
\end{tabularx}
\end{table}

\section{Reproducibility checklist}
\label{sec:si-reproducibility}

\begin{enumerate}[leftmargin=1.6em,itemsep=3pt]

\item \textbf{Locked before unblinding.}
Endpoint specification
(\Cref{tab:si-endpoint-lock}); comparator covariates and model family
(\Cref{tab:si-comparator}); cross-fitting folds; frozen comparator object
\(\mathcal{G}_{\mathrm{frz}}\); participant eligibility and estimability rules;
minimum retained-trial and route-specific denominator-leverage requirements;
the resolution-floor rule \(\beta_{\min}\) and its calculation from the
label-blind residual scale, retained yield and route-specific effective
assignment leverage in \eqref{eq:si-beta-min}; \(N_{\min}\);
assignment-isolation randomisation-replicate count \(R\) and confirmatory level
\(\alpha\); the sequential martingale/e-value grid where that route is used;
bootstrap settings; the selection-model class; and the conditional
selection-sensitivity gate in \eqref{eq:si-selection-gate}.

\item \textbf{Operational scheduler record.}
Scheduler \(P_{\mathrm{sch}}(\cdot\mid\mathcal{R})\) and, where applicable,
current-trial conditional scheduler laws are recorded. The record also includes
declared scheduler inputs and stratum structure, assignment seeds, logs and
timestamps, the concealment mechanism, records linking assignment to delivery,
and the plan for implementation swaps. These records qualify the scheduler
mechanism in (R1); they do not establish the full-filtration factorisation
\eqref{eq:si-exogeneity}.

\item \textbf{Pipeline integrity.}
Strictly causal preprocessing; temporal-leakage audit result with tolerance
\(\delta_{\mathrm{leak}}\); frozen residual array
\eqref{eq:si-residual}; source-data fingerprint; and confirmation that neither
the comparator nor the residual array is refitted inside the assignment-test
loop.

\item \textbf{Delivery, retention and estimability.}
Assigned-versus-measured latency record; stratum-specific delivery, rejection,
missingness and inclusion; retained-versus-excluded endpoint summaries recorded
before assigned-delay labels are accessed; marginal retention audit;
endpoint-dependent inclusion checks used to qualify
\(\mathrm{(R3^\star)}\); participant-level retained-delay leverage; the number
of eligible and estimable participants; and the declared rule for datasets in
which the population slope is not estimable.

\item \textbf{Inference.}
Calibration route, either assignment isolation or sequential
martingale/e-value, together with its route-specific diagnostics;
the assignment-isolation randomisation value in \eqref{eq:si-pvalue} where
that route is used; the e-value-derived sequential quantity
\(p_{\mathrm{seq}}\) where the carryover-sensitive route is used; the
participant-bootstrap \(\UCB\) in \eqref{eq:si-ucb}; and the BCa and
\(t\)-interval sensitivity analyses. The negative tail is the sole
confirmatory level-\(\alpha\) hypothesis. The positive tail is retained as a
prespecified opposite-direction diagnostic: it blocks a forward-only-adequate
classification but is not an alternative route to Level~II-A support or to a
confirmatory class-insufficiency claim. The unadjusted committed-endpoint slope, its sign and its relation to the
frozen-residual slope are reported as an adjustment-sensitivity diagnostic that
can reveal substantial attenuation or disagreement but does not guarantee that
every form of shared-structure absorption will be detected. It does not enter
the executable classifier.

\item \textbf{Decision.}
The non-compensatory hierarchy comprising operational scheduler and
implementation audits, retained-sample qualification, participant estimability,
the route-specific inferential component, the participant-level magnitude bound,
\(N_{\min}\), the conditional selection-sensitivity gate and the
endpoint-by-delay collider diagnostic. Record the resulting mutually exclusive
outcome class: directional support, forward-only adequate, diagnostic failure,
selection-limited, opposite-direction departure or inconclusive. Audit failure,
failed retained-sample qualification, failed estimability requirements and a
collider-diagnostic fire are support-blocking according to the declared
hierarchy. The scalar trial-level selection-sensitivity gate is applicable only
to an already resolved material departure; failure of that applicable gate
produces a selection-limited outcome and blocks support, but the gate is not a
requirement for a clean affirmative null. Failure to establish the selected
route's validity condition is routed to the inconclusive class. Disagreement
between the inferential and magnitude components is likewise routed to the
inconclusive class rather than allowed to fall through to forward-only adequate.

\item \textbf{Benchmarks.}
Benchmarks include the generator code and planning grid
(\Cref{tab:si-power}), clean-null and adversarial-null false-support rates,
their forward-only-adequate and inconclusive rates, injected-residual recovery,
audit-firing rates, participant estimability, qualification of retained
leverage, and calibration of the selection gate. They also include the
endpoint-by-delay collider scope test and the sweep over collider strength
(\Cref{tab:si-oc-outcomes,tab:si-collider-sweep}), opposite-direction
classification, the route-matched null comparison
(\Cref{tab:si-route-matched-null-comparison}), and both route-specific adequacy
operating characteristics
(\Cref{tab:si-adequacy-assignment-isolation,tab:si-adequacy-sequential-evalue}).
Record \(\delta^\star_{r,-}\) and \(\delta^\star_{r,+}\) for each qualified
inference route rather than collapsing them to one universal magnitude.

\item \textbf{Replication.}
For programme-level support, require an independent sample analysed with the
same locked endpoint, comparator family, frozen-residual construction,
resolution-floor rule, participant-estimability requirements, inference route,
retained-sample qualification logic, decision hierarchy and audit battery.
Numerical values that are prospectively recomputed from label-blind residual
scale, retained yield or retained leverage are recalculated by the same locked
rule rather than copied from the original dataset.

\item \textbf{Code, certified outputs and permanent archive.}
The executable benchmark package (source code, configuration and environment
files, run scripts, tests, generated tables, figure source data, raw Monte Carlo
rows, summaries and metadata) is maintained in the public development repository
at \url{https://github.com/geosop/LEVEL-IIA} and permanently archived on Zenodo.
The version-specific DOI for the release used here is
\url{https://doi.org/10.5281/zenodo.21887583}; the concept DOI representing the
software across releases is \url{https://doi.org/10.5281/zenodo.21804380}.
Reproducibility claims in this manuscript refer to the version-specific archive,
not to the moving development state of the repository.
The archived release is Git tag \texttt{v1.2.1}, release commit
\texttt{ea2e4f89583b703780caf8ba07ffe228b31aa028}. Release~v1.2.1 is a
rendering/provenance correction release over the same certified benchmark:
the executable benchmark package version used to generate the certified run
remains \texttt{1.2.0}, and no Monte Carlo benchmark was rerun.
Within the archived release, the certified seven-scenario parent run is
\LevelIIARunHash, generated under source-fingerprint commit
\path{e45455e359646c4784b1d7b847ef44dd8f3499fd} and stored under
\path{outputs/0cd4cac11153c546/}. The release commit identifies the archived
repository snapshot and the source-fingerprint commit records the code state that
generated the certified run; the two provenance roles are distinct.
The parent run contains \(M=1200\) datasets per scenario, \(P=24\) participants,
24 planned trials per assigned-delay bin, \(R=999\) assignment-isolation draws
where that route is used and \(B=999\) participant-bootstrap resamples; the
adversarial carryover null uses the sequential e-value route. The mutually
exclusive manuscript counts are locked in
\path{manuscript/certified_run_counts.json}; the canonical parent summary is
\path{outputs/0cd4cac11153c546/summary/operating_characteristics.csv}, and the
displayed \(\pm60\,\mu\mathrm{V\,s^{-1}}\) false-adequacy points are in
\path{outputs/0cd4cac11153c546/summary/false_adequacy_rates.csv}.
The adequacy certificate \texttt{adequacy\_498657101acbb4e6} evaluates
\(\delta\in\{5,10,15,20,30,40,50,60,75,90\}\,\mu\mathrm{V\,s^{-1}}\) in both
directions and under both inference routes, giving 40 declared cells, with
certified boundaries of \(15\,\mu\mathrm{V\,s^{-1}}\) for assignment isolation and
\(30\,\mu\mathrm{V\,s^{-1}}\) for the sequential e-value route; its summary is
\path{outputs/0cd4cac11153c546/summary/adequacy_operating_characteristic.csv}.
The route-matched auxiliary experiment \texttt{route\_match\_1be69ec6cd081a58}
(generating commit \path{9c4fdbb14454cfb3587e3925ef0d5b20520cfef6}) has its
summary and contrasts under the parent run's auxiliary directories; that commit
applies only to the route-matched experiment, and the parent source-fingerprint
commit remains the provenance anchor for the seven-scenario run.
Manuscript-facing tables and Figure~2 are copied to \path{manuscript/tables/} and
\path{manuscript/figures/}; the README gives exact verification and
separate-output regeneration commands; historical hash-addressed directories are
provenance records only. The Zenodo archive contains the tagged repository
snapshot and committed benchmark artefacts but not the Perspective manuscript or
Supplementary Information source.
\end{enumerate}

\begingroup
\small
\setlength{\bibsep}{2pt}
\bibliographystyle{unsrtnat}
\bibliography{LeveLIIArefs}
\endgroup